\documentclass[letterpaper,USenglish,cleveref,numberwithinsect,thm-restate]{no-lipics-v2022}
\usepackage[utf8]{inputenc}
\usepackage{color}
\usepackage{csquotes}
\usepackage{xspace}
\usepackage{amsmath}
\usepackage{amssymb}
\usepackage{mathtools}
\usepackage{bbm}
\usepackage[linesnumbered, noend]{algorithm2e}
\usepackage[draft]{fixme}
\usepackage{tikz}
\usetikzlibrary{arrows,arrows.meta,decorations.pathmorphing}
\usepackage{bbding}
\usepackage{xcolor,pifont}
\usepackage{microtype}
\usepackage{verbatimbox}
\usepackage{MnSymbol}

\nolinenumbers

\newcommand{\floor}[1]{\lfloor{#1}\rfloor}

\newcommand{\set}[1]{\{#1\}}
\newcommand{\setof}[2]{\set{#1\colon\,#2}}

\newcommand{\Oh}{\ensuremath{\mathcal{O}}\xspace}
\newcommand{\addtilde}[1]{\ensuremath{\smash{\rlap{\raisebox{-0.2ex}{$\widetilde{\phantom{#1}}$}}#1}}\xspace}
\newcommand{\Ohtilde}{\addtilde{\Oh}}
\newcommand{\tOh}{\Ohtilde}

\newcommand{\RR}{\mathbb{R}\xspace}

\renewcommand{\emptyset}{\varnothing}
\newcommand{\ie}{i.e.\xspace}

\newcommand{\on}[1]{\operatorname{#1}}

\DeclareMathOperator{\polylog}{polylog}

\newcommand{\dd}{.\,.}
\newcommand{\fragmentco}[2]{[#1\dd #2)}

\newcommand{\fragmentcc}[2]{[#1\dd #2]}

\newcommand{\row}[2]{#1_1,\ldots,#1_{#2}}

\crefname{observation}{Observation}{Observations}
\crefname{hypothesis}{Hypothesis}{Hypotheses}

\newif\ifpatternlabels

\declaretheorem[sibling=theorem, style=plainstyle, name=Computer-Aided Lemma, refname={Computer-Aided Lemma,Computer-Aided Lemmas}, Refname={Computer-Aided Lemma,Computer-Aided Lemmas}]{calemma}
\declaretheorem[sibling=theorem, style=plainstyle, name=Computer-Aided Theorem, refname={Computer-Aided Theorem,Computer-Aided Theorems}, Refname={Computer-Aided Theorem,Computer-Aided Theorems}]{catheorem}

\DeclareMathOperator{\clemb}{clemb}
\DeclareMathOperator{\subw}{subw}
\DeclareMathOperator{\sharpw}{\#subw}
\newcommand{\somewidth}{\textup{-width}}
\newcommand{\mwidth}{\ensuremath{\textup{m}\somewidth}\xspace}
\newcommand{\mwidthH}[1]{\ensuremath{\textup{m}\somewidth_{#1}}}
\newcommand{\mpwidth}{\ensuremath{\textup{m}'\somewidth}\xspace}

\newcommand{\dswidth}[1]{\ensuremath{\textup{ds}_{#1}\somewidth}\xspace}

\newcommand{\dslwidth}{\dswidth{\ell}}
\newcommand{\dslogwidth}{\dswidth{\log}}
\makeatletter
\newcommand{\lipicsEndNotRestated}{\ifthmt@thisistheone\lipicsEnd\fi}
\makeatother

\DeclareMathOperator{\wed}{wed}
\DeclareMathOperator{\tw}{tw}
\newcommand{\bu}{\overline{u}}
\newcommand{\bv}{\overline{v}}
\newcommand{\partenc}{\mathcal{E}}
\newcommand{\bE}{\overline{\partenc}}

\newcommand{\cliquelistconj}{Clique Listing Hypothesis\xspace}

\newcommand{\combcliqueconj}{Combinatorial Clique Hypothesis\xspace}
\newcommand{\miniso}[1]{Min-Weight-$#1$-Subgraph\xspace}
\newcommand{\listiso}[1]{$#1$-Listing\xspace}
\newcommand{\enumiso}[1]{$#1$-Enumeration\xspace}
\newcommand{\countiso}[1]{$#1$-Counting\xspace}
\newcommand{\enciso}[1]{Full-$#1$-Encoding\xspace}
\newcommand{\relenciso}[1]{Relaxed-$#1$-Encoding\xspace}

\newcommand{\listH}{\listiso{H}}
\newcommand{\enumH}{\enumiso{H}}
\newcommand{\countH}{\countiso{H}}
\newcommand{\encH}{\enciso{H}}
\newcommand{\Hminiso}{\miniso{H}}
\newcommand{\Hlistiso}{\listiso{H}}
\newcommand{\Henumiso}{\enumiso{H}}
\newcommand{\Hcountiso}{\countiso{H}}
\newcommand{\Henciso}{\enciso{H}}
\newcommand{\Hrelenciso}{\relenciso{H}}

\newcommand{\lo}{\textup{lo}}
\newcommand{\hi}{\textup{hi}}
\newcommand{\subf}{\varphi}

\newcommand{\mfun}{\xi}
\newcommand{\hmfun}{\hat{\mfun}}
\newcommand{\hf}{\hat{f}}
\newcommand{\hS}{\hat{S}}

\newcommand{\fwidth}{\subf\somewidth}
\newcommand{\funwidth}[1]{#1\somewidth}
\newcommand{\submodset}{\mathsf{Sub}}
\newcommand{\sharpmodset}{\mathsf{\#Sub}}
\newcommand{\msubmodset}{\textup{m-}\mathsf{Sub}}
\newcommand{\td}{\textup{TD}}
\newcommand{\tdmin}{\textup{TD}_{\min}}

\newcommand{\prj}[2]{\pi_{#2}(#1)}
\newcommand{\attr}{\textup{attr}}
\newcommand{\prog}{\on{Prog}}
\newcommand{\cA}{\mathcal{A}}
\newcommand{\cB}{\mathcal{B}}
\newcommand{\cC}{\mathcal{C}}
\newcommand{\cF}{\mathcal{F}}

\newcommand{\conf}{\cC_2}

\newcommand{\cG}{\mathcal{G}}

\newcommand{\cc}{\textup{cc}}

\newcommand{\val}{\mathsf{val}}
\newcommand{\subp}[1]{V_{#1}^{i_{#1}}}
\newcommand{\PM}{\mathsf{PM}}

\newcommand{\coloneqq}{:=}

\title{Separations Between Submodular-Like Graph Widths}

\author{Karl Bringmann}{ETH Zürich}{karl.bringmann@inf.ethz.ch}{https://orcid.org/0000-0003-1356-5177}{}%

\author{Egor Gorbachev}{ETH Zürich}{peltorator@pm.me}{https://orcid.org/0009-0005-5977-7986}{}

\authorrunning{K. Bringmann and E. Gorbachev}

\Copyright{Karl Bringmann and Egor Gorbachev}

\acknowledgements{}

\begin{document}

\maketitle

\setcounter{page}{0}

\begin{abstract}
    Many algorithmic frameworks exist for evaluating join queries in relational databases.
They range from simple ones, such as dynamic programming over a single tree decomposition, high-degree-low-degree splitting from graph algorithms, and the degree uniformization of Joglekar and R\'e, to sophisticated ones based on information-theoretic inequalities, such as the PANDA framework for submodular width.
Each of these frameworks except high-degree-low-degree splitting comes with a width notion of the query that determines its running-time exponent.
We fill this gap by formalizing high-degree-low-degree splitting with $\ell$ degree classes, which gives rise to the \emph{degree-splitting width} $\dslwidth$.
These width notions form a hierarchy, but it was not known whether the more complicated frameworks achieve better worst-case exponents than the simpler ones on any graph query.

We study this question for full self-join-free join queries with binary relations, which amounts to colored subgraph isomorphism for a pattern graph $H$.
We show that every inequality in the hierarchy $\subw(H) \le \sharpw(H) \le \mwidth(H) \le \dslwidth(H)$ can be strict.
For instance, $\mwidth(K_{3,3}) = 2 < \dslwidth(K_{3,3})$ for every constant $\ell$, so \emph{no constant number of degree classes reaches the exponent of degree uniformization}.
We find the smallest separating patterns, with $8$, $9$, and $6$ nodes, respectively, by searching through all small patterns with a software tool that computes these widths via mixed integer linear programming.
Along the way, we give a degree-based algorithm that removes the polylogarithmic overhead of degree uniformization when evaluating full join queries.
Furthermore, we show that with logarithmically many degree classes, the degree-splitting width coincides with the \mwidth of Joglekar and R\'e.
Finally, we discuss several open problems.

\end{abstract}

\newpage

\tableofcontents

\newpage

\section{Introduction} \label{sec:introduction}

Join queries are the central operation of relational databases, and determining how fast they can be evaluated is one of the core questions of database theory.
A join query $Q = R_1 \bowtie \cdots \bowtie R_k$ combines relations $R_1, \ldots, R_k$ of a database; each relation is a table whose columns are labeled by attributes.
An \emph{answer} of $Q$ assigns a value to every attribute such that for every relation $R_i$, the values of the attributes of $R_i$ form a row of $R_i$.
For example, the answers of the triangle query $R_1(x, y) \bowtie R_2(y, z) \bowtie R_3(z, x)$ are the triples $(x, y, z)$ with $(x, y) \in R_1$, $(y, z) \in R_2$, and $(z, x) \in R_3$.
The structure of a query is captured by its \emph{hypergraph}, which has one node per attribute and one hyperedge per relation, consisting of the attributes of this relation.
We consider queries that are \emph{full}, that is, all attributes appear in the answers (there are no projections), and \emph{self-join-free}, that is, no relation occurs twice.
For such a query, one asks whether an answer exists (the Boolean query), how many answers there are, or one wants to output all answers (evaluating the query).
The input size~$N$ is the total number of rows of all relations.
The goal is to determine, for every query hypergraph, the smallest exponent $c$ such that these tasks can be solved in time $\tOh(N^c)$\footnote{Throughout the paper, we write $\tOh(T) \coloneqq \bigcup_{k \ge 0} \Oh(T \log^k T)$ and $\tilde{\Omega}(T) \coloneqq \bigcup_{k \ge 0} \Omega(T / \log^k T)$.}, plus the number of answers when all answers are output.

In this paper we study full self-join-free join queries in which every relation is binary, that is, has exactly two attributes.
Then the hypergraph of the query is a graph, the \emph{pattern graph} $H$, with one node per attribute and one edge per relation.
The database becomes a \emph{host graph} $G$ whose nodes are colored by attributes and whose edges are the rows of the relations, and the answers to the query are exactly the color-respecting copies of $H$ in $G$, which we call \emph{$H$-subgraphs}.
Evaluating the query is thus the same task as finding all $H$-subgraphs of $G$, a colored variant of the problem known in graph algorithms as subgraph isomorphism listing.
The input size $N$ becomes the number $m$ of edges of $G$, and we write $t$ for the number of $H$-subgraphs.
Correspondingly, we consider three tasks: \emph{decision}, where we ask whether some $H$-subgraph exists (the Boolean query), \emph{listing} all $H$-subgraphs (evaluating the query), and \emph{counting} them.
For decision and counting, we search for the smallest exponent $c$ such that the problem is solvable in $\tOh(m^c)$ time, and for listing, the time complexity we aim for is $\tOh(m^c + t)$.
Throughout the paper, we use this graph terminology rather than database terminology because it suits the special case of the join queries we consider.

The achievable exponent depends heavily on the structure of $H$.
Acyclic queries (trees) can be handled in linear time: decision and counting take time $\Oh(m)$, and listing takes time $\Oh(m + t)$~\cite{Yannakakis81,CarmeliZBCKS22}.
In contrast, already for the triangle no linear-time algorithm is known, even for decision~\cite{ItaiR78,AlonYZ97}, and it has been conjectured that time $m^{1+\delta}$ is necessary for some constant~$\delta > 0$~\cite{AbboudW14}.
This motivates general frameworks that aim for good exponents for all patterns at once.
Most of them follow a common roadmap, which we describe next.

\begin{figure}[ht]
    \centering
    \begin{tikzpicture}[
        vtx/.style={circle, draw, fill=white, inner sep=0pt, minimum size=17pt, font=\small},
        bag/.style={draw, rounded corners=6pt, inner sep=3.5pt, font=\small}
    ]
        \node[vtx] (u) at (0,0) {$u$};
        \node[vtx] (v) at (3,0) {$v$};
        \foreach \i/\h in {1/0.85, 2/1.7, 3/2.55} {
            \node[vtx] (x\i) at (1,\h) {$x_\i$};
            \node[vtx] (y\i) at (2,\h) {$y_\i$};
            \draw (u) -- (x\i) -- (y\i) -- (v);
        }
        \draw (u) -- (v);
        \node at (1.5,-0.85) {(a)};
        \begin{scope}[xshift=5cm]
            \node[bag] (C) at (0,1.275) {$u, v$};
            \foreach \i/\h in {1/2.55, 2/1.275, 3/0} {
                \node[bag] (L\i) at (2.1,\h) {$u, v, x_\i$};
                \node[bag] (R\i) at (4.5,\h) {$v, x_\i, y_\i$};
                \draw (C.east) -- (L\i.west);
                \draw (L\i.east) -- (R\i.west);
            }
            \node at (2.25,-0.85) {(b)};
        \end{scope}
    \end{tikzpicture}
    \caption{(a) A pattern consisting of three $4$-cycles that share the edge $\set{u,v}$. (b) A tree decomposition of the pattern.}
    \label{fig:intro-td}
\end{figure}
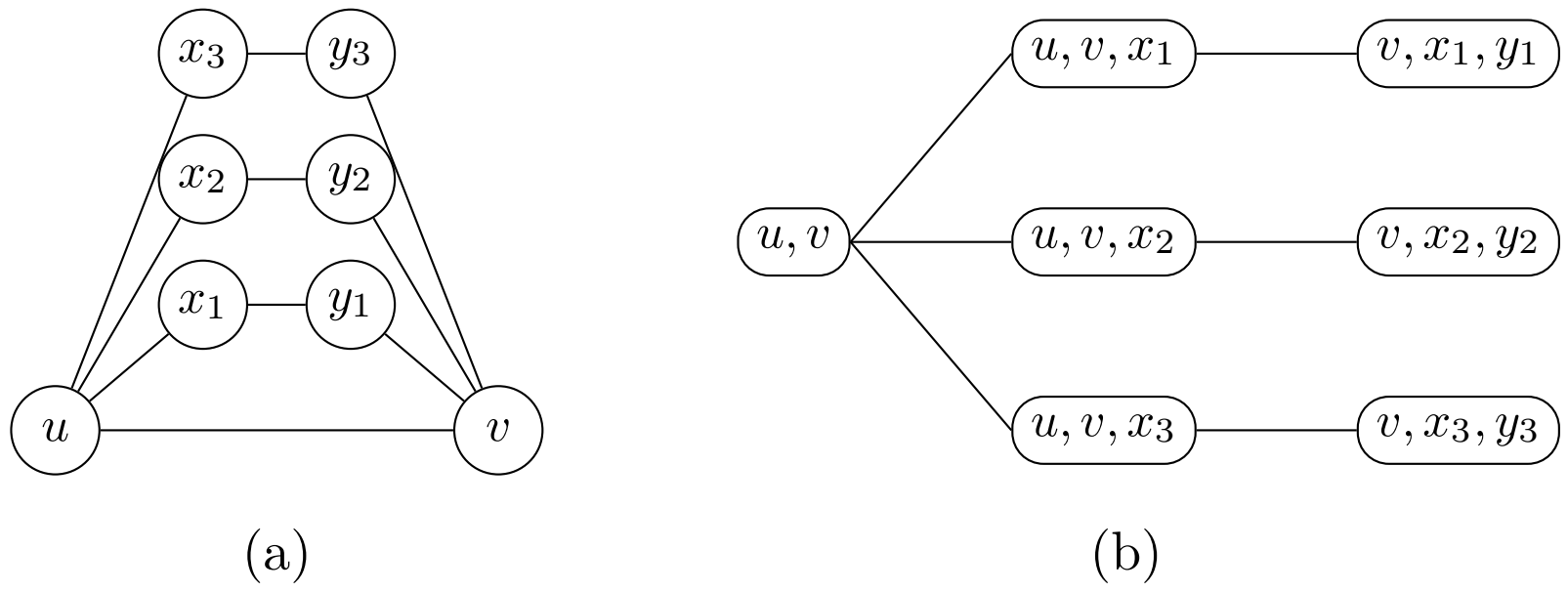

\subparagraph{From trees to tree decompositions.}
Linear-time algorithms for acyclic queries use dynamic programming~\cite{Yannakakis81}.
Root the tree and process it bottom-up: for every host node $v$ of color $a$, compute whether, or in how many ways, the subtree rooted at $a$ can be mapped into $G$ with $a$ mapped to $v$.
To handle an arbitrary pattern, we would like to group its nodes into \emph{bags} that are arranged in a tree, so that the same dynamic programming can run over the tree of bags.
This is exactly what a \emph{tree decomposition} provides.
It is a tree whose nodes are bags of pattern nodes such that
(1) every edge of $H$ lies inside some bag, so that every relation can be checked within a single bag, and
(2) for every node of $H$, the bags containing it form a connected subtree, so that the value of this node can be kept consistent by comparing adjacent bags only.
See \cref{fig:intro-td} for an example.

Suppose that for every bag $B$ we have computed a table $S_B$ of tuples that assign a host node to every node of $B$, such that $S_B$ contains the restriction of every $H$-subgraph to $B$; we say that we have \emph{materialized} $B$.
Then each bag behaves like a single node of a tree pattern whose possible values are the tuples in its table $S_B$, and two adjacent bags are compatible if their tuples agree on the common nodes.
This turns the problem into an instance with a tree pattern, of size linear in the total size of the tables, which is solved by the dynamic programming for trees (see, e.g.,~\cite[Lemma 7.1]{BG24}).
The idea goes back to tree clustering for constraint satisfaction~\cite{DechterP89} and underlies the classical algorithms for queries of bounded hypertree width and fractional hypertree width~\cite{GottlobLS02,GroheM14}.
Hence, the main cost lies in materializing the bags, and the task is to choose a good tree decomposition and to materialize its bags quickly.

\subparagraph{Treewidth and fractional hypertree width.}\label{sec:intro-fec}
The simplest way to materialize a bag $B$ is to try all $|B|$-tuples of host nodes, which takes time $\Oh(n^{|B|})$, where $n$ is the number of host nodes.
The best tree decomposition is then the one minimizing the largest bag size.
This leads to the \emph{treewidth} $\tw(H)$, which is the minimum over all tree decompositions of $H$ of the largest bag size minus one, and to the running time $\Oh(n^{\tw(H)+1})$, plus $\Oh(t)$ for listing.

A bag can be materialized faster by joining the relations inside it.
By the AGM bound~\cite{AtseriasGM13}, the restrictions of the $H$-subgraphs to $B$ take at most $m^{\rho^*(B)}$ distinct values, where $\rho^*(B)$ is the \emph{fractional edge cover number} of $B$ in $H$, and worst-case optimal join algorithms compute these tuples in time $\tOh(m^{\rho^*(B)})$~\cite{NgoPRR18,Veldhuizen12}.
Minimizing the largest value $\rho^*(B)$ over all tree decompositions gives the \emph{fractional hypertree width} $\textup{fhtw}(H)$~\cite{GroheM14}, and materializing every bag in this way gives the running time $\tOh(m^{\textup{fhtw}(H)})$.
For example, the triangle takes time $\Oh(n^3) \subseteq \Oh(m^3)$ via treewidth but only $\tOh(m^{3/2})$ via fractional hypertree width.

\subparagraph{Several tree decompositions and encodings.}
A single tree decomposition is not optimal, however.
The $4$-cycle has fractional hypertree width $2$, yet it can be listed in time $\Oh(m^{3/2} + t)$ by a variant of the algorithm of Alon, Yuster, and Zwick~\cite{AlonYZ97}.
This algorithm splits the host nodes of each color into high-degree and low-degree nodes and handles different combinations with different tree decompositions.
The same idea applies to the pattern in \cref{fig:intro-td}.
Its fractional hypertree width is also $2$, so any single tree decomposition gives time $\Oh(m^2 + t)$, whereas handling each of its three $4$-cycles separately with its own high-degree and low-degree cases gives time $\Oh(m^{3/2} + t)$.
The frameworks below therefore compute an \emph{encoding} of the solutions: a collection of tree decompositions with a table for each of their bags, such that every $H$-subgraph is \emph{encoded} by at least one of them, meaning that its restrictions to all bags of this tree decomposition appear in the corresponding tables.
Encodings are a variant of factorized representations of query results~\cite{OlteanuZ15}.
Given an encoding, decision and listing can be solved in time linear in its size, plus $t$ for listing, by running the dynamic programming above for every tree decomposition and removing duplicates.
The question thus becomes: what is the cheapest strategy to compute an encoding?

\subparagraph{Submodular width.}
Marx~\cite{Marx13} introduced the \emph{submodular width} $\subw(H)$, together with a decision algorithm that runs in time polynomial in $m^{\subw(H)}$.
Running time $\tOh(m^{\subw(H)})$ was first achieved by the PANDA algorithm of Abo Khamis, Ngo, and Suciu~\cite{Khamis0S17,KhamisNS24}, which computes an encoding by turning proofs of information-theoretic inequalities into query plans.
PANDAExpress~\cite{KhamisNS25} reduces the hidden polylogarithmic factor to a single logarithm.

\subparagraph{Counting and full encodings.}
The encodings above are \emph{relaxed}: several of their tree decompositions may encode the same $H$-subgraph.
This is harmless for decision and listing (where duplicates can be removed), but it is incompatible with counting and other non-idempotent aggregates.
These require a \emph{full encoding}, in which every $H$-subgraph is encoded in exactly one tree decomposition.
Given a full encoding, $H$-subgraphs can be counted in time linear in the size of the encoding (see, e.g.,~\cite[Lemma B.2]{BG24}).
To compute full encodings, the authors of~\cite{KhamisCMNNOS20} introduced the \emph{\#-submodular width} $\sharpw(H)$, together with an algorithm running in time $\tOh(m^{\sharpw(H)})$.
They also show $\subw(H) \le \sharpw(H) \le \textup{fhtw}(H)$.

\subparagraph{Degree-based algorithms.}
PANDA-type algorithms are complicated: they build intermediate tables and split cases according to degrees in these tables~\cite{KhamisNS24,KhamisC26}.
Joglekar and R\'e~\cite{JoglekarR18} proposed a much simpler approach that uses only the degrees in the input relations, which database systems commonly maintain anyway.
They partition the host nodes of each color into logarithmically many classes of nodes with roughly equal degrees, a technique called \emph{degree uniformization}, and handle each of the resulting instances separately with its own tree decomposition.
To capture the time complexity of the algorithms based on degree uniformization, Joglekar and R\'e define the \emph{\mwidth} of an instance.
We propose a variant of their definition that does not depend on the host graph: by taking the worst case over all host graphs, we obtain a measure $\mwidth(H)$ that only depends on the pattern $H$ and thus is more in line with other width notions.
Since the degree classes are disjoint, this approach computes a full encoding, in time $\tOh(m^{\mwidth(H)})$.

\paragraph*{Our Results}

\subparagraph{High-degree-low-degree splitting.}
A different perspective on degree-based algorithms comes from graph algorithms, where the most common technique for subgraph isomorphism problems is \emph{high-degree-low-degree splitting}.
Instead of logarithmically many degree classes, it uses a single degree threshold, or a constant number of them, to split the host nodes into high-degree and low-degree nodes, and it handles each combination of high and low nodes separately.
For example, triangles can be listed in time $\Oh(m^{3/2})$~\cite{ItaiR78,ChibaN85}, and more generally, Alon, Yuster, and Zwick~\cite{AlonYZ97} list $k$-cycles in time $\Oh(m^{2-1/\lceil k/2 \rceil}+t)$, by distinguishing whether a cycle contains a high-degree node.
Variants of this idea underlie most combinatorial subgraph isomorphism algorithms, but it had not been formalized in general into a width notion.
To compare it to the other width notions, we introduce the \emph{degree-splitting width} $\dslwidth(H)$ as the exponent of the best algorithm that splits the host nodes of each color into $\ell$ degree classes at fixed thresholds and then proceeds like degree uniformization; see \cref{sec:high-low}.
The case $\ell = 2$ corresponds to classic high-degree-low-degree splitting, and for every constant $\ell$, this approach computes a full encoding in time $\Oh(m^{\dslwidth(H)})$, without logarithmic factors.

\subparagraph{Do the time complexity exponents of these frameworks differ?}
On small patterns, degree-splitting width and \mwidth coincide: we show that $\dswidth{2} = \mwidth$ holds for all patterns on at most five nodes.
In general, however, they do not.
We prove that for a bipartite clique $K_{3, 3}$ with three nodes in each part we have $\mwidth(K_{3,3}) = 2$ while $\dslwidth(K_{3,3}) > 2$ for every $\ell \ge 1$.
Hence, within the degree-splitting framework, the logarithmic overhead of degree uniformization is necessary: for $K_{3,3}$, no constant number of degree classes reaches the exponent of \mwidth.
See \cref{sec:high-low} and Lemma~\ref{thm:ca-k33-min-nodes-edges} for the proofs of these results.

Similarly, we want to separate the other width notions.
The frameworks above become more and more general: high-degree-low-degree splitting uses a constant number of degree classes, degree uniformization uses logarithmically many, and the polymatroid-based frameworks underlying $\subw$ and $\sharpw$ may also split according to degrees in intermediate tables.
Correspondingly, their exponents satisfy $\subw(H) \le \sharpw(H) \le \mwidth(H)$ for every pattern $H$.
But do these frameworks actually differ?
On small patterns, they do not: we show that $\subw = \sharpw$ holds for all patterns on at most $7$ nodes, and $\sharpw = \mwidth$ holds for all patterns on at most $8$ nodes.
If they all coincided, there would be no reason to run a complicated PANDA-type algorithm instead of a simple degree-based one.
To answer this question, we develop a software tool~\cite{WidthCalculator} that computes all these parameters for a given pattern and use it to search through all small patterns.
It turns out that both inequalities can be strict; the smallest separating patterns are listed in \cref{tab:intro-separations}.
We are not aware of any prior separation between these parameters in the literature.
We show all these results in \cref{sec:examples}.

\begin{table}[!htbp]
    \centering
    \renewcommand{\arraystretch}{1.2}
    \newcommand{\tabpic}[2]{\rule[-1.45cm]{0pt}{2.9cm}\raisebox{-0.5\height}{\tikzset{every picture/.append style={scale=#2}}\input{pictures/#1}}}
    \newcommand{\tabcell}[1]{\begin{tabular}{@{}c@{}}#1\end{tabular}}
    \begin{tabular}{c|c|c|c}
        Gap & Pattern & Widths & Smallest with respect to \\
        \hline
        $\subw < \sharpw$
            & \rule[-1.45cm]{0pt}{2.9cm}\raisebox{-0.5\height}{\tikzset{every picture/.append style={scale=0.48}}\begin{tikzpicture}[x=1.4pt, y=1.4pt, every node/.style={circle, fill, inner sep=0pt, minimum size=5.5pt}]
    \node (n0) at (0,96) {};
    \node (n1) at (0,48) {};
    \node (n2) at (0,0) {};
    \node (n3) at (64,96) {};
    \node (n4) at (64,48) {};
    \node (n5) at (64,0) {};
    \node (n6) at (96,32) {};
    \node (n7) at (96,16) {};
    \foreach \a/\b in {0/3, 0/4, 0/5, 1/3, 1/4, 1/5, 2/3, 2/4, 2/5, 4/6, 5/7, 6/7} {
        \draw[line width=1.1pt] (n\a) -- (n\b);
    }
    \ifpatternlabels
        \node[left=3.5pt, fill=none, minimum size=0pt, font=\small] at (n0) {$a_1$};
        \node[left=3.5pt, fill=none, minimum size=0pt, font=\small] at (n1) {$a_2$};
        \node[left=3.5pt, fill=none, minimum size=0pt, font=\small] at (n2) {$a_3$};
        \node[right=3.5pt, fill=none, minimum size=0pt, font=\small] at (n3) {$b_1$};
        \node[above right=2pt, fill=none, minimum size=0pt, font=\small] at (n4) {$b_2$};
        \node[below right=2pt, fill=none, minimum size=0pt, font=\small] at (n5) {$b_3$};
        \node[right=3.5pt, fill=none, minimum size=0pt, font=\small] at (n6) {$c_1$};
        \node[right=3.5pt, fill=none, minimum size=0pt, font=\small] at (n7) {$c_2$};
    \fi
\end{tikzpicture}}
            & \tabcell{$\subw = 9/4$\\$\sharpw = 25/11$}
            & \tabcell{nodes ($8$)\\and edges ($12$)} \\
        \hline
        $\subw < \sharpw$
            & \rule[-1.45cm]{0pt}{2.9cm}\raisebox{-0.5\height}{\tikzset{every picture/.append style={scale=0.41}}\begin{tikzpicture}[x=1.4pt, y=1.4pt, every node/.style={circle, fill, inner sep=0pt, minimum size=5.5pt}]
    \node (n0) at (34.3,114.1) {};
    \node (n1) at (61.5,114.2) {};
    \node (n2) at (84.6,99.4) {};
    \node (n3) at (96,74.7) {};
    \node (n4) at (92.1,47.5) {};
    \node (n5) at (73.9,26.7) {};
    \node (n6) at (48,19.2) {};
    \node (n7) at (48,0) {};
    \node (n8) at (21.8,26.9) {};
    \node (n9) at (3.9,47.6) {};
    \node (n10) at (0,74.6) {};
    \node (n11) at (11.4,99.4) {};
    \foreach \a/\b in {0/1, 0/11, 1/2, 2/3, 3/4, 4/5, 5/6, 5/7, 6/8, 7/8, 8/9, 9/10, 10/11} {
        \draw[line width=1.1pt] (n\a) -- (n\b);
    }
\end{tikzpicture}}
            & \tabcell{$\subw = 15/8$\\$\sharpw \ge 17/9$}
            & value of $\subw$ \\
        \hline
        $\sharpw < \mwidth$
            & \rule[-1.45cm]{0pt}{2.9cm}\raisebox{-0.5\height}{\tikzset{every picture/.append style={scale=0.86}}\begin{tikzpicture}[scale=0.6, every node/.style={circle, fill, inner sep=0pt, minimum size=5.5pt}]
    \node (n8) at (0,2) {};
    \node (n9) at (0,-2) {};
    \node (n2) at (0,0) {};
    \node (n3) at (-1.2,0.7) {};
    \node (n4) at (-1.2,-0.7) {};
    \node (n1) at (1.2,0.7) {};
    \node (n5) at (1.2,-0.7) {};
    \node (n6) at (-2.6,0) {};
    \node (n7) at (2.6,0) {};
    \foreach \u in {1,2,3,4,5} {\draw[line width=1.1pt] (n8) -- (n\u); \draw[line width=1.1pt] (n9) -- (n\u);}
    \foreach \a/\b in {6/2, 6/3, 6/4, 7/1, 7/2, 7/5, 1/5} {\draw[line width=1.1pt] (n\a) -- (n\b);}
    \draw[line width=1.1pt] (n6) .. controls (-2.6,-3.3) and (2.6,-3.3) .. (n7);
\end{tikzpicture}}
            & \tabcell{$\sharpw = 8/3$\\$\mwidth = 27/10$}
            & nodes ($9$) \\
        \hline
        \tabcell{$\mwidth < \dslwidth$\\for all $\ell \ge 1$}
            & \rule[-1.45cm]{0pt}{2.9cm}\raisebox{-0.5\height}{\tikzset{every picture/.append style={scale=0.73}}\begin{tikzpicture}[x=1.4pt, y=1.4pt, every node/.style={circle, fill, inner sep=0pt, minimum size=5.5pt}]
    \node (n0) at (0,64) {};
    \node (n1) at (0,32) {};
    \node (n2) at (0,0) {};
    \node (n3) at (48,64) {};
    \node (n4) at (48,32) {};
    \node (n5) at (48,0) {};
    \foreach \a/\b in {0/3, 0/4, 0/5, 1/3, 1/4, 1/5, 2/3, 2/4, 2/5} {
        \draw[line width=1.1pt] (n\a) -- (n\b);
    }
\end{tikzpicture}}
            & \tabcell{$\mwidth = 2$\\$\dslwidth \ge 2 + \frac{1}{144\ell}$}
            & \tabcell{nodes ($6$)\\and edges ($9$)} \\
    \end{tabular}
    \caption{Smallest patterns separating the width notions.}
    \label{tab:intro-separations}
\end{table}

\subparagraph{A degree-based algorithm with no polylogarithmic overhead.}
Although a constant number of degree classes does not reach \mwidth in general, degree information alone suffices to avoid the polylogarithmic overhead when a relaxed encoding is enough.
Motivated by PANDAExpress, which partially removes the polylogarithmic overhead of PANDA by replacing axis-parallel partitioning with arbitrary hyperplane cuts~\cite{KhamisNS25}, we show that for every pattern $H$, a relaxed $H$-encoding can be computed in time $\Oh(m^{\mwidth(H)})$; see \cref{lm:new-mwidth-algo}.
Our algorithm still uses only degree information, but instead of uniformizing degrees, it compares them directly: it extends partial solutions and discards one as soon as a cost determined by the degrees of its nodes exceeds $m^{\mwidth(H)}$.

In particular, $\subw$ and \mwidth coincide for every pattern on at most $7$ nodes, so on these patterns this elementary algorithm reaches the exponent of PANDA without any logarithmic factor, using only degrees of the input relations.
Patterns of this size are practically relevant, as queries in real-world logs tend to be small~\cite{BonifatiMT20}.
Unlike degree uniformization, this algorithm may encode the same $H$-subgraph by several tree decompositions, so it cannot be used for counting.
Whether a \emph{full} encoding can be computed in time $\Oh(m^{\mwidth(H)})$ remains an interesting open problem.

\subparagraph{What the separations tell us.}
The gap between \mwidth and $\sharpw$ shows that degree information is not all you need: to reach $\sharpw$, and thus $\subw$, in general, an algorithm must split cases by statistics beyond the degrees in the input relations.
The gap between $\sharpw$ and $\subw$ shows that the $\sharpw$ framework in general does not run in time $\tOh(m^{\subw(H)})$.
However, whether counting in time $\tOh(m^{\subw(H)})$ is possible remains an open problem~\cite{KhamisC26,KhamisNS26}.
As an interesting data point, we analyze the pattern $H$ from \cref{fig:subw-vs-sharpw:min-vert}, which separates $\subw < \sharpw$, and show that this pattern can be counted in time $\tOh(m^{\subw(H)}) = \tOh(m^{9/4})$; see \cref{lm:better-counting-algo-for-specific-graph}.
The main ingredient of this specialized algorithm is \emph{hyper-degree splitting}~\cite{BG24}: instead of splitting single nodes by their degrees, it splits tuples of nodes, such as pairs, by their number of common neighbors, or more generally by their number of extensions in an already computed table.
PANDA-type algorithms also split according to degrees in intermediate tables, which is a related idea.
Splitting only by the degrees of single nodes keeps the encoding full for free: it partitions the host graph into disjoint instances, each of which contains exactly the $H$-subgraphs of one case.
A split of tuples does not partition the host graph, so more care is needed.
Every tree decomposition used for such a case must contain the split tuple in one of its bags, and the table of this bag must be restricted to the tuples of this case, so that the tree decomposition encodes exactly the $H$-subgraphs of this case.
Bringmann and Gorbachev used this technique to characterize all patterns with $\subw(H) < 2$, and their algorithms also allow counting in time $\Oh(m^{\subw(H)})$ for all of these patterns~\cite[Theorems B.1 and D.1]{BG24}.
We are not aware of any pattern for which degree and hyper-degree splitting do not suffice to count in time $\tOh(m^{\subw(H)})$.
So the general problem remains open: is counting always possible in time $\tOh(m^{\subw(H)})$?

\subparagraph{Is submodular width optimal?}
For detection, submodular width can be beaten using fast matrix multiplication.
For example, triangles can be detected in time $\Oh(m^{2\omega/(\omega+1) + o(1)}) \subseteq \Oh(m^{1.41})$ instead of $\Oh(m^{3/2})$~\cite[Theorem 3.5]{AlonYZ97}, and the $\omega$-submodular width~\cite{KhamisHS25} extends such improvements to general queries.
These improvements rely on fast matrix multiplication, whose asymptotically fastest variants are impractical.
For combinatorial algorithms (that is, algorithms that do not use the algebraic techniques underlying fast matrix multiplication), we are not aware of any pattern for which submodular width can be beaten~\cite{FanKZ23,KhamisNS26}.
This might simply be because not enough patterns have been tried, and small patterns are again the natural place to look.
To know where to look, we need to understand the limits of combinatorial techniques.

These limits are usually stated for cliques.
Already the triangle is closely tied to fast matrix multiplication: triangle detection has a truly subcubic combinatorial algorithm if and only if Boolean matrix multiplication has one~\cite{WilliamsW18}.
More generally, finding $k$-cliques is one of the central problems of fine-grained complexity.
The \combcliqueconj states that for no $k \ge 3$ and $\varepsilon > 0$ is there a combinatorial algorithm that detects a $k$-clique in an $N$-node graph in time~$\Oh(N^{k-\varepsilon})$~\cite{AbboudBW18}.
The \cliquelistconj states that for no $k \ge 3$ and $\varepsilon > 0$ is there an algorithm, combinatorial or not, that lists all $k$-cliques of an $N$-node graph in time $\tOh(N^{k-\varepsilon} + t)$ (see, e.g.,~\cite[Hypothesis~5.1]{BG24}).
The \emph{clique embedding power} $\clemb(H)$ of Fan, Koutris, and Zhao~\cite{FanKZ23} transfers these barriers from cliques to arbitrary patterns: it measures how efficiently a $k$-clique instance can be turned into a host graph for $H$.
Assuming the \combcliqueconj, no combinatorial algorithm solves decision or counting in time $\Oh(m^{\clemb(H) - \varepsilon})$ for any~$\varepsilon > 0$~\cite[Proposition~13]{FanKZ23}, and assuming the \cliquelistconj, no algorithm solves listing in time $\tOh(m^{\clemb(H) - \varepsilon} + t)$.\footnote{These listing lower bounds were previously derived from the Zero-Clique Hypothesis~\cite{BG24}, which was recently refuted together with the 3SUM and APSP Hypotheses~\cite{AlmanW26}. The \cliquelistconj is not known to be affected, but it has lost this support; see \cref{sec:example:open-problems} for a discussion.}
Fan, Koutris, and Zhao show $\clemb(H) \le \subw(H)$ and identify several classes of patterns on which the two coincide, including cycles and chordal graphs.
In general, however, they differ: a pattern they call \emph{boat} ($P(3,3,3)$ in our terminology; see \cref{def:par-graphs}) has $\clemb = 17/9$ but $\subw = 2$.

Listing the copies of a boat was shown to require quadratic time under the 3XOR and 3SUM Hypotheses~\cite{JoglekarR18,BG24}.
Both hypotheses have recently been refuted~\cite{AlmanW26}, and for 3SUM there even is a refutation that does not require fast matrix multiplication~\cite{ChanIMY26}.
Still, no truly subquadratic combinatorial algorithm is known for finding a boat.
Since $\clemb < 2$ for the boat, reductions from cliques via clique embeddings cannot show its quadratic hardness, and since $\subw = 2$ for the boat, a truly subquadratic algorithm would have to go beyond all frameworks discussed above.
Bringmann and Gorbachev show that $\clemb(H) = \subw(H)$ for every pattern with $\subw(H) < 2$~\cite[Theorem D.1]{BG24}, and it follows from their results that every pattern with $\subw(H) \ge 2$ either has $\clemb(H) \ge 2$ or contains the boat as an induced minor~\cite[Theorem D.1 and Corollary 1.9]{BG24}.
Consequently, every pattern with $\clemb(H) < 17/9$ satisfies $\clemb(H) = \subw(H)$, so clique embeddings certify combinatorial optimality of submodular width for all exponents below~$17/9$.
Beyond the boat, our search finds further patterns with $\clemb < \subw$; the smallest one is listed in \cref{tab:intro-open}.

\begin{table}[!htbp]
    \centering
    \renewcommand{\arraystretch}{1.2}
    \newcommand{\tabpic}[2]{\rule[-1.45cm]{0pt}{2.9cm}\raisebox{-0.5\height}{\tikzset{every picture/.append style={scale=#2}}\input{pictures/#1}}}
    \newcommand{\tabcell}[1]{\begin{tabular}{@{}c@{}}#1\end{tabular}}
    \begin{tabular}{c|c|c|c}
        Gap & Pattern & Values & Smallest with respect to \\
        \hline
        $\clemb < \subw$
            & \rule[-1.45cm]{0pt}{2.9cm}\raisebox{-0.5\height}{\tikzset{every picture/.append style={scale=0.84}}
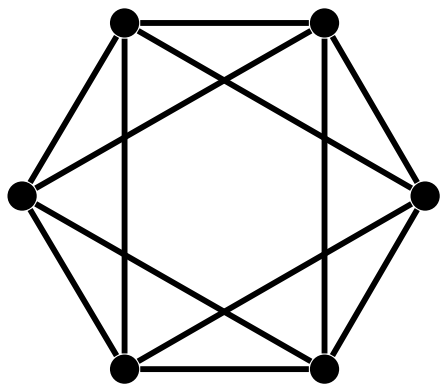
\begin{tikzpicture}[x=1.4pt, y=1.4pt, every node/.style={circle, fill, inner sep=0pt, minimum size=5.5pt}]
    \node (n0) at (16.4,55.4) {};
    \node (n1) at (48.4,55.4) {};
    \node (n2) at (16.4,0) {};
    \node (n3) at (48.4,0) {};
    \node (n4) at (64.5,27.7) {};
    \node (n5) at (0,27.7) {};
    \foreach \a/\b in {0/1, 0/2, 0/4, 0/5, 1/3, 1/4, 1/5, 2/3, 2/4, 2/5, 3/4, 3/5} {
        \draw[line width=1.1pt] (n\a) -- (n\b);
    }
\end{tikzpicture}}
            & \tabcell{$\clemb = 9/4$\\$\subw = 5/2$}
            & \tabcell{nodes ($6$)} \\
        \hline
        $\clemb < \subw$
            & \rule[-1.45cm]{0pt}{2.9cm}\raisebox{-0.5\height}{\tikzset{every picture/.append style={scale=0.73}}\begin{tikzpicture}[x=1.4pt, y=1.4pt, every node/.style={circle, fill, inner sep=0pt, minimum size=5.5pt}]
    \node (n0) at (0,32) {};
    \node (n1) at (32,64) {};
    \node (n2) at (64,64) {};
    \node (n3) at (96,32) {};
    \node (n4) at (32,32) {};
    \node (n5) at (64,32) {};
    \node (n6) at (32,0) {};
    \node (n7) at (64,0) {};
    \foreach \a/\b in {0/1, 0/4, 0/6, 1/2, 2/3, 3/5, 3/7, 4/5, 6/7} {
        \draw[line width=1.1pt] (n\a) -- (n\b);
    }
\end{tikzpicture}}
            & \tabcell{$\clemb = 17/9$\\$\subw = 2$}
            & \tabcell{edges ($9$)\\and value of $\clemb$} \\
    \end{tabular}
    \caption{Smallest patterns with $\clemb < \subw$, for which the optimal combinatorial exponent is unknown: the octahedron (top) and the boat $P(3,3,3)$ (bottom).}
    \label{tab:intro-open}
\end{table}

\subparagraph{Computing the parameters on small examples.}
Computing $\subw$, $\sharpw$, and \mwidth is challenging even for small patterns.
Their definitions alternate between a maximization over infinitely many functions or degree vectors and minimizations over tree decompositions and forests.
We formulate each of them, as well as $\clemb$~\cite{FanKZ23}, as a mixed integer linear program (MILP) and solve it with an off-the-shelf solver.
The size of these programs grows quickly with the pattern, and we propose several ideas to keep them manageable (see \cref{sec:implementation}):
\begin{itemize}
    \item \textbf{Clique separators.} If a pattern is disconnected, its components can be handled independently, and each parameter of the pattern is the maximum of its values on the components. More generally, a \emph{clique separator} of $H$ is a set of pairwise adjacent nodes whose removal disconnects~$H$. Each of $\clemb$, $\subw$, $\sharpw$, and \mwidth splits over clique separators: its value on $H$ is the maximum of its values on the parts obtained by cutting $H$ along a clique separator, where each part keeps a copy of the separator (see \cref{lm:everything-splits-over-clique-separators}). This extends an observation of~\cite[Lemma D.5]{BG24} for $\subw$. Hence, it suffices to consider clique-separator-free patterns. For example, only $2792$ of the $12346$ graphs on $8$ nodes are clique-separator-free. For $\dslwidth$, the same reduction works after increasing $\ell$; see \cref{lm:ds-clique-refinement}.
    \item \textbf{Few tree decompositions and bags.} Since all functions involved are monotone, only tree decompositions that are minimal under inclusion of bags matter, and we enumerate them via elimination orderings. Moreover, every bag whose fractional edge cover number is at most $\clemb(H)$ can be discarded; see \cref{lm:filter-out-fec-easy-bags}.
    \item \textbf{Using the hierarchy.} We compute \mwidth first, since in our examples it was the easiest to compute. Its optimal degree vector yields a candidate function that on many examples certifies $\subw = \mwidth$ or $\sharpw = \mwidth$ without solving the MILP for $\subw$ or $\sharpw$. Moreover, on many examples, we have $\clemb = \mwidth$, which certifies that all parameters coincide.
\end{itemize}
We provide a software tool~\cite{WidthCalculator} that supports calculating $\tw$, $\clemb$, $\subw$, $\sharpw$, and \mwidth over all pattern graphs on up to $8$ nodes and on up to $12$ edges; all computer-assisted statements in this paper can be reproduced with it.
We state the results that rely on computations with our tool as \emph{Computer-Aided Lemmas} and \emph{Computer-Aided Theorems}.
The proof of each computer-aided lemma and theorem describes the computation and gives a command that reproduces it.

\subparagraph{Extreme values.}
Furthermore, we determine the range of the aforementioned pattern graph parameters in terms of treewidth; see \cref{sec:extreme-bounds}.
For every pattern $H$, we prove 
    \[
        (\tw(H) + 1)/2 \le \subw(H), \sharpw(H), \mwidth(H), \dslwidth(H) \le \tw(H) + 1.
    \]
    For every treewidth $\tw \ge 2$ and $\ell \ge 2$, both bounds are attainable.
    Moreover, $\clemb(H) \in \widetilde{\Omega}(\sqrt{\tw(H)})$ and there are patterns of arbitrarily large treewidth with $\clemb(H) \in \Oh(\sqrt{\tw(H)})$.

Hence, reductions from clique problems via clique embeddings cannot certify submodular width on all patterns of large treewidth, and the gap between $\clemb$ and $\subw$ on some families of patterns is polynomial rather than constant.
Moreover, the exponent $\tw(H) + 1$, which is achieved by simply trying all tuples of a bag, is conditionally necessary for some patterns of every treewidth $\tw \ge 2$, even when running time is measured in $m$ rather than in the number of host nodes (see \cref{lm:extreme-widths}).
In terms of the number of host nodes, such patterns were known for every treewidth $\tw \ge 3$ under hyperclique hypotheses~\cite{BringmannS21}, but their hard host graphs are dense, so these bounds do not carry over to $m$.

\paragraph*{Further Related Work}

Further related work includes output-sensitive algorithms for queries with projections~\cite{Hu24,DeepZFK24,BringmannFW26}.
For the fine-grained complexity of further query-evaluation tasks, see, e.g.,~\cite{CarmeliK21,BringmannCM22,BringmannC22}.
We measure running time in terms of $m$, the input size of the query, while much of the work on subgraph finding problems uses the number of nodes of the host graph instead~\cite{Marx10,BringmannS21}.
We study the colored problem variant, which is the natural one for join queries, while the uncolored variant is widely studied in graph algorithms~\cite{YusterZ97,DahlgaardKS17}.
Color coding~\cite{AlonYZ95} transfers algorithms to the uncolored variant up to logarithmic factors, but no transfer is known in the other direction.
For example, uncolored $4$-cycles can be listed in time $\tOh(m^{4/3} + t)$~\cite{AbboudKLS23}, faster than colored ones.
We search for subgraphs that are not necessarily induced; induced subgraph isomorphism behaves quite differently (see, e.g.,~\cite{WilliamsWWY15,DalirrooyfardVW21,DalirrooyfardW22}).
Further work considers restrictions on the host graph, such as bounded arboricity~\cite{ChibaN85} or planarity~\cite{Kowalik03}, and parameters other than the input size (see, e.g.,~\cite{MarxP14}).

Lanzinger, Merkl, and Suciu~\cite{LanzingerMS26} also separate width notions, but different ones and by different means: they give a single graph on $32$ nodes whose adaptive, normal, linear, entropic, and submodular widths are pairwise distinct.

\paragraph*{Organization}

After introducing the problem formulations, encodings, and width notions in \cref{sec:preliminaries}, we define \mwidth, give its submodular formulation, and prove $\clemb \le \subw \le \sharpw \le \mwidth$ in \cref{sec:m-width}.
In \cref{sec:high-low}, we formalize degree splitting, define degree-splitting width, prove $\mwidth \le \dslwidth$, show that this inequality is strict for $K_{3,3}$ and every constant $\ell$, and show that with logarithmically many degree ranges degree-splitting width coincides with \mwidth.
\Cref{sec:m-width-algo} gives the degree-based algorithm that computes a relaxed encoding in time $\Oh(m^{\mwidth(H)})$.
In \cref{sec:width-hierarchy}, we show that all these width notions split over clique separators (degree-splitting width after increasing $\ell$), and in \cref{sec:extreme-bounds}, we bound them in terms of treewidth.
\Cref{sec:implementation} describes our software tool, and \cref{sec:examples} presents the separating examples and the open problems.
Finally, \cref{sec:appendix-m-width} contains deferred proofs.

\section{Preliminaries} \label{sec:preliminaries}

\subsection{Basic Notation and Problem Formulations}
\label{sec:prelims-problem-formulations}

All graphs in this paper are simple and undirected unless specified otherwise.
In addition to standard notation from graph theory, we use the following notation.
We write $[n] \coloneqq \set{1,\ldots,n}$ for a positive integer $n$, and $\fragmentco{p}{q}=\set{p,p+1,\ldots,q-1}$ and $\fragmentcc{p}{q} = \set{p, p+1, \ldots,q}$ for integers $p\le q$.

Throughout the paper we distinguish two kinds of graphs:
constant-sized \emph{pattern graphs} $H$ that specify the particular variant of $H$-subgraph isomorphism problems and the input \emph{host graphs} $G$ in which we search for copies of $H$.

Formally, a \emph{pattern graph} is a connected graph $H$ on at least two nodes.
A host graph is a graph $G$ together with a partition $V(G) = \bigcup_{a \in V(H)} V_a$ of its nodes.
We stress the fact that the graph $H$ is always assumed to be fixed and in particular has constant size.
An \emph{$H$-subgraph} of $G$ is a tuple of nodes $\bv = (v_a)_{a \in V(H)} \in \prod_{a \in V(H)} V_a$ such that for each $\{a,b\} \in E(H)$ we have $\{v_a, v_b\} \in E(G)$.
We denote by $m \coloneqq |E(G)|$ the input size and write $n \coloneqq |V(G)|$.
Throughout the paper, we assume that the nodes of $G$ are represented by integers bounded by a polynomial in $n$, so that tuples of nodes can be sorted by radix sort in linear time.

The assumption that the pattern $H$ is connected and has at least two nodes is without loss of generality. If $H$ is disconnected, each occurrence of $H$ in $G$ is obtained by combining occurrences of its connected components, so the problems split into independent subproblems on those components. If $|V(H)| = 1$, the problems are trivial.

We always do some basic preprocessing on the host graph $G$ in $\Oh(n + m)$ time.
Discard all edges of $G$ whose endpoint colors do not form an edge of $H$.
Such edges are irrelevant to the problem at hand.
Furthermore, remove all isolated nodes of $G$.
Such nodes cannot be part of an $H$-subgraph.
In particular, we now have $n \in \Oh(m)$.

For a pattern graph $H$ and a subset $V' \subseteq V(H)$ of its nodes, let $H' = H[V']$ be the subgraph of $H$ induced by $V'$.
Furthermore, let $G[H']$ be the subgraph of $G$ induced by all nodes in $G$ that are colored by nodes in $H'$.
For every $e = \set{a, b} \in E(H)$, let $E_e(G)$ be the edges of $G$ between $V_a$ and $V_b$.

Having set up all the basic notation and conventions we use, we are ready to define the central problems of the paper.

\begin{itemize}
    \item In the \emph{\Hcountiso} problem, the task is to compute the total number of $H$-subgraphs of a given host graph $G$.
    \item In the \emph{\listH} problem, the task is to output all $H$-subgraphs of a given host graph $G$. The number of such $H$-subgraphs (i.e., the output size) is denoted by $t$.
    \item In the \emph{\enumH} problem, the task is the same as in the \listH problem. However, we measure the time complexity differently. We measure the \emph{preprocessing time} before printing the first solution, and the \emph{delay} between printing consecutive solutions and after printing the last solution.
\end{itemize}

For a vector $\bv$ we denote by $\prj{\bv}{B}$ the projection to the entries with indices in $B$.
Furthermore, for a set $S$ of vectors we denote by $\prj{S}{B}$ the set $\setof{\prj{\bv}{B}}{\bv \in S}$.
For $B \subseteq V(H)$, we call a set $S \subseteq \prod_{a \in B} V_a$ a \emph{materialization} of $B$ if for every $H$-subgraph $\bv$ of $G$ we have $\prj{\bv}{B} \in S$.
We say that $B$ is \emph{materialized} if we computed a materialization of $B$.

\subsection{Tree Decompositions}
\label{sec:prelims-treedecomp}

A \emph{tree decomposition} of a graph $H$ consists of a tree $T = ([\ell], E_T)$ and \emph{bags} $B(1),\ldots,B(\ell) \subseteq V(H)$ such that:
(1) every node of $H$ is contained in some bag;
(2) every edge of $H$ has both endpoints in a common bag; and
(3) for every node $v \in V(H)$, the bags containing $v$ induce a connected subtree of~$T$.

The \emph{treewidth} $\tw(H)$ of $H$ is the minimum, over all tree decompositions of $H$, of the largest size of a bag minus one.
We call a tree decomposition \emph{simple} if its bags are distinct and no bag is contained in another bag.
Every tree decomposition can be made simple by removing redundant bags and adjusting the underlying tree.
As the precise tree structure of a tree decomposition will not matter in this paper, we identify two simple tree decompositions when their sets of bags coincide.
That is, we view a simple tree decomposition as its set of bags, and we denote the finite set of all simple tree decompositions of $H$ by $\td(H)$.

For $\mathcal{T}, \mathcal{T}' \in \td(H)$, write $\mathcal{T} \succeq \mathcal{T}'$ if every bag of $\mathcal{T}'$ is contained in some bag of $\mathcal{T}$.
This is a partial order on simple tree decompositions.\footnote{Indeed, reflexivity and transitivity are immediate.
For antisymmetry, suppose that $\mathcal{T} \succeq \mathcal{T}' \succeq \mathcal{T}$ and fix a bag $B \in \mathcal{T}$.
There are bags $B' \in \mathcal{T}'$ and $B'' \in \mathcal{T}$ with $B \subseteq B' \subseteq B''$.
Since the bags of $\mathcal{T}$ form an antichain, $B = B''$, and hence $B = B'$.
Thus every bag of $\mathcal{T}$ occurs in $\mathcal{T}'$, and the symmetric argument proves equality.}
We denote the set of minimal elements of this order by $\tdmin(H)$.

\subsection{Encodings}

Solutions to different variants of $H$-subgraph isomorphism frequently follow a common roadmap.
We define two intermediate problems that are used to treat the different variants of $H$-subgraph isomorphism in a unified way.

\begin{definition}[Full and Relaxed Encoding Problem, Variation of {\cite[Definition 2.9]{BG24}}]
	Let $H$ be a pattern and $G$ be a host graph with partition $V = \bigcup_{a \in V(H)} V_a$.
    A \emph{partial $H$-encoding} $\partenc$ of $G$ is a simple tree decomposition $\mathcal{T}$ of~$H$ and, for every bag $B$ of $\mathcal{T}$, a set $S_B \subseteq \prod_{a \in B} V_a$ called a \emph{submaterialization}.
    We say that an $H$-subgraph $\bv$ of $G$ is \emph{encoded} by $\partenc$ if $\prj{\bv}{B} \in S_B$ holds for every bag $B$ of $\mathcal{T}$.
    The \emph{size} of $\partenc$ is $1 + \sum_{B\in\mathcal T}|S_B|$.
	
    A \emph{full $H$-encoding} $\bE$ of $G$ is a 
    collection of partial $H$-encodings $\partenc_1, \ldots, \partenc_k$ such that every $H$-subgraph $\bv$ of $G$ is encoded by \emph{exactly} one $\partenc_i$.\footnote{Note that compared to \cite[Definition 2.9]{BG24}, we do not require the number of partial $H$-encodings to be constant.}
    The \emph{size} of a full $H$-encoding is the sum of all sizes of its partial encodings $\partenc_1, \ldots, \partenc_k$.
    
    The \emph{\Henciso} problem asks to compute a full $H$-encoding of a given host graph~$G$.

    \smallskip

    Furthermore, a \emph{relaxed $H$-encoding} $\bE$ of $G$ is a collection of partial $H$-encodings $\partenc_1, \ldots, \partenc_k$ such that every $H$-subgraph $\bv$ of $G$ is encoded by \emph{at least} one $\partenc_i$.
    The \emph{size} of a relaxed $H$-encoding is the sum of all sizes of its partial encodings $\partenc_1, \ldots, \partenc_k$.

    The \emph{\Hrelenciso} problem asks to compute a relaxed $H$-encoding of a given host graph~$G$.
    \lipicsEnd
\end{definition}

The only difference between the two variants of the problem is that in \Henciso, partial encodings are required to encode disjoint sets of $H$-subgraphs of $G$.
This requirement is crucial for \Hcountiso to ensure that solutions are not overcounted.
The following lemma shows that without loss of generality we may assume that the output of \Hrelenciso consists of a constant number of partial encodings.
We do not know of a comparable result for \Henciso.

\begin{lemma}\label{lm:constant-relaxed-encoding}
    Let $H$ be a pattern.
    There is an algorithm that given a host graph $G$ and a relaxed $H$-encoding $\bE$ of $G$ of size $T$, computes a relaxed $H$-encoding $\bE'$ of $G$ with a constant number of partial encodings (\ie, the number may depend on $H$ but not on $G$) in $\Oh(T + m)$ time.
\end{lemma}

\begin{proof}
    We group the partial encodings in $\bE$ according to the underlying tree decompositions.
    For a tree decomposition $\mathcal{T}$, let $\row{\partenc}{t}$ be the partial encodings with $\mathcal{T}$ as the underlying tree decomposition.
    We create a single partial encoding $\partenc'_{\mathcal{T}}$ by merging for each bag of $\mathcal{T}$ the submaterializations from $\row{\partenc}{t}$.
    To merge the submaterializations, we unite the lists, use radix sort to lexicographically sort them, and then delete repetitions.
    (Here we use that all nodes of $G$ are represented by integers bounded by a polynomial in $n$.)
    
    The partial encoding $\partenc'_{\mathcal{T}}$ encodes all $H$-subgraphs that any of $\row{\partenc}{t}$ encode.
    By repeating the merging for all tree decompositions $\mathcal{T}$, we compute a relaxed $H$-encoding.
    The number of partial $H$-encodings we get is limited by the number of tree decompositions of $H$ and is thus constant.
    The total time complexity is $\Oh(T + m)$.
\end{proof}

In the following two lemmas we show how the solution of \Henciso or \Hrelenciso leads to the solutions of the variants of $H$-subgraph isomorphism we are interested in.
Consequently, we may focus on developing algorithms for \Henciso and \Hrelenciso.

\begin{lemma}[Extension of {\cite[Lemma 2.11]{BG24}}] \label{lm:all-reductions}
    For a pattern $H$, if \Hrelenciso can be solved in time $T(m)$ then 
    \begin{itemize}
    \item \Hlistiso can be solved in time $\Oh(T(m) + t)$, where $t$ is the output size and
    \item \Henumiso can be solved in preprocessing time $\Oh(T(m))$ and delay $\Oh(1)$.
    \end{itemize}
\end{lemma}

\begin{proof}
    Compute a relaxed $H$-encoding and apply \cref{lm:constant-relaxed-encoding}.
    We obtain constantly many partial encodings $\partenc_1,\ldots,\partenc_\ell$ of total size $\Oh(T(m))$.

    By \cite[Lemma 7.1]{BG24}, a partial encoding of size $s$ can be transformed in $\Oh(m+s)$ time into a tree pattern $H'$ and a host graph $G'$ of size $\Oh(m+s)$.
    The $H'$-subgraphs of $G'$ are in bijection with the $H$-subgraphs encoded by the partial encoding, and each $H'$-subgraph can be converted to the corresponding $H$-subgraph in constant time.
    Apply this reduction to all $\ell$ partial encodings.
    Their total construction time and size are $\Oh(T(m))$.

    For enumeration, preprocess the tree instances for constant-delay enumeration using \cite{BaganDG07}, and run their enumerators consecutively.
    Converting their outputs back to $H$-subgraphs takes constant time per output.
    Every solution appears at least once and at most $\ell$ times.
    Thus, after $\Oh(T(m))$ preprocessing, the time to produce the first $r$ outputs, counted with repetitions, is $\Oh(r)$.
    The Cheater's Lemma \cite[Lemma 7]{CarmeliK21} removes these bounded repetitions while preserving $\Oh(T(m))$ preprocessing and constant delay.
    Running this enumeration algorithm to completion also gives listing time $\Oh(T(m)+t)$.
\end{proof}

\begin{fact}[{\cite[Lemma B.2]{BG24}}]\label{fct:counting-reduction}
    For a pattern $H$, if \Henciso can be solved in time $T(m)$ then \Hcountiso can be solved in time $\Oh(T(m))$.
    \lipicsEnd
\end{fact}

Note that \cite[Lemma B.2]{BG24} is formulated for a constant number of partial encodings. The same proof works in the general case if the radix sorts for all partial encodings are performed jointly, so that the additive $\Oh(m)$ term of \cite[Lemma 7.1]{BG24} is paid only once.

\subsection{Clique Embeddings}

Fan, Koutris, and Zhao \cite{FanKZ23} introduced the notion of clique embedding power to obtain conditional lower bounds for $H$-subgraph isomorphism based on the common hypotheses from fine-grained complexity (see \cref{sec:example:open-problems}).

\begin{definition}[Clique Embedding, \cite{FanKZ23}]
    Let $H$ be a graph. We say that sets $A,B \subseteq V(H)$ \emph{touch} in $H$ if $A \cap B \neq \emptyset$ or there exists an edge $\set{a, b} \in E(H)$ with $a \in A$ and $b \in B$.
    
    Let $H$ be a pattern graph and $k$ be a positive integer. An \emph{embedding} from a clique $K_k$ to $H$ is a function~$\psi$ that maps every node $u \in V(K_k)$ to a non-empty subset $\psi(u) \subseteq V(H)$ such that $H[\psi(u)]$ is connected for every $u \in V(K_k)$, and $\psi(u)$ and $\psi(v)$ touch in $H$ for all pairs of nodes $u,v \in V(K_k)$.

    For an embedding $\psi$, the \emph{weak depth} of an edge $e = \set{a, b} \in E(H)$ is \[d_{\psi}(e) \coloneqq |\setof{u \in V(K_k)}{\set{a,b} \cap \psi(u) \ne \emptyset}|.\] The \emph{weak edge depth} of the embedding $\psi$ is \[\wed(\psi) \coloneqq \max_{e \in E(H)} d_{\psi}(e).\] Finally, the \emph{clique embedding power} of $H$ is \[\clemb(H) \coloneqq \sup_{k \ge 3} \max_{\psi \colon K_k \to H} \frac{k}{\wed(\psi)},\] where the maximum is taken over all clique embeddings $\psi$ from $K_k$ to $H$.
    \lipicsEnd
\end{definition}

\subsection{Submodularity and Width Notions}\label{sec:submodular}

We now introduce the main width notions that encapsulate the time complexity exponents of different algorithmic frameworks for $H$-subgraph isomorphism.

\begin{definition}[\cite{Marx13}]\label{def:f-width}
For a pattern $H$ and a function $\subf \colon 2^{V(H)} \to \RR$, let \[\fwidth(H) \coloneqq \min_{\mathcal{T} \in \td(H)} \max_{B \in \mathcal{T}} \subf(B). \lipicsEnd\]
\end{definition}

In other words, $\fwidth(H)$ is the smallest value $w$ such that $H$ has a tree decomposition in which every bag $B$ satisfies $\subf(B) \le w$.

\begin{definition}[\cite{Marx13}]\label{def:submodularity}
    For a pattern $H$, a function $\subf \colon 2^{V(H)} \to \RR$ is called \emph{submodular} if \[\subf(X) + \subf(Y) \ge \subf(X \cap Y) + \subf(X \cup Y)\] holds for all $X, Y \subseteq V(H)$.
    Furthermore, $\subf$ is called \emph{monotone} if $\subf(X) \le \subf(X')$ holds for all $X, X' \subseteq V(H)$ with $X \subseteq X'$.
    Finally, $\subf$ is called \emph{edge-dominated} if $\subf(e) \le 1$ holds for every $e \in E(H)$.
    \lipicsEnd
\end{definition}

\begin{observation}
    For a pattern $H$ and a monotone function $\subf \colon 2^{V(H)} \to \RR$, we have \[\fwidth(H) = \min_{\mathcal{T} \in \boldsymbol{\tdmin}(H)} \max_{B \in \mathcal{T}} \subf(B). \lipicsEnd\]
\end{observation}

Based on the above definitions, we are now ready to define submodular width.

\begin{definition}[Submodular Width, \cite{Marx13}] \label{def:subw}
     For a pattern $H$, let $\submodset(H)$ be the set of all submodular monotone edge-dominated functions $\subf \colon 2^{V(H)} \to \RR_{\ge 0}$ with $\subf(\emptyset) = 0$.
     The \emph{submodular width} of $H$, denoted by $\subw(H)$, is $\sup\setof{\fwidth(H)}{\subf \in \submodset(H)}$.
    \lipicsEnd
\end{definition}

The results of \cite{KhamisNS24} imply an $\tOh(m^{\subw(H)})$-time algorithm for computing a \Hrelenciso.
For \Henciso, a different width notion is required.

\begin{definition}[\#-Submodular Width, \cite{KhamisCMNNOS20}] \label{def:sharpw}
     For a pattern $H$, let $\sharpmodset(H)$ be the set of all monotone edge-dominated functions $\subf \colon 2^{V(H)} \to \RR_{\ge 0}$ with $\subf(\emptyset) = 0$ that satisfy $\subf(X) + \subf(Y) \ge \subf(X \cap Y) + \subf(X \cup Y)$ for all $X, Y \subseteq V(H)$ with $X \cap Y \subseteq e$ for some $e \in E(H)$.
     The \emph{\#-submodular width} of $H$, denoted by $\sharpw(H)$, is $\sup\setof{\fwidth(H)}{\subf \in \sharpmodset(H)}$.
    \lipicsEnd
\end{definition}

The results of \cite{KhamisCMNNOS20} imply an $\tOh(m^{\sharpw(H)})$-time algorithm for computing a \Henciso.

The third width notion we consider, \mwidth, is defined in \cref{sec:m-width}.

\section{\texorpdfstring{\mwidth}{m-width}} \label{sec:m-width}

Joglekar and R\'e~\cite{JoglekarR18} introduced \mwidth to capture the time complexity of algorithms based on degree uniformization.
The original definition of \mwidth is instance-dependent.
It first splits the host graph into degree classes and then asks how cheaply each bag can be materialized using the resulting degree bounds.

\begin{definition}[\mwidth of a Host Graph, {\cite[Section 5.4]{JoglekarR18}}, Reformulated for Graphs] \label{def:mwidth_general}
    Fix a pattern $H$ and a host graph $G$.
    Let $m$ be an arbitrary parameter such that $|E_e(G)| \le m$ holds for every $e \in E(H)$.
    A relation $R$ with coordinate set $\attr(R)$ is a set of tuples indexed by $\attr(R)$.
    For $X\subseteq\attr(R)$ and a tuple $\bv\in\prod_{a\in X}V_a$, let $\deg(\bv,R,X)$ count its extensions in $R$.
    Formally, for nonempty $R$ and sets $X, Y\subseteq\attr(R)$ with $X \subseteq Y$, define
    \[
        \deg(\bv,R,X)\coloneqq|\setof{\bu\in R}{\prj{\bu}{X}=\bv}|,
        \qquad
        d(X,Y,R)\coloneqq\log_m\max_{\bv\in\prj{R}{X}}
            \deg(\bv,\prj{R}{Y},X).
    \]
    Thus $m^{d(X,Y,R)}$ is the largest number of extensions from fixed $X$-coordinates to $Y$-coordinates.

    For a host graph $J$ and $B\subseteq V(H)$, let $q_J(B)$ be the optimum of the following linear program $\prog_J(B)$:
    \begin{align*}
        &\max s_B&&&\\
        &\text{s.t. }&\text{(i) }&s_X\in\RR&\forall X\subseteq V(H)\\
        &&\text{(ii) }&s_\emptyset=0&\\
        &&\text{(iii) }&s_X\le s_Y&\forall X\subseteq Y\subseteq V(H)\\
        &&\text{(iv) }&s_{Y\cup Z}\le s_{X\cup Z}+d(X,Y,E_e(J))
            &\forall e\in E(H),\ X\subseteq Y\subseteq e,\ Z\subseteq V(H).
    \end{align*}

    A \emph{degree configuration} $c$ assigns an integer in $\fragmentcc{0}{\floor{\log_2|E_e(G)|}}$ to each pair $(e,a)$ with $e\in E(H)$ and $a \in e$.
    Let
    \[
        V_a(c)\coloneqq\setof{v\in V_a}{2^{c(e,a)}\le\deg(v,E_e(G),\set{a})<2^{c(e,a)+1}
        \text{ for every }e\ni a},
    \]
    and set
    $G(c)\coloneqq G[\bigcup_a V_a(c)]$.
    Write $\conf(G)$ for the set of configurations.
    The \mwidth of $G$ with respect to $H$ is
    \[
        \mwidthH{H}(G)\coloneqq
        \max_{c\in\conf(G)}\funwidth{q_{G(c)}}(H).
    \lipicsEnd\]
\end{definition}

The results of \cite{JoglekarR18} imply an $\tOh(m^{\mwidthH{H}(G)})$-time algorithm for \Henciso.

The definition of \mwidth from \cref{def:mwidth_general} depends on both a pattern graph $H$ and a host graph~$G$, whereas $\subw$ and $\sharpw$ are defined for a pattern graph alone.
To compare these width notions, we propose a variant of \mwidth that only depends on the pattern by taking the worst case over all host graphs.

\begin{definition}[\mwidth of a Pattern] \label{def:mwidth1}
    Let $H$ be a pattern and, for every positive integer~$m$, let~$\cG_m(H)$ be the set of all host graphs $G$ with $|E_e(G)| \le m$ for all $e \in E(H)$.
    We define the \mwidth of $H$ as \[\mwidth(H) \coloneqq \limsup_{m \to +\infty} \max_{G \in \cG_m(H)} \mwidthH{H}(G). \lipicsEnd\]
\end{definition}

This definition is rather complicated and very different from the definitions of $\subw$ and $\sharpw$.
We therefore propose an alternative definition that is similar to the definitions of these submodular widths.
The only difference is that we require the submodular inequality only when $Y \subseteq e$ for some edge $e$.

\begin{definition}[\mwidth, Submodular Formulation] \label{def:mwidth4}
    For a pattern $H$, let $\msubmodset(H)$ be the set of all monotone edge-dominated functions $\subf \colon 2^{V(H)} \to \RR_{\ge 0}$ that satisfy $\subf(\emptyset) = 0$ and $\subf(X) + \subf(Y) \ge \subf(X \cap Y) + \subf(X \cup Y)$ for all $X, Y \subseteq V(H)$ with $Y \subseteq e$ for some $e \in E(H)$.
     Let \[\mpwidth(H) \coloneqq \sup\setof{\fwidth(H)}{\subf \in \msubmodset(H)}. \lipicsEnd\]
\end{definition}

The two definitions coincide.
We defer the proof to \cref{sec:appendix-m-width}.

\begin{restatable}{lemma}{mwidthsubmodular}\label{lm:mwidth-submodular}
    For every pattern $H$, we have $\mwidth(H) = \mpwidth(H)$.
    \lipicsEndNotRestated
\end{restatable}

Functions in $\msubmodset(H)$ satisfy the following increment bounds, which we use repeatedly.

\begin{lemma}\label{lm:msub-increments}
    Let $H$ be a pattern, $\subf \in \msubmodset(H)$, $X \subseteq V(H)$, and $a \in V(H) \setminus X$.
    Then $\subf(X \cup \set{a}) \le \subf(X) + \subf(\set{a})$.
    If moreover $b \in X$ and $\set{a,b} \in E(H)$, then $\subf(X \cup \set{a}) \le \subf(X) + 1 - \subf(\set{b})$.
\end{lemma}

\begin{proof}
    Since $H$ is connected and has at least two nodes, $a$ lies in some edge.
    Applying the submodularity constraint to $X$ and $\set{a}$ gives $\subf(X \cup \set{a}) \le \subf(X) + \subf(\set{a}) - \subf(\emptyset) = \subf(X) + \subf(\set{a})$.
    If $b \in X$ and $\set{a,b} \in E(H)$, applying the submodularity constraint to $X$ and $\set{a,b}$ gives $\subf(X \cup \set{a}) \le \subf(X) + \subf(\set{a,b}) - \subf(\set{b}) \le \subf(X) + 1 - \subf(\set{b})$, where the last step uses edge domination.
\end{proof}

The submodular formulation directly gives the ordering of the width notions defined so far.

\begin{lemma}\label{lm:width-hierarchy}
    For every pattern $H$, we have
    \[\clemb(H) \le \subw(H) \le \sharpw(H) \le \mwidth(H).\]
\end{lemma}

\begin{proof}
    For $\clemb(H) \le \subw(H)$, see \cite[Corollary 9]{FanKZ23}.
    The next two inequalities follow from $\submodset(H) \subseteq \sharpmodset(H) \subseteq \msubmodset(H)$ and \cref{lm:mwidth-submodular}: each step imposes submodularity on fewer pairs of sets.
\end{proof}

\section{Degree-Splitting Width} \label{sec:high-low}

In graph algorithms, the most common technique for subgraph isomorphism problems is high-degree-low-degree splitting.
We want to understand the capabilities and limitations of such algorithms.
However, this technique has not been formalized in general.
In this section, we formalize it by defining the \emph{degree-splitting width} $\dslwidth$, which captures the time complexity exponent of algorithms that split the host nodes of each part into $\ell$ degree ranges.
We compare it with \mwidth, and we show that with logarithmically many degree ranges, degree splitting captures exactly \mwidth.

\paragraph*{Warm-Up: The 4-Cycle} \label{sec:c4-example}

We start with the classic algorithm for 4-cycles~\cite{AlonYZ97}.
It is usually described without any explicit reference to tree decompositions: it distinguishes whether a 4-cycle contains a high-degree node, and in each case it enumerates short paths and combines them.
However, we show that implicitly this algorithm uses the tree-decomposition-based framework.
We describe the same algorithm in the language of tree decompositions, materializations, and encodings from \cref{sec:preliminaries}.
We use the following simple observation about high-degree nodes.

\begin{observation} \label{obs:high-degree-small-cnt-nodes}
    Let $G$ be a $k$-partite graph for $k \ge 2$ with parts $V_1, V_2, \ldots, V_k$. For every $i \in [k]$ and $\Delta > 0$, the number of nodes in $V_i$ with degree at least $\Delta$ is at most $|E(G)| / \Delta$.
    \lipicsEnd
\end{observation}

\begin{example}[4-Cycle] \label{ex:c4}
    Let $H = C_4$ with nodes $a_1, a_2, a_3, a_4$ in cyclic order, and let $G$ be a host graph.
    We call a node of $G$ \emph{high-degree} if its degree is at least $\sqrt m$, and \emph{low-degree} otherwise.
    By \cref{obs:high-degree-small-cnt-nodes}, every part of $G$ contains at most $\sqrt m$ high-degree nodes.
    We compute a full $H$-encoding consisting of three partial encodings.

    First, we encode all $H$-subgraphs in which the node in $V_{a_1}$ is high-degree.
    We use the tree decomposition with bags $\set{a_1, a_2, a_3}$ and $\set{a_1, a_3, a_4}$.
    We materialize $\set{a_1, a_2, a_3}$ by listing all pairs of a high-degree node of $V_{a_1}$ and an edge between $V_{a_2}$ and $V_{a_3}$, in time $\Oh(\sqrt m \cdot m)$.
    The bag $\set{a_1, a_3, a_4}$ is materialized analogously.

    Second, we encode all $H$-subgraphs in which the node in $V_{a_1}$ is low-degree and the node in $V_{a_3}$ is high-degree.
    This works in the same way, using the same tree decomposition and pairing the high-degree nodes of $V_{a_3}$ with the edges between $V_{a_1}$ and $V_{a_2}$ and between $V_{a_1}$ and $V_{a_4}$, respectively.

    Finally, we encode all $H$-subgraphs in which the nodes in $V_{a_1}$ and $V_{a_3}$ are both low-degree.
    We use the tree decomposition with bags $\set{a_1, a_2, a_4}$ and $\set{a_2, a_3, a_4}$.
    We materialize $\set{a_1, a_2, a_4}$ by listing all edges between $V_{a_1}$ and $V_{a_2}$ and extending each of them by all neighbors in $V_{a_4}$ of its low-degree endpoint in $V_{a_1}$, in time $\Oh(m \cdot \sqrt m)$.
    The bag $\set{a_2, a_3, a_4}$ is materialized analogously, using the low-degree node in $V_{a_3}$.

    These three partial encodings encode disjoint sets of $H$-subgraphs, and together they encode all of them.
    Hence, they form a full $H$-encoding, which we compute in time $\Oh(m^{3/2})$.
    \lipicsEnd
\end{example}

In this example, the degree of a node determines the cost of materializing a bag in two ways.
A part restricted to high-degree nodes has few nodes, which helps when a node is chosen freely, as~$a_1$ in the first tree decomposition.
A low-degree node has few neighbors, which helps when a bag is materialized by extending along an edge, as from $a_1$ to $a_4$ in the last case.
More generally, a high-degree-low-degree algorithm for subgraph isomorphism typically consists of a case distinction on the degrees of the nodes, followed by enumerating partial solutions in each case and combining them.
Translated as above, each case becomes a partial encoding with its own tree decomposition, and the enumeration steps materialize its bags.
We now formalize this view.

\paragraph*{Degree Splitting and Degree-Splitting Width} \label{sec:degree-splitting}

We consider the following generalization of this idea: thresholds may differ between parts, and each part may be split into more than two degree ranges.
Within each resulting instance, we choose a tree decomposition and incrementally materialize its bags.
Formally, for each $a \in V(H)$, we choose a sequence \[1 = D_{a, 0} < D_{a, 1} < \cdots < D_{a, \ell_a} = m + 1\] and partition $V_a$ into $V_a^0, V_a^1, \ldots, V_a^{\ell_a - 1}$, where \[\subp{a} \coloneqq \setof{v_a \in V_a}{\deg_G(v_a) \in \fragmentco{D_{a, i_a}}{D_{a, i_a+1}}}\] for all $i_a \in \fragmentco{0}{\ell_a}$.
We call this process \emph{degree splitting}.
We then iterate over all $i \in \prod_{a \in V(H)} \fragmentco{0}{\ell_a}$ and solve \Henciso for each $G_i \coloneqq G[\bigcup_{a \in V(H)} \subp{a}]$ independently.
This reduction is valid as every $H$-subgraph of $G$ belongs to exactly one $G_i$.
For each $i_a \in \fragmentco{0}{\ell_a}$, all nodes in $V_a^{i_a}$ have degree in $G_i$ at most $D_{a, i_a+1}$, and $|V_a^{i_a}| \le m / D_{a, i_a}$ holds due to \cref{obs:high-degree-small-cnt-nodes}.
The algorithm then uses this information to solve \Henciso for a fixed graph $G_i$ using a single partial $H$-encoding.
The bags of this partial $H$-encoding are materialized via the following iterative process.
For a single-node bag $\set{a}$, we materialize it in time $\Oh(m / D_{a, i_a})$ by listing all elements of $\subp{a}$.
For a bag $B$ with more than one node, we start by materializing a pair of adjacent nodes $a$ and $b$ from $B$ in time $\Oh(m)$ by listing all edges between $\subp{a}$ and $\subp{b}$ in $G_i$.
We then iteratively add more and more nodes to the bag by searching for each new node as a neighbor of an already fixed node in some other part.
As a result, we get a tree rooted at the edge $\set{a, b}$, and each node is used to generate all its children.
Let $T$ be the resulting tree.
Each child of node $c$ contributes a factor of at most $D_{c,i_c+1}$, so this process takes time \[\Oh\bigl(m \prod_{c \in B} D_{c, i_c+1}^{\deg_T(c) - 1}\bigr),\]  independently of the chosen root edge.
In general, we pick some forest that spans some superset $B'$ of the whole bag $B$, consecutively materialize each of its connected components, and then project the resulting materialization of $B'$ onto $B$.

To describe the time complexity of materializing bags via forests, we use the following notation.
For every $X \subseteq V(H)$, let $\cF_H(X)$ be the set of all forests $F$ on the node set $X$ that are subgraphs of $H$.
That is, $V(F) = X$ and $E(F) \subseteq E(H)$.
For a forest $F$, let $\cc(F)$ be the number of connected components in $F$.
Furthermore, to capture the idea of materializing a superset of a bag and then projecting it onto the bag, we consider the following notion of monotonic closure.

\begin{definition}[Monotonic Closure]
    Let $S$ be a finite set and $f \colon 2^S \to \RR$ be a function.
    The \emph{monotonic closure} of $f$ is a function $\hf$ defined as \[\hf(X) \coloneqq \min\setof{f(Y)}{X \subseteq Y \subseteq S}\] for all $X \subseteq S$.
    Thus, $\hf$ is the largest monotone function bounded above by $f$.
    \lipicsEnd
\end{definition}

For a constant number $\ell$ of degree ranges in each part, we choose the thresholds as powers $m^{d_a^j}$ of $m$ with exponents $d_a^j \in [0,1]$.
The following definition captures the optimal time complexity exponent of such algorithms.

\begin{definition}[Degree-Splitting Width] \label{def:dswidth}
    Let $H$ be a pattern and $\ell$ be a positive integer.
    For every $a \in V(H)$, choose \emph{degree-splitting exponents}
    $0=d_a^0\le d_a^1\le\cdots\le d_a^{\ell}=1$.
    A vector $i\in\fragmentco{0}{\ell}^{V(H)}$ selects one degree range in each part.
    Write $d=(d_a^j)_{a\in V(H),\,j\in\fragmentcc{0}{\ell}}$ for the collection of all splitting exponents.
    For this choice, write $L_a=d_a^{i_a}$ and $U_a=d_a^{i_a+1}$, and for $X\subseteq V(H)$ define
    \[
        f_d^i(X)\coloneqq\min_{F\in\cF_H(X)}
        \left(\cc(F)+\sum_{\substack{a\in X\\\deg_F(a)\ge1}}
        (\deg_F(a)-1)U_a-\sum_{\substack{a\in X\\\deg_F(a)=0}}L_a\right).
    \]
    The \emph{degree-splitting width} is defined as
    \[
        \dslwidth(H)\coloneqq\min_d\max_{i\in\fragmentco{0}{\ell}^{V(H)}}
        \funwidth{\hf_d^i}(H).
    \]
    Here $\hf_d^i$ is the monotonic closure of $f_d^i$.
    \lipicsEnd
\end{definition}
Note that the minimum exists because the objective is continuous in $d$ and the set of splitting exponents is compact.

The two range endpoints have different roles: a lower degree bound gives $\Oh(m^{1-L_a})$ choices for an isolated node, while an upper degree bound gives at most $m^{U_a}$ choices when extending a tree from a node in part $a$.
Thus a nontrivial tree $T$ takes time \[\Oh\bigl(m^{1+\sum_{a\in V(T)}(\deg_T(a)-1)U_a}\bigr)\] to materialize, and the formula above adds the exponents of its components.
Following the discussion above, for every constant $\ell \ge 1$ there is an $\Oh(m^{\dslwidth(H)})$-time algorithm for \Henciso.

Here, $\dswidth{2}(H)$ captures the high-degree-low-degree approach in the most strict sense: like the algorithms inspired by \cite{AlonYZ97}, it splits the nodes in each part into two groups, high-degree nodes and low-degree nodes.
To highlight this, instead of $i \in \fragmentco{0}{2}^{V(H)}$ in the definition of $\dswidth{2}(H)$, we write $i \in \set{\lo, \hi}^{V(H)}$, where $\lo$ represents $0$ and $\hi$ represents $1$.

Degree-splitting width is never smaller than \mwidth, and it can only decrease when the number of degree ranges grows.

\begin{lemma}\label{lm:width-hierarchy-ds}
    For every pattern $H$ and integers $k \ge \ell \ge 1$, we have
    \[\mwidth(H) \le \dswidth{k}(H) \le \dswidth{\ell}(H).\]
\end{lemma}

\begin{proof}
    For the first inequality, by \cref{lm:mwidth-submodular} it suffices to show $\fwidth(H) \le \dswidth{k}(H)$ for every $\subf \in \msubmodset(H)$.
    Fix such a function $\subf$ and splitting exponents $d$ for $k$ degree ranges.
    For every $a \in V(H)$, let $e_a \coloneqq 1 - \subf(\set{a})$; as $\subf$ is monotone and edge-dominated and every node lies in an edge, we have $e_a \in [0,1]$.
    Choose a splitting box $i$ that contains $e$, that is, $L_a \le e_a \le U_a$ for all $a \in V(H)$.
    We show $\subf(X) \le f_d^i(X)$ for all $X \subseteq V(H)$.
    Fix a forest $F \in \cF_H(X)$, and add the nodes of $X$ one at a time, component by component, such that every component with at least two nodes starts with the two endpoints of an edge and then grows by attaching leaves.
    We bound the increments of $\subf$ by \cref{lm:msub-increments}.
    Adding an isolated node $a$ increases $\subf$ by at most $\subf(\set{a}) = 1 - e_a \le 1 - L_a$.
    Adding the two endpoints $a$ and $b$ of the first edge of a component increases $\subf$ by at most $\subf(\set{a}) + (1 - \subf(\set{a})) = 1$.
    Attaching a leaf to a node $p$ increases $\subf$ by at most $1 - \subf(\set{p}) = e_p \le U_p$.
    In a tree $T$ with at least two nodes rooted at an arbitrary edge, exactly $\deg_T(p) - 1$ children are attached to every node $p$, so the total increment for $T$ is at most $1 + \sum_{p \in V(T)} (\deg_T(p) - 1) U_p$.
    Summing over all components, $\subf(X)$ is at most the cost of $F$ in the definition of $f_d^i$, and minimizing over $F$ gives $\subf(X) \le f_d^i(X)$.
    Since $\subf$ is monotone, this implies $\subf \le \hf_d^i$, and hence $\fwidth(H) \le \funwidth{\hf_d^i}(H)$.
    As $d$ was arbitrary, $\fwidth(H) \le \dswidth{k}(H)$.

    By using duplicates, splitting exponents allowed by $\dswidth{\ell}$ are a subset of those allowed by~$\dswidth{k}$.
    This proves the second inequality.
\end{proof}

\paragraph*{Is a Constant Number of Degree Ranges Sufficient?} \label{sec:ds-vs-mwidth}

If for some $\ell \ge 1$, we have $\dslwidth(H) = \mwidth(H)$, degree splitting removes the polylogarithmic overhead of degree uniformization for full $H$-encodings.
This raises the question whether more than a constant number of degree splits is ever necessary.
That is, does every pattern $H$ admit a constant $\ell \ge 1$ with $\mwidth(H) = \dslwidth(H)$, which would give an $\Oh(m^{\mwidth(H)})$-time algorithm for \Henciso?
A strong version of this question asks whether even $\mwidth(H) = \dswidth{2}(H)$ holds for every pattern $H$.

For small patterns, a constant number of degree splits indeed suffices: every pattern $H$ on at most five nodes satisfies $\mwidth(H) = \dswidth{2}(H)$, as we show in \cref{thm:ca-k33-min-nodes-edges}.
However, as shown in the following lemmas, there is a graph $H$ on six nodes for which $\mwidth(H)$ is strictly smaller than $\dslwidth(H)$ for every constant $\ell \ge 1$.

\begin{lemma}\label{lm:k33-widths}
    We have $\clemb(K_{3, 3}) = \subw(K_{3, 3}) = \sharpw(K_{3, 3}) = \mwidth(K_{3, 3}) = 2$.
\end{lemma}

\begin{proof}
    Let the nodes of one part of $K_{3, 3}$ be $a_1, a_2, a_3$ and the nodes of the other part be $b_1, b_2, b_3$.
    Consider the following embedding of $K_4$ into $K_{3, 3}$: Embed one node of $K_4$ into $a_1$, one node into $b_1$, one node into $\{a_2, b_2\}$, and one node into $\{a_3, b_3\}$.
    As the embedding has weak edge depth $2$, $\clemb(K_{3, 3}) \ge 2$ holds.
    This was proven in \cite[Proposition 20]{FanKZ23}.

    We now prove that $\mwidth(K_{3, 3}) \le 2$.
    By \cref{lm:mwidth-submodular}, it suffices to show $\fwidth(K_{3,3}) \le 2$ for every $\subf \in \msubmodset(K_{3,3})$.
    Fix such a function $\subf$, and assume without loss of generality that $\subf(\set{b_1}) \le \subf(\set{v})$ for all nodes $v$.
    We choose the tree decomposition of $K_{3,3}$ consisting of the bags $\set{a_i, b_1, b_2, b_3}$ for $i \in [3]$, connected in a line.
    For such a bag, add its nodes in the order $a_i, b_2, b_3, b_1$.
    By \cref{lm:msub-increments}, the four increments of $\subf$ are at most $\subf(\set{a_i})$, $1 - \subf(\set{a_i})$, $1 - \subf(\set{a_i})$, and $\subf(\set{b_1})$, respectively.
    Hence, $\subf(\set{a_i, b_1, b_2, b_3}) \le 2 - \subf(\set{a_i}) + \subf(\set{b_1}) \le 2$.
    Therefore, $\fwidth(K_{3,3}) \le 2$ and $\mwidth(K_{3, 3}) \le 2$.

    The reverse inequalities follow from $\clemb(K_{3,3}) \ge 2$ and \cref{lm:width-hierarchy}.
\end{proof}

We now prove that for every constant $\ell$, the degree-splitting width $\dslwidth$ of $K_{3, 3}$ is strictly greater than $2$.

\begin{lemma}\label{lm:k33-ds-width}
    For every integer $\ell \ge 1$, we have
    \[
        \dslwidth(K_{3, 3}) \ge 2 + \frac{1}{144 \ell} > 2 = \mwidth(K_{3, 3}).
    \]
\end{lemma}

\begin{proof}
    Consider the following two tree decompositions:
    \[
    \cA=\setof{\set{a_j,b_1,b_2,b_3}}{j\in[3]},
    \qquad
    \cB=\setof{\set{a_1,a_2,a_3,b_j}}{j\in[3]}.
    \]
    These are the only minimal tree decompositions of $K_{3,3}$.
    Indeed, consider a tree decomposition of $K_{3,3}$ with tree $T$, and for every node $v$ of $K_{3,3}$ let $T_v$ be the subtree of $T$ formed by the bags containing $v$.
    First assume that some bag contains $a_1$, $a_2$, and $a_3$.
    For every $j \in [3]$, the subtrees $T_{a_1}, T_{a_2}, T_{a_3}, T_{b_j}$ then pairwise intersect, since every edge $\set{a_i,b_j}$ lies in some bag.
    Hence, some bag contains $\set{a_1,a_2,a_3,b_j}$, and the decomposition $\succeq$-dominates $\cB$.
    (If some connected subtrees of a tree pairwise intersect, then they have a nonempty common intersection.)
    Otherwise, no bag contains $a_1$, $a_2$, and $a_3$, and thus two of them, say $a_1$ and $a_2$, never lie in a common bag.
    Then $T_{a_1}$ and $T_{a_2}$ are disjoint, and every path in $T$ from $T_{a_1}$ to $T_{a_2}$ passes through the node $x$ of $T_{a_1}$ closest to $T_{a_2}$.
    Every subtree $T_{b_j}$ intersects both $T_{a_1}$ and $T_{a_2}$, since $b_j$ is adjacent to $a_1$ and $a_2$, and thus contains $x$.
    So the bag at node $x$ contains $b_1$, $b_2$, and $b_3$, and the argument of the first case with the two parts exchanged shows that the decomposition $\succeq$-dominates $\cA$.
    Consequently, for every monotone function $f$,
\begin{equation}\label{eq:k33-two-decompositions}
    \funwidth{f}(K_{3,3})
    =\min\left\{\max_{X\in\cA}f(X),\max_{X\in\cB}f(X)\right\}.
\end{equation}

    Fix the splitting exponents for \dslwidth.
    There are $6(\ell-1)$ internal thresholds, each cutting the degree hypercube perpendicular to one coordinate.
    We construct $6(\ell-1)+1$ boxes such that each threshold cuts the interior of at most one box.

    Put $N=6\ell-5$ and $\varepsilon=1/(24N)$.
    For $j\in[N]$, let $\delta_j=2j\varepsilon$ and let $P_j$ be a box with coordinate intervals
    \[
        \begin{cases}
            (1/2+\delta_j,1/2+\delta_j+\varepsilon)&\text{at }a_1\text{ and }b_1,\\
            (1/2-\delta_j-\varepsilon,1/2-\delta_j)&\text{at all the other nodes}.
        \end{cases}
    \]
    The intervals are disjoint in each coordinate, so some $P_j$ is not cut by any threshold.
    It is therefore entirely contained in some splitting box $\prod_a[L_a,U_a]$.
    Write $\delta=\delta_j$ and let $i$ select that splitting box.
    In particular,
    \[
        U_{a_1},U_{b_1}\ge1/2+\delta+\varepsilon,
        \qquad
        1-L_c\ge1/2+\delta+\varepsilon\quad\text{for all }c\notin\set{a_1,b_1},
    \]
    and both $U_c$ and $1-L_c$ are at least $1/2-\delta$ at every node.

    Consider $B=\set{a_1,b_1,b_2,b_3}$.
    We first show that any forest $F$ on a superset $X\supsetneq B$ costs at least $|X|(1/2-\delta)$.
    It suffices to show this separately for every connected component of $F$.
    An isolated node $c$ costs $1-L_c\ge1/2-\delta$.
    A tree $T$ on $r\ge2$ nodes costs $1+\sum_{c\in V(T)}(\deg_T(c)-1)U_c$.
    Since $\sum_{c\in V(T)}(\deg_T(c)-1)=r-2$ and $U_c\ge1/2-\delta$, this is at least $1+(r-2)(1/2-\delta)\ge r(1/2-\delta)$.
    Summing over the components gives the claim.
    As $|X|\ge5$ and $\delta\le\delta_N=1/12$, this lower bound is at least
    $5(1/2-\delta)\ge25/12>2+\varepsilon$.
    On $B$ itself, the only edges of $K_{3,3}$ are the three edges at $a_1$, so every forest on $B$ is a subgraph of the star centered at $a_1$.
    With zero, one, two, or three edges, its cost is at least
    \[
        2+2\varepsilon,\qquad 2+\varepsilon,\qquad
        2+\varepsilon,\qquad 2+2\delta+2\varepsilon,
    \]
    respectively.
    Thus $\hf_d^i(B)\ge2+\varepsilon$.
    Exchanging the two parts gives the same bound for $\set{a_1,a_2,a_3,b_1}$.
    By \eqref{eq:k33-two-decompositions},
    $\funwidth{\hf_d^i}(K_{3,3})\ge2+\varepsilon$.
    Since $\varepsilon=1/(24(6\ell-5)) > 1 / (144 \ell)$, this proves the claim.
\end{proof}

The lemmas above show that the degree-splitting framework captured by $\dslwidth$ does not yield an $\Oh(m^{\mwidth(K_{3,3})})$-time algorithm for \enciso{K_{3, 3}} for any constant $\ell$.
So by applying a width-based algorithm, all we get is time $\tOh(m^2)$, with a polylogarithmic overhead.
In \cref{sec:m-width-algo}, we show that nevertheless, \enciso{K_{3, 3}} can be solved in time $\Oh(m^2)$ by an algorithm that compares degrees directly instead of breaking them into groups.

\paragraph*{Logarithmically Many Degree Ranges} \label{sec:degree-uniformization}

Note that the exponent lower bound $2 + 1/(144\ell)$ from \cref{lm:k33-ds-width} decreases with $\ell$ and disappears when the number of degree ranges grows logarithmically in $m$.
In fact, if we allow logarithmically many degree ranges, degree splitting captures exactly \mwidth.

Consider degree splitting with $\ell_a \coloneqq \floor{\log_2 m} + 1$ degree ranges in every part and the thresholds $D_{a, j} \coloneqq 2^{j}$ for all $j \in \fragmentco{0}{\ell_a}$.
We call such degree splitting \emph{degree uniformization} because for every $i$, the nodes in each part of $G_i$ have degrees in $G$ that differ by at most a factor of two.
It reduces \Henciso on an arbitrary graph $G$ to $\Oh(\log^{|V(H)|} m)$ instances $G_i$, and each of them is handled exactly as in the degree-splitting framework.

What changes compared to \cref{def:dswidth} is that neighboring thresholds differ only by a constant factor: $D_{a, i_a} \le D_{a, i_a+1} \le 2 D_{a, i_a}$ for all $a$ and $i_a$.
Hence, it does not matter whether we bound the cost of a part by its lower or by its upper threshold: both bounds agree up to a constant factor.
So both $L_a$ and $U_a$ can be replaced by the same value $d_a \coloneqq \log_m D_{a, i_a} \in [0, 1]$.
An isolated node then costs $1 - d_a = 1 + (\deg_F(a) - 1) d_a$, so it follows the same formula as the nodes of larger components, and the cost of a forest $F$ in \cref{def:dswidth} simplifies to
\[
    \cc(F) + \sum_{\substack{a \in V(F)\\\deg_F(a) \ge 1}} (\deg_F(a) - 1) d_a - \sum_{\substack{a \in V(F)\\\deg_F(a) = 0}} d_a
    = \cc(F) + \sum_{a \in V(F)} (\deg_F(a) - 1) d_a.
\]
Moreover, the exponents $d_a$ are no longer fixed in advance: as $m$ grows, the vectors $(d_a)_{a \in V(H)}$ that occur come arbitrarily close to every point of $[0, 1]^{V(H)}$.
Thus, the minimum over splitting exponents and the maximum over splitting boxes in \cref{def:dswidth} turn into a maximum over all vectors $d \in [0, 1]^{V(H)}$.
This leads to the following definition.

\begin{definition}[\dslogwidth] \label{def:mwidth5}
    Let $H$ be a pattern and $d \in [0, 1]^{V(H)}$ be a real-valued vector.
    For a forest $F \in \cF_H(X)$ with $X \subseteq V(H)$, define its \emph{cost} as
    \[
        \val_d(F) \coloneqq \cc(F) + \sum_{a \in V(F)} (\deg_F(a) - 1) d_a,
    \]
    and let
    \[
        \mfun_d(X) \coloneqq \min_{F \in \cF_H(X)} \val_d(F).
    \]
    We define \[\dslogwidth(H) \coloneqq \max_{d\in[0,1]^{V(H)}}\funwidth{\hmfun_d}(H),\] where $\hmfun_d$ is the monotonic closure of $\mfun_d$.
    \lipicsEnd
\end{definition}
    Note that the maximum above exists: forest costs are affine in $d$, and taking finitely many minima and maxima preserves continuity on the compact cube $[0,1]^{V(H)}$.

Every forest can be built by repeatedly adding a new node $a$, either as an isolated node or as a leaf attached to a node $b$ that is already in the forest.
Adding $a$ as an isolated node increases the cost by $1-d_a$.
Attaching $a$ as a leaf to $b$ increases the cost by $d_b$.
This gives the following recurrence, which avoids enumerating forests.

\begin{lemma} \label{lm:m-equiv-4}
    Let $H$ be a pattern and $d \in [0, 1]^{V(H)}$ be a real-valued vector.
    Define a function $\mfun'_d \colon 2^{V(H)} \to \RR$ inductively.
    Let $\mfun'_d(\emptyset) \coloneqq 0$ and for every nonempty $X \subseteq V(H)$, let
    \[
        \mfun'_d(X) \coloneqq \min\set{
            \min_{a \in X} \bigl(\mfun'_d(X \setminus \set{a}) + 1 - d_a\bigr),
            \min_{\substack{a,b \in X\\\set{a,b} \in E(H)}} \bigl(\mfun'_d(X \setminus \set{a}) + d_b\bigr)
        }.
    \]

    We have $\mfun_d(X) = \mfun'_d(X)$ for all $X \subseteq V(H)$.
\end{lemma}

\begin{proof}
    Every nonempty forest has an isolated node or a leaf.
    Deleting an isolated node $a$ decreases its cost by $1-d_a$; deleting a leaf $a$ with neighbor $b$ decreases its cost by $d_b$.
    Conversely, adding $a$ to any forest on $X\setminus\set{a}$ as an isolated node or as a leaf adjacent to $b$ increases its cost by exactly the corresponding amount.
    Thus $\mfun_d$ satisfies the same recurrence as $\mfun'_d$.
    Since both functions are zero at the empty set, they agree on every set.
\end{proof}

The same two operations also characterize which monotone functions are bounded by $\hmfun_d$.

\begin{lemma}\label{lm:forest-increments}
    Fix $d\in[0,1]^{V(H)}$.
    The function $\hmfun_d$ is the pointwise largest monotone function $\psi\colon 2^{V(H)}\to\RR$ with $\psi(\emptyset)=0$ satisfying
    \begin{align*}
        \psi(X\cup\set{a})&\le\psi(X)+1-d_a
            &&\text{for every }a\notin X,\\
        \psi(X\cup\set{a})&\le\psi(X)+d_b
            &&\text{for every }a\notin X,\ b\in X,\ \set{a,b}\in E(H).
    \end{align*}
    Moreover, for every node $a$ and every edge $\set{a,b}$,
    \[
        \hmfun_d(\set{a})=1-d_a,
        \qquad
        \hmfun_d(\set{a,b})=\min\set{1,2-d_a-d_b}.
    \]
\end{lemma}

\begin{proof}
    Let $\psi$ satisfy the stated conditions.
    Build a forest $F$ one node at a time by adding isolated nodes and leaves.
    Summing the corresponding increment bounds gives $\psi(V(F))\le\val_d(F)$.
    If $F$ covers $X$, monotonicity gives $\psi(X)\le\val_d(F)$.
    Minimizing over covering forests proves $\psi(X)\le\hmfun_d(X)$.

    Conversely, $\hmfun_d$ is monotone by definition and is zero at the empty set.
    Take the forest attaining $\hmfun_d(X)$.
    If it already contains $a$, it also covers $X\cup\set{a}$ at no extra cost.
    Otherwise, add $a$ as an isolated node, or as a leaf adjacent to $b$ when $b\in X$ and $\set{a,b}\in E(H)$.
    These constructions prove the two increment bounds for $\hmfun_d$.

    Finally, an isolated node $a$ costs $1-d_a$, and every component containing an edge costs at least $1$.
    Thus every forest covering $a$ costs at least $1-d_a$, with equality attained by the isolated node.
    A forest covering both $a$ and $b$ either contains an edge, and costs at least one, or consists only of isolated nodes, and costs at least $2-d_a-d_b$.
    The one-edge forest and the two isolated nodes attain these bounds.
\end{proof}

Based on the above observations, the following lemma, whose proof is deferred to \cref{sec:appendix-m-width}, shows that \dslogwidth coincides with \mwidth as defined in \cref{def:mwidth1}.

\begin{restatable}{lemma}{mequivthree}\label{lm:m-equiv-3}
    For every pattern $H$, we have $\dslogwidth(H) = \mwidth(H)$.
    \lipicsEndNotRestated
\end{restatable}

Thus, degree splitting with $\floor{\log_2 m}+1$ degree ranges at the thresholds $2^{i}$ has time complexity exponent exactly $\mwidth(H)$, at the cost of an $\Oh(\log^{|V(H)|} m)$ overhead.
From now on, we also use \cref{def:mwidth5} as a characterization of \mwidth.

\section{\texorpdfstring{New Algorithm in \mwidth Time}{New Algorithm in m-width Time}} \label{sec:m-width-algo}

Degree uniformization pays a polylogarithmic factor because it treats the degree ranges separately.
In this section, we avoid this factor by comparing degrees directly.
As a warm-up, we do this for $K_{3,3}$, for which no constant number of degree ranges suffices by \cref{lm:k33-ds-width}.

\paragraph*{Warm-Up: \texorpdfstring{$K_{3,3}$}{K33}} \label{sec:k33-warmup}

In the proof of \cref{lm:k33-widths}, the node $b_1$ with the smallest value $\subf(\set{b_1})$ plays the role of a node of largest degree.
Choosing a node of largest degree can be implemented directly, without grouping degrees into ranges.

\begin{lemma}\label{lm:k33-opt-algo}
    There is an algorithm that solves \enciso{K_{3, 3}} in time $\Oh(m^2)$.
\end{lemma}

\begin{proof}
    Fix an order on the pattern nodes and compare host nodes $v_c$ by their \emph{scores} $(\deg_G(v_c),c)$, lexicographically.
    We build one partial encoding $\partenc_c$ for each $c\in V(K_{3,3})$, encoding precisely the $H$-subgraphs $\bv$ in which $v_c$ has the largest score.
    These six classes partition all occurrences, so their union is a full encoding.

    It suffices to describe $\partenc_{a_1}$; exchanging the two parts and permuting nodes within a part gives the other five.
    Use the decomposition $\cB$ from \cref{lm:k33-ds-width}.
    For each bag $B_j=\set{a_1,a_2,a_3,b_j}$, we construct the submaterialization consisting of the tuples satisfying
    \[
        \begin{gathered}
            \set{v_{a_2},v_{b_j}},\set{v_{a_3},v_{b_j}}\in E(G),\\
            (\deg_G(v_{a_1}),a_1)>
            (\deg_G(v_c),c)\quad(c\in B_j\setminus\set{a_1}).
        \end{gathered}
    \]
    An $H$-subgraph belongs to all three submaterializations exactly when its largest score is at $a_1$, as required.

    To build the submaterialization of $B_j$, fix $v_{a_1}$ and scan the edges $\set{v_{a_2},v_{b_j}}$.
    If both endpoints have smaller scores, enumerate the neighbors $v_{a_3}$ of $v_{b_j}$ in $V_{a_3}$ and retain those with smaller scores as well.
    For each scanned edge, this procedure takes $\Oh(\deg_G(v_{a_1}))$ time, since we enumerate neighbors only when $\deg_G(v_{b_j})\le\deg_G(v_{a_1})$.
    Hence, for a fixed $v_{a_1}$, the time is $\Oh(m\deg_G(v_{a_1}))$.
    Summing over $v_{a_1}$ gives time
    \[
        \Oh\left(m\sum_{v_{a_1}\in V_{a_1}}\deg_G(v_{a_1})\right)
        \subseteq\Oh(m^2).
    \]
    There are only three bags and six partial encodings, so the total running time is $\Oh(m^2)$.
\end{proof}

The above algorithm uses only degree information to guide its choices but avoids the idea of degree splitting and circumvents the lower bound of \cref{lm:k33-ds-width}.

\paragraph*{General Patterns} \label{sec:m-width-algo-general}

Inspired by \cref{lm:k33-opt-algo}, we now avoid the polylogarithmic factor for every pattern.
This is achieved by generating forest copies directly and stopping a branch as soon as its cost exceeds $m^{\mwidth(H)}$.
The resulting guarantees in the general case are weaker than in \cref{lm:k33-opt-algo}: the same $H$-subgraph may be captured by several tree decompositions, and thus we compute a \emph{relaxed} encoding.

\begin{lemma}\label{lm:new-mwidth-algo}
    Let $H$ be a pattern.
    There is an algorithm that, given a host graph $G$, solves \Hrelenciso on $G$ in time $\Oh(m^{\mwidth(H)})$.
\end{lemma}

\begin{proof}
    Write $w=\mwidth(H)$ and $M=m^w$.
    Since $H$ is fixed, we may precompute $w$, all forests in $H$, and all simple tree decompositions of $H$.
    Store the adjacency lists by color and order every part by nonincreasing degree in $\Oh(m)$ time using radix sort.
    (Here we use that all nodes of $G$ are represented by integers bounded by a polynomial in $n$.)

    \subparagraph{Generating forest copies.}
    Fix a forest $F\subseteq H$ with node set $X$.
    Root each component arbitrarily, direct its edges away from the root, and order the nodes so that every parent precedes its children.
    For a copy $\bv$ of $F$ in $G$, define its cost by
    \[
        P_F(\bv)\coloneqq
        \prod_{\substack{a\in X\\a\text{ a root}}}\frac{m}{\deg_G(v_a)}
        \prod_{a\to b\in E(\overrightarrow F)}\deg_G(v_a)
        =m^{\cc(F)}\prod_{a\in X}\deg_G(v_a)^{\deg_F(a)-1}.
    \]
    The second expression shows that the cost does not depend on the chosen roots.
    We build the set $S_{X,F}$ of all such copies with $P_F(\bv)\le M$.

    Generate the coordinates recursively in the chosen order, carrying the product $\Phi$ of the factors already introduced.
    Initially $\Phi=1$.
    For a root $a$, try the nodes $v_a\in V_a$ in nonincreasing degree order and recurse with cost $\Phi\cdot m/\deg_G(v_a)$ while this cost is at most $M$.
    For a non-root $b$ with parent $a$, recurse over the neighbors of the already chosen $v_a$ in $V_b$ with cost $\Phi\cdot\deg_G(v_a)$, provided this cost is at most $M$.
    At a complete tuple, add the tuple to $S_{X,F}$.
    Every factor in the definition of $P_F$ is at least $1$, so every copy of cost at most $M$ passes all its prefix tests.

    To bound the running time, assign weight $1/\Phi$ to a recursive call.
    The total weight of its children is at most its own weight.
    Indeed, at a root $a$ it is at most
    \[
        \sum_{v_a\in V_a}\frac{\deg_G(v_a)}{\Phi\cdot m}\le\frac1\Phi,
    \]
    since the sum of degrees in a single part is at most $m$.
    At a non-root, there are at most $\deg_G(v_a)$ children, each of weight $1/(\Phi\deg_G(v_a))$.
    Thus the total weight at every recursion level is at most one.
    Every generated call has $\Phi\le M$, and hence weight at least $1/M$, so each level has at most $M$ calls.
    There are $|X|+1\in\Oh(1)$ levels.
    Apart from its children, a call spends constant time.
    Consequently, $S_{X,F}$ is generated in $\Oh(M)$ time.

    \subparagraph{Building the encoding.}
    For each $B\subseteq V(H)$, form
    \[
        \hS_B\coloneqq
        \bigcup_{\substack{X\supseteq B\\F\in\cF_H(X)}}\prj{S_{X,F}}{B}.
    \]
    Use radix sort to remove duplicates.
    For every $\mathcal T\in\td(H)$, create the partial encoding with tree decomposition $\mathcal T$ and submaterializations $\hS_B$ for $B\in\mathcal T$.
    Return all these partial encodings.

    \medskip

    To prove that every copy of $H$ is encoded, fix an $H$-subgraph $\bu$ and set $d_a=\log_m\deg_G(u_a)$.
    By \cref{def:mwidth5}, there is some $\mathcal T\in\td(H)$ for which every bag $B$ has a covering forest $F$ with cost $\hmfun_d(B)\le w$.
    
    The tuple $\prj{\bu}{V(F)}$ is a copy of $F$ with cost
    $P_F(\prj{\bu}{V(F)})=m^{\hmfun_d(B)}\le M$.
    It is therefore generated, so $\prj{\bu}{B}\in\hS_B$ for every bag of $\mathcal T$.
    Hence $\bu$ is encoded by the partial encoding for $\mathcal T$.

    There are only constantly many forests, subsets, and tree decompositions, since $H$ is fixed.
    Generating all $S_{X,F}$, forming their projected unions, and building the partial encodings takes $\Oh(M+m)\subseteq\Oh(m^w)$ time.
\end{proof}

We leave as an open question the possibility of an $\Oh(m^{\mwidth(H)})$-time algorithm for \Henciso.

\section{Splitting over Clique Separators} \label{sec:width-hierarchy}

In this section we show that in many cases the computation of $\clemb$, $\subw$, $\sharpw$, and $\mwidth$ on a pattern graph $H$ can be reduced to the computation for smaller subgraphs of $H$.
Concretely, we show that this holds when $H$ contains a \emph{clique separator}.
Therefore, clique separators are an essential component in understanding the optimal time complexities of the different variants of $H$-subgraph isomorphism.

\begin{definition}[Clique Separator, \cite{Gavril77,Tarjan85}]
    Let $H$ be a graph and $C \subsetneq V(H)$ be a (possibly empty) subset of its nodes. We say that $C$ is a \emph{clique separator} in $H$ if $C$ forms a clique in $H$ and $H[V(H) \setminus C]$ is disconnected.
    \lipicsEnd
\end{definition}

\begin{definition}[Clique Separator Decomposition, {\cite[Definition 1.3]{BG24}}] \label{def:clique-separator-decomp}
  Let $H$ be a graph. We define $D(H)$ as the set of all induced subgraphs $H[V']$ of $H$ such that $H[V']$ has no clique separator and $V' \subseteq V(H)$ is an inclusionwise maximal subset with this property.
    \lipicsEnd
\end{definition}

\cite[Corollary 2.14]{BG24} states that to solve \Henciso, it is sufficient to solve \enciso{H'} for all $H' \in D(H)$.
In this paper we further show that to compute many graph parameters $p(H)$ related to $H$-subgraph isomorphism, it is sufficient to compute $p(H')$ for all $H' \in D(H)$.
We will require the following fact about clique separators from \cite{BG24}.

\begin{fact}[{\cite[Lemma 1.4]{BG24}}] \label{lm:d-set-computation}
    Let $H$ be a graph and $C$ be a clique separator in $H$ which is minimal, \ie, no proper subset of $C$ is a clique separator. Let $S_1, S_2, \ldots, S_t \subseteq V(H)$ be the connected components of $H[V(H) \setminus C]$. Then $D(H) = \bigcup_{i \in [t]} D(H[S_i \cup C])$.
    \lipicsEnd
\end{fact}

The following useful fact states that the clique embedding power of an induced minor gives a lower bound on the clique embedding power of the whole pattern graph.

\begin{fact}[Induced Minors for $\clemb$, {\cite[Lemma 2.5]{BG24}}]  \label{lm:clemb-of-induced-minors}
    Let $H$ and $H'$ be patterns such that $H'$ is an induced minor of~$H$. Then $\clemb(H) \ge \clemb(H')$.
    \lipicsEnd
\end{fact}

We now formally define what it means for a graph parameter to \emph{split over clique separators}.

\begin{definition}\label{def:split-over-clique-separators}
    We say that a real-valued graph parameter $p$ \emph{splits over clique separators} if $p(H) = \max_{H' \in D(H)} p(H')$ for all patterns $H$.
    \lipicsEnd
\end{definition}

That is, to compute a parameter $p$ that splits over clique separators for all patterns $H$, it is sufficient to compute it for clique-separator-free patterns.
\cref{lm:d-set-computation} implies that such a parameter $p$ can be computed recursively by finding a minimal clique separator in $H$.

As shown in the next lemma, $\clemb$, $\subw$, $\sharpw$, and $\mwidth$ split over clique separators.

\begin{lemma}\label{lm:everything-splits-over-clique-separators}
    The graph parameters $\clemb, \subw, \sharpw, \mwidth$ split over clique separators.
\end{lemma}

\begin{proof}
    If $H$ has no clique separator, then $D(H)=\set{H}$ and there is nothing to prove.
    Otherwise, let $C$ be a minimal clique separator of $H$, let $S_1,\ldots,S_t$ be the connected components of $H[V(H) \setminus C]$, and put $H_i \coloneqq H[S_i \cup C]$.
    We prove $p(H)=\max_i p(H_i)$ for each parameter in the statement.
    Applying this equality recursively proves the lemma by \cref{lm:d-set-computation}.

    First consider $p \in \set{\subw,\sharpw,\mwidth}$, using \cref{def:mwidth4} and \cref{lm:mwidth-submodular} for $\mwidth$.
    For a monotone function $f \colon 2^{V(H)} \to \RR$, let $f_i$ be its restriction to $2^{V(H_i)}$.
    We have
    \begin{equation}\label{eq:function-width-clique-separator}
        \funwidth{f}(H)
        = \max_i \funwidth{f_i}(H_i).
    \end{equation}
    Indeed, intersecting all bags of a tree decomposition of $H$ with $V(H_i)$ gives a tree decomposition of $H_i$ and cannot increase $f$.
    Conversely, take an optimal decomposition of each $H_i$.
    Each has a bag containing the clique $C$; connect these bags in a star to obtain a decomposition of $H$ with the same maximum bag value.

    Restricting an admissible function on $H$ to $H_i$ preserves all its defining conditions.
    Thus \cref{eq:function-width-clique-separator} gives $p(H) \le \max_i p(H_i)$.
    For the reverse inequality, extend an admissible function $g$ on $H_i$ by $f(X) \coloneqq g(X \cap V(H_i))$.
    This extension is admissible on $H$: intersection preserves unions and intersections, and the intersection of any edge of $H$ with $V(H_i)$ is contained in an edge of $H_i$.
    For the latter fact, note that $H_i$ is connected and has at least two nodes, so even a singleton intersection is contained in an edge.
    Equation~\eqref{eq:function-width-clique-separator} now gives $p(H) \ge p(H_i)$ for every $i$.
    This argument for $\subw$ already appears in \cite[Lemma D.5]{BG24}.

    \smallskip

    For $\clemb$, the inequality $\clemb(H) \ge \max_i \clemb(H_i)$ follows from \cref{lm:clemb-of-induced-minors}.
    To prove the other direction, consider an embedding $\psi$ of $K_k$ into $H$.
    If some set $\psi(u)$ avoids $C$, it lies in a single component $S_i$.
    Every other set touches $\psi(u)$ and hence intersects with $S_i \cup C$.
    If instead every set intersects with $C$, choose any $i$.
    In either case define
    \[
        \psi'(v) \coloneqq \psi(v) \cap (S_i \cup C).
    \]
    These sets are nonempty and connected: any part of a connecting path outside $S_i \cup C$ can be replaced by an edge of the clique $C$.
    They also pairwise touch.
    Two sets intersecting with $C$ touch inside $C$.
    A set $\psi(v)$ that avoids $C$ can only occur in the first case; it is connected and touches $\psi(u) \subseteq S_i$, so it lies inside $S_i$.
    Then $\psi'(v)=\psi(v)$, and its ``contact'' with every other set is preserved.
    Thus $\psi'$ is an embedding into $H_i$, and $\wed(\psi') \le \wed(\psi)$ because we only removed nodes from the sets.
    Consequently $k/\wed(\psi) \le \clemb(H_i)$.
    Taking the supremum over $\psi$ gives the remaining inequality.
\end{proof}

Even though $\dswidth{\ell}$ need not split over clique separators for a fixed $\ell$ (different subgraphs may want to choose different splitting degrees for the common clique nodes), this can be circumvented by increasing the parameter $\ell$.

\begin{lemma}\label{lm:ds-clique-refinement}
    Let $D(H)=\set{H_1,\ldots,H_t}$ and choose integers $\ell_1,\ldots,\ell_t\ge1$.
    For $L\coloneqq1+\sum_{i=1}^t(\ell_i-1)$, we have
    \begin{equation}\label{eq:ds-clique-refinement}
        \dswidth{L}(H)\le\max_{i\in[t]}\dswidth{\ell_i}(H_i).
    \end{equation}
    In particular, if $\dswidth{\ell_i}(H_i)=\mwidth(H_i)$ for every $i$, then $\dswidth{L}(H)=\mwidth(H)$.
\end{lemma}

\begin{proof}
    Choose optimal splitting exponents for each $H_i$.
    At each node $a$ of $H$, take the union of all internal thresholds chosen for $a$ in the pieces containing it.
    There are at most $\sum_i(\ell_i-1)=L-1$ such thresholds.
    Together with $0$ and $1$, they define a splitting into at most $L$ ranges; repeat the final threshold $1$ to obtain exactly $L$ ranges at every node.

    Fix a box of this global splitting.
    Its restriction to each $H_i$ is contained in a box of the chosen local splitting.
    A forest in $H_i$ has no larger cost in the global box: the lower degree bounds only increase and the upper degree bounds only decrease.
    Moreover, every forest in $H_i$ covering a local bag is also a forest in $H$ covering that bag.
    Thus the global monotonic closure on each local bag is at most the corresponding local monotonic closure.
    Choose a local tree decomposition of width at most $\dswidth{\ell_i}(H_i)$ for each piece and glue them recursively along clique separators, as in the preceding proof.
    Every bag of the resulting decomposition has global value at most $\max_i\dswidth{\ell_i}(H_i)$.
    This holds for every global box and proves the first claim.

    If each local width equals $\mwidth(H_i)$, the right-hand side of \eqref{eq:ds-clique-refinement} is $\mwidth(H)$ by \cref{lm:everything-splits-over-clique-separators}.
    The reverse inequality $\mwidth(H)\le\dswidth{L}(H)$ follows from \cref{lm:width-hierarchy-ds}.
\end{proof}

Therefore, when searching for minimal patterns $H$ satisfying $\mwidth(H) < \dswidth{\ell}(H)$ for every $\ell \ge 1$, it is sufficient to consider clique-separator-free patterns $H$.

\section{Extreme Bounds} \label{sec:extreme-bounds}

In this section we bound the graph parameters defined in this paper in terms of treewidth.
The following lemma shows that $\subw$, $\sharpw$, $\mwidth$, and $\dslwidth$ are within a factor of two of each other.

\begin{lemma} \label{lm:tw-vs-widths}
    For a pattern $H$ and an integer $\ell \ge 1$, we have
    \[(\tw(H) + 1) / 2 \le \subw(H) \le \sharpw(H) \le \mwidth(H) \le \dswidth{\ell}(H) \le \tw(H) + 1.\]
\end{lemma}

\begin{proof}
    The function $\subf(X) \coloneqq |X|/2$ is monotone, edge-dominated, and submodular, with $\subf(\emptyset)=0$.
    Its width is exactly $(\tw(H)+1)/2$, so $\subw(H) \ge (\tw(H)+1)/2$.

    For the upper bound, take a tree decomposition with bags of size at most $\tw(H)+1$.
    For arbitrary splitting exponents $d$ and a choice $i$ of degree classes, the edgeless forest on a bag $B$ gives
    \[
        \hf_d^i(B) \le f_d^i(B)
        \le \sum_{a \in B}(1-d_a^{i_a})
        \le |B| \le \tw(H)+1.
    \]
    Thus $\dslwidth(H) \le \tw(H)+1$.
    The remaining inequalities follow from \cref{lm:width-hierarchy,lm:width-hierarchy-ds}.
\end{proof}

To show that the upper bound is attained, we use the following family of \emph{parallel-path graphs}.

\begin{definition}[$P$-graphs, {\cite[Definition 1.5]{BG24}}]\label{def:par-graphs}
    For positive integers $\ell_1, \ell_2, \ldots, \ell_k$, the \emph{parallel-path graph} (or \emph{$P$-graph} for short) $P(\ell_1, \ell_2, \ldots, \ell_k)$ is constructed by starting from path graphs of lengths $\ell_1, \ell_2, \ldots, \ell_k$ and then identifying all their start nodes and identifying all their end nodes. Note that at most one of $\ell_1, \ell_2, \ldots, \ell_k$ can be~1 (as otherwise we would not obtain a simple graph).
    \lipicsEnd
\end{definition}

The following lemma shows that both the upper and the lower bounds from \cref{lm:tw-vs-widths} can be achieved for every treewidth $t \ge 2$.

\begin{lemma}\label{lm:extreme-widths}
    For every $t \ge 1$ and $\ell \ge 2$, we have $\tw(K_{t+1}) = t$ and \[\clemb(K_{t+1}) = \subw(K_{t+1}) = \sharpw(K_{t+1}) = \mwidth(K_{t+1}) = \dswidth{\ell}(K_{t+1}) = (t + 1) / 2.\]
    Furthermore, for every $t \ge 2$, there is a pattern $H^{\star}$ with $\tw(H^{\star}) = t$ and \[\clemb(H^{\star}) = \subw(H^{\star}) = \sharpw(H^{\star}) = \mwidth(H^{\star}) = \dswidth{\ell}(H^{\star}) = t + 1\] for all $\ell \ge 1$.
\end{lemma}

\begin{proof}
    Write $n=t+1$.
    Clearly, $\tw(K_n) = t$.
    A trivial embedding of $K_n$ into $K_n$ gives $\clemb(K_n) \ge n / 2$.

    For the upper bound on $\dswidth{2}$, use the one-bag decomposition $B=V(K_{t+1})$ and split every degree at exponent $1/2$.
    Fix any choice of high- and low-degree nodes.
    If $n$ is even, a perfect matching on $B$ has forest cost $n/2$.
    If $n$ is odd, choose a node $a$ and match the other $n-1$ nodes.
    When $a$ is high-degree, leaving it isolated adds cost $1/2$.
    When $a$ is low-degree, replace one matched edge $\set{b,c}$ by the two edges $\set{a,b}$ and $\set{a,c}$; this three-node tree has cost $1+1/2$.
    In both cases the total cost is $n/2$.
    Thus $\dswidth{2}(K_{t+1}) \le n/2$, and all claimed equalities for $K_{t+1}$ follow from \cref{lm:width-hierarchy,lm:width-hierarchy-ds}.

    \medskip

    For $H^{\star}$, start with nodes $v_1,\ldots,v_n$ and connect each pair by four internally disjoint paths of length three.
    In other words, $H^{\star}$ is a clique $K_n$ in which every edge is replaced by a copy of $P(3,3,3,3)$; see \cref{fig:clique-with-p3333-edges}.
    For each $i,j \in [n]$ with $i < j$, index the four paths between $v_i$ and $v_j$ by $z\in\set{a,b,c,d}$ and write the path indexed by $z$ as
    \[
        v_i,\ z_{i,j},\ z_{j,i},\ v_j.
    \]
    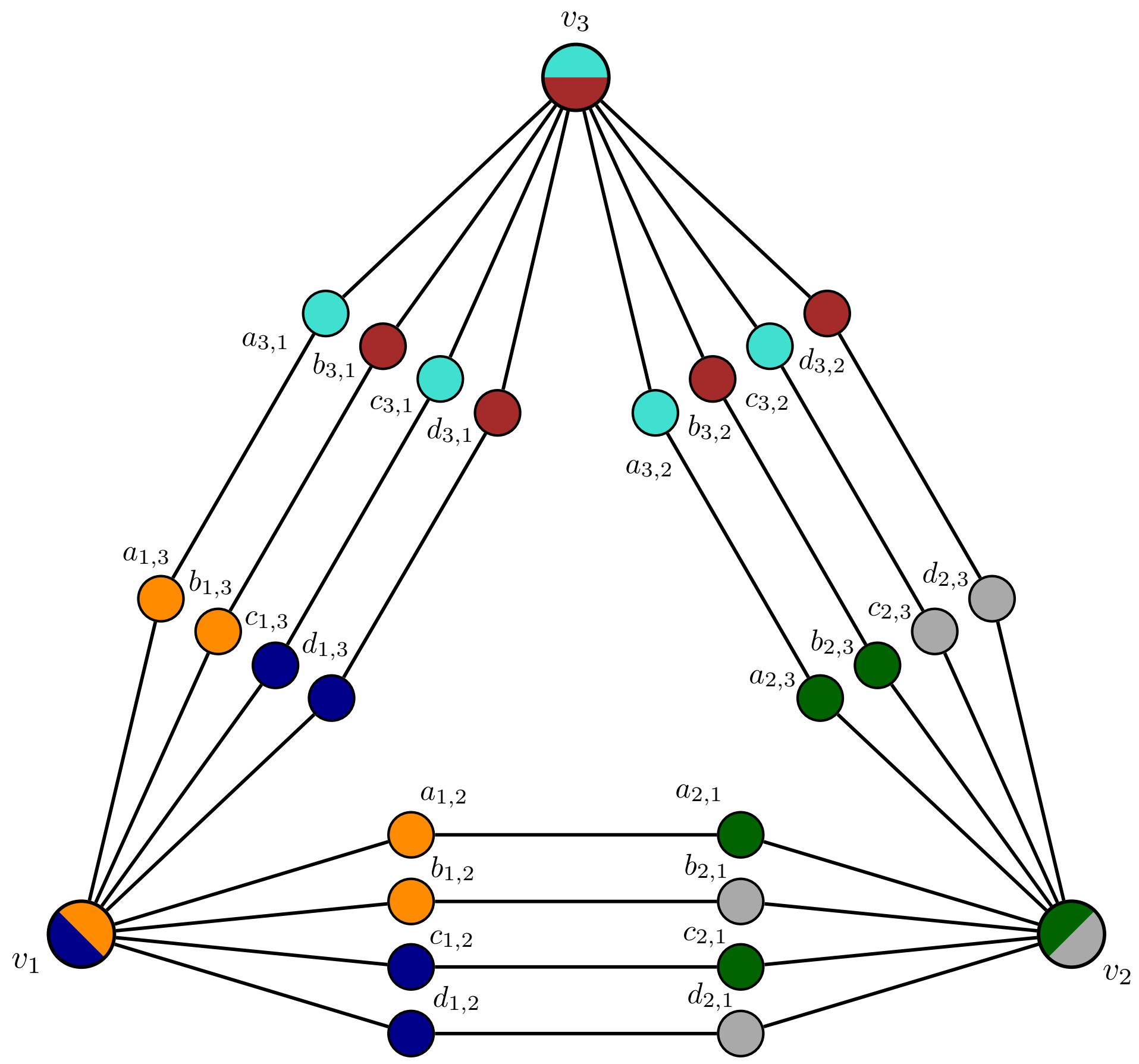
\begin{figure}
        \begin{center}
            \begin{tikzpicture}[scale=1.3,
        sub/.style={circle, draw, line width=0.8pt, inner sep=0pt, minimum size=14.3pt},
        corner/.style={circle, draw, line width=1.1pt, inner sep=0pt, minimum size=20.8bp},
        lab/.style={inner sep=0pt, font=\small},
        ed/.style={line width=1.1pt}]
    \definecolor{cturquoise}{rgb}{0.251,0.878,0.816}
    \definecolor{cbrown}{rgb}{0.647,0.165,0.165}
    \definecolor{cdarkblue}{rgb}{0,0,0.545}
    \definecolor{cdarkorange}{rgb}{1,0.549,0}
    \definecolor{cdarkgreen}{rgb}{0,0.392,0}
    \definecolor{cdarkgray}{rgb}{0.663,0.663,0.663}
    \foreach \i/\p/\a/\ca/\cb in {1/{(0,0)}/45/cdarkorange/cdarkblue, 2/{(8.47,0)}/135/cdarkgreen/cdarkgray, 3/{(4.23,7.33)}/90/cturquoise/cbrown} {
        \fill[\ca] \p ++(\a-90:8bp) arc[start angle=\a-90, delta angle=180, radius=8bp] -- cycle;
        \fill[\cb] \p ++(\a+90:8bp) arc[start angle=\a+90, delta angle=180, radius=8bp] -- cycle;
        \node[corner] (v\i) at \p {};
    }
    \node[lab, font=\normalsize, anchor=north east] at (-0.33,-0.16) {$v_1$};
    \node[lab, font=\normalsize, anchor=north west] at (8.73,-0.26) {$v_2$};
    \node[lab, font=\normalsize, anchor=south] at (4.23,7.72) {$v_3$};
    \node[sub, fill=cdarkorange] (a12) at (2.82,0.85) {}; \node[lab, anchor=south west] at (2.89,1.08) {$a_{1,2}$};
    \node[sub, fill=cdarkorange] (b12) at (2.82,0.28) {}; \node[lab, anchor=south west] at (2.98,0.42) {$b_{1,2}$};
    \node[sub, fill=cdarkblue] (c12) at (2.82,-0.28) {}; \node[lab, anchor=south west] at (2.97,-0.15) {$c_{1,2}$};
    \node[sub, fill=cdarkblue] (d12) at (2.82,-0.85) {}; \node[lab, anchor=south west] at (3,-0.69) {$d_{1,2}$};
    \node[sub, fill=cdarkgreen] (a21) at (5.64,0.85) {}; \node[lab, anchor=south east] at (5.5,1.1) {$a_{2,1}$};
    \node[sub, fill=cdarkgray] (b21) at (5.64,0.28) {}; \node[lab, anchor=south east] at (5.55,0.44) {$b_{2,1}$};
    \node[sub, fill=cdarkgreen] (c21) at (5.64,-0.28) {}; \node[lab, anchor=south east] at (5.54,-0.12) {$c_{2,1}$};
    \node[sub, fill=cdarkgray] (d21) at (5.64,-0.85) {}; \node[lab, anchor=south east] at (5.6,-0.66) {$d_{2,1}$};
    \node[sub, fill=cdarkorange] (a13) at (0.68,2.87) {}; \node[lab, anchor=south east] at (0.77,3.13) {$a_{1,3}$};
    \node[sub, fill=cdarkorange] (b13) at (1.17,2.59) {}; \node[lab, anchor=south] at (1.11,2.87) {$b_{1,3}$};
    \node[sub, fill=cdarkblue] (c13) at (1.66,2.3) {}; \node[lab, anchor=south] at (1.59,2.58) {$c_{1,3}$};
    \node[sub, fill=cdarkblue] (d13) at (2.14,2.02) {}; \node[lab, anchor=south] at (2.09,2.34) {$d_{1,3}$};
    \node[sub, fill=cturquoise] (a31) at (2.09,5.31) {}; \node[lab, anchor=north east] at (1.79,5.15) {$a_{3,1}$};
    \node[sub, fill=cbrown] (b31) at (2.58,5.03) {}; \node[lab, anchor=north east] at (2.37,4.98) {$b_{3,1}$};
    \node[sub, fill=cturquoise] (c31) at (3.07,4.75) {}; \node[lab, anchor=north east] at (2.86,4.61) {$c_{3,1}$};
    \node[sub, fill=cbrown] (d31) at (3.56,4.46) {}; \node[lab, anchor=east] at (3.37,4.3) {$d_{3,1}$};
    \node[sub, fill=cdarkgreen] (a23) at (6.32,2.02) {}; \node[lab, anchor=east] at (6.13,2.17) {$a_{2,3}$};
    \node[sub, fill=cdarkgreen] (b23) at (6.81,2.3) {}; \node[lab, anchor=south east] at (6.63,2.36) {$b_{2,3}$};
    \node[sub, fill=cdarkgray] (c23) at (7.3,2.59) {}; \node[lab, anchor=east] at (7.12,2.75) {$c_{2,3}$};
    \node[sub, fill=cdarkgray] (d23) at (7.79,2.87) {}; \node[lab, anchor=south east] at (7.61,2.94) {$d_{2,3}$};
    \node[sub, fill=cturquoise] (a32) at (4.91,4.46) {}; \node[lab, anchor=north] at (4.86,4.06) {$a_{3,2}$};
    \node[sub, fill=cbrown] (b32) at (5.4,4.75) {}; \node[lab, anchor=north] at (5.38,4.45) {$b_{3,2}$};
    \node[sub, fill=cturquoise] (c32) at (5.89,5.03) {}; \node[lab, anchor=north] at (5.87,4.63) {$c_{3,2}$};
    \node[sub, fill=cbrown] (d32) at (6.38,5.31) {}; \node[lab, anchor=north] at (6.34,5.02) {$d_{3,2}$};
    \foreach \z in {a,b,c,d} {
        \draw[ed] (v1) -- (\z12) -- (\z21) -- (v2);
        \draw[ed] (v1) -- (\z13) -- (\z31) -- (v3);
        \draw[ed] (v2) -- (\z23) -- (\z32) -- (v3);
    }
\end{tikzpicture}
        \end{center}
    
        \caption{A $K_3$ with every edge replaced by a copy of $P(3,3,3,3)$. This pattern has $\clemb = \subw = \sharpw = \mwidth = \dslwidth = \tw+1=3$. The colors give an embedding of $K_6$ with weak edge depth two: each color represents one node of $K_6$ and the set it is embedded into.}
        \label{fig:clique-with-p3333-edges}
    \end{figure}

    We now prove that $\tw(H^{\star})=t$.
    A tree decomposition starts with the bag $\set{v_1,\ldots,v_n}$.
    For each path $v_i,z_{i,j},z_{j,i},v_j$, attach a bag $\set{v_i,v_j,z_{i,j}}$ to this central bag and then a bag $\set{v_j,z_{i,j},z_{j,i}}$ to the new bag.
    Every bag has size at most $n$, so $\tw(H^{\star}) \le t$.
    Also $K_n$ is a minor of $H^{\star}$, so $\tw(H^{\star})=t$.

    Next, we prove $\clemb(H^\star) \ge n$.
    Let
    \[
        L_0=\set{a,b},\quad L_1=\set{c,d},\qquad
        R_0=\set{a,c},\quad R_1=\set{b,d}.
    \]
    Now label the nodes of $K_{2n}$ by $(i,s) \in [n] \times \set{0,1}$ and set
    \[
        \psi(i,s) \coloneqq \set{v_i}
        \cup \setof{z_{i,j}}{j>i,\ z\in L_s}
        \cup \setof{z_{i,j}}{j<i,\ z\in R_s}.
    \]
    Each set is a star centered at $v_i$.
    The two sets at the same center intersect there.
    For $i<j$ and $s,r\in\set{0,1}$, let $z$ be the unique letter in $L_s\cap R_r$; then $\psi(i,s)$ and $\psi(j,r)$ touch along the edge $\set{z_{i,j},z_{j,i}}$.
    Hence $\psi$ is a clique embedding.
    It is clear that $\wed(\psi)=2$.
    Therefore, $\clemb(H^{\star}) \ge 2n/2=n$.
    Together with \cref{lm:tw-vs-widths,lm:width-hierarchy,lm:width-hierarchy-ds}, this proves all claimed equalities for $H^{\star}$.
\end{proof}

For $\clemb$, only a polynomially weaker lower bound compared to \cref{lm:tw-vs-widths} is possible.

\begin{lemma} \label{lm:tw-vs-clemb}
    For every pattern $H$, we have
    \[
        \clemb(H) \in \widetilde{\Omega}\bigl(\sqrt{\tw(H)}\bigr)
        \qquad\text{and}\qquad
        \clemb(H)\le\tw(H)+1.
    \]
\end{lemma}

\begin{proof}
    The upper bound follows from \cref{lm:tw-vs-widths,lm:width-hierarchy}.
    Write $t=\tw(H)$.
    A \emph{bramble} is a family of connected subgraphs that pairwise touch.
    Its \emph{order} is the minimum size of a node set intersecting every subgraph of the bramble, and its \emph{congestion} is the maximum number of subgraphs containing one node.
    By \cite[Theorem~1.1]{HatzelKPS22}, $H$ has a bramble $\mathcal{B}$ of congestion two and order
    \[
        r\in \widetilde{\Omega}(\sqrt{t}).
    \]

    Consider a clique with one node for each member of $\mathcal{B}$, and embed each of its nodes into the node set of the corresponding member.
    This is a clique embedding because the members are connected and pairwise touch.
    Moreover, $|\mathcal{B}|\ge r$: choosing one node from each member gives a hitting set of size at most $|\mathcal{B}|$.
    Finally, each endpoint of an edge belongs to at most two members, so every edge meets at most four images.
    Thus
    \[
        \clemb(H)\ge\frac{|\mathcal{B}|}{4}
        \ge\frac r4
        \in \widetilde{\Omega}(\sqrt{t}).\qedhere
    \]
\end{proof}

The following two lemmas show that the square-root dependence in \cref{lm:tw-vs-clemb} is tight up to polylogarithmic factors.

\begin{lemma} \label{lm:bound-on-clemb-from-num-of-edges}
    For a pattern $H$, we have $\clemb(H) \le \sqrt{|E(H)|}$.
\end{lemma}

\begin{proof}
    Consider an embedding $\psi$ of $K_k$ into $H$.
    Every ordered pair $(u,v) \in V(K_k)^2$, including $u=v$, is witnessed by an edge of $H$ intersecting both $\psi(u)$ and $\psi(v)$.
    Indeed, for disjoint sets, use an edge witnessing that they touch; for intersecting sets, use any edge incident to a node in their intersection.
    An edge $e$ witnesses exactly $d_\psi(e)^2$ ordered pairs, so
    \[
        k^2 \le \sum_{e \in E(H)} d_\psi(e)^2
        \le |E(H)|\,\wed(\psi)^2.
    \]
    Hence $k/\wed(\psi) \le \sqrt{|E(H)|}$, and taking the supremum over all embeddings proves the lemma.
\end{proof}

\begin{lemma} \label{lm:expander}
    There are patterns $H$ of arbitrarily large treewidth such that
    $\clemb(H)\in\Oh(\sqrt{\tw(H)})$.
\end{lemma}

\begin{proof}
    Fix constants $\Delta\ge3$ and $\varepsilon>0$ and an infinite family of $\Delta$-regular $\varepsilon$-vertex-expanders; such families exist, see, e.g., \cite{HooryLW06}.

    Write $n=|V(H)|$ and $t=\tw(H)$.
    By \cite[Proposition~1]{GroheM09}, vertex expansion at least $\varepsilon$ implies
    $t\ge \varepsilon n/4-1$.
    Therefore
    \[
        |E(H)|=\frac{\Delta n}{2}
        \le \frac{2\Delta}{\varepsilon}(t+1)
        \in\Oh(t).
    \]
    By \cref{lm:bound-on-clemb-from-num-of-edges},
    $\clemb(H)\le\sqrt{|E(H)|}\in\Oh(\sqrt{t})$.
\end{proof}

\section{Implementation} \label{sec:implementation}

In this section we discuss how we implement the algorithms computing $\clemb, \subw, \sharpw, \mwidth$, and $\tw$.
We first discuss basic implementations, and then provide further optimizations.

\subparagraph{Usage.}
Our tool is available at~\cite{WidthCalculator}.
To run it from the root of its repository, use
\begin{center}
    \verb+python width-calculator.py [options]+,
\end{center}
replacing \verb+[options]+ with the desired flags.
The flags \verb+--clemb+, \verb+--subw+, \verb+--sharp_subw+, \verb+--m_width+, and \verb+--treewidth+ select the parameters to compute.
The flags \verb+-n+ and \verb+-m+ select ranges of node and edge counts, respectively, while \verb+-a+ supplies explicit graphs.
Repeating \verb+-v+ controls the amount of console output.
The README at~\cite{WidthCalculator} gives installation instructions, a complete description of the flags, and further examples.

\subsection{Linear Programs and Mixed Integer Linear Programs}
\label{sec:prelims-milp}

A \emph{linear program} (LP) optimizes a linear objective function over real-valued variables subject to finitely many linear equality and inequality constraints.
An assignment to the variables is \emph{feasible} if it satisfies all constraints, and an optimal solution is a feasible assignment attaining the minimum or maximum objective value, depending on the direction of optimization.
A \emph{mixed integer linear program} (MILP) is a linear program in which some variables are restricted to be integers.
If the ranges of all variables in a mixed integer linear program are bounded, one can simulate constraints of the form $y = \min\set{x_1, \ldots, x_k}$ or $y = \max\set{x_1, \ldots, x_k}$, where $y$ is a variable and $x_1, \ldots, x_k$ are variables or constants.
Specifically, suppose that all variables of a mixed integer linear program are bounded in $[L, R]$.
To emulate the constraint $y = \min\set{x_1, \ldots, x_k}$, we introduce $k$ integer variables $c_1, \ldots, c_k$ and set their ranges to be $\set{0, 1}$.
We add an equality $c_1 + \cdots + c_k = 1$ and inequalities $y \le x_i$ and $y \ge x_i - (R - L) \cdot (1 - c_i)$ for all $i \in \fragmentcc{1}{k}$.
The inequalities $y \le x_i$ give $y \le \min_i x_i$, while the unique index $i$ with $c_i = 1$ also gives $y \ge x_i$; hence they enforce $y = \min_i x_i$.
Similarly, we can encode $y = \max_i x_i$ by replacing $y \le x_i$ by $y \ge x_i$ and $y \ge x_i - (R-L)(1-c_i)$ by $y \le x_i + (R-L)(1-c_i)$.

We solve mixed integer linear programs with the solver SCIP~\cite{BolusaniEtAl24} through the pywraplp interface of Google OR-Tools~\cite{ortools}.

\subsection{Basic Implementations}

For computing all minimal tree decompositions, we rely on elimination orders.
An \emph{elimination order} is a permutation $v_1,\ldots,v_h$ of the nodes of the pattern, where $h=|V(H)|$ (see, e.g.,~\cite{BodlaenderK10} for the correspondence between elimination orders and tree decompositions).
For each order, we construct a tree decomposition of $H$.
Start with $H_1 = H$ and iterate over $v_1,\ldots,v_h$.
When considering $v_i$, create a bag $B_i$ consisting of $v_i$ and all its neighbors in $H_i$.
After that, create a graph $H_{i+1}$ from $H_i$ by connecting all pairs of neighbors of $v_i$ and removing node $v_i$ from the graph.
At the end of this process, the bags form a tree decomposition: in reverse order, $B_i$ can be attached to the bag of the first subsequently eliminated neighbor of $v_i$, which contains all the other such neighbors because they were made into a clique.
At the same time, for each edge $\set{v_i, v_j} \in E(H)$, both nodes lie in $B_{\min\set{i, j}}$.
We remove bags contained in other bags and, after processing all orders, retain only the minimal collections under $\succeq$ (see \cref{sec:prelims-treedecomp}).
To prove that all minimal simple tree decompositions occur in the final list, consider one such decomposition and a leaf bag $L$ with neighboring bag $P$.
Every node in $L\setminus P$ occurs only in $L$, so we may eliminate all these nodes and then delete $L$; throughout these eliminations, all current neighbors of an eliminated node lie in $L$, including neighbors introduced by earlier eliminations.
Repeating this from the leaves inward, and eliminating the nodes of the last bag arbitrarily, produces an order whose generated bags are contained in bags of the original decomposition.
The resulting simple tree decomposition is therefore $\succeq$-dominated by the original one and, by minimality, must coincide with it.
Given the tree decompositions, we can compute treewidth as the minimum, over the decompositions, of the maximum bag size minus one.

For computing $\clemb$, we use the mixed integer linear program from \cite[Section 5.2]{FanKZ23}.
For computing $\subw$, we first recall the following classical alternative characterization of submodular functions.

\begin{observation}\label{obs:alt-submodularity}
    Let $H$ be a pattern.
A function $\subf \colon 2^{V(H)} \to \RR$ is submodular if and only if \[\subf(X \cup \set{v}) + \subf(X \cup \set{u}) \ge \subf(X) + \subf(X \cup \set{v, u})\] holds for all $X \subseteq V(H)$ and $v, u \in V(H) \setminus X$ with $v \neq u$.
    \lipicsEnd
\end{observation}
Having the observation above, we use the following mixed integer linear program implicit in \cref{def:subw} to compute $\subw$:
\begin{align*}
        \noalign{$\max \fwidth(H)$}\\
        &\text{s.t. }&\text{(i) }&\subf(X) \in [0, |V(H)|]&\forall X \subseteq V(H)\\
        &&\text{(ii) }&\subf(\emptyset) = 0&\\
        &&\text{(iii) }&\subf(X) \le \subf(Y)&\forall X \subseteq Y \subseteq V(H)\\
        &&\text{(iv) }&\subf(e) \le 1&\forall e \in E(H)\\
        &&\text{(v) }&\subf(X \cup \{v\}) + \subf(X \cup \{u\}) \ge \subf(X) + \subf(X \cup \{v, u\})&\forall X \subseteq V(H), v \neq u \in V(H) \setminus X.
\end{align*}
Condition~(i) does not change the feasible objective values: monotonicity, submodularity, and edge domination imply $0 \le \subf(X) \le |X| \le |V(H)|$, since every node of a pattern lies in some edge.
The explicit definition of $\fwidth(H)$ contains alternating $\min$ and $\max$ operators.
These can be encoded in a mixed integer linear program as explained in \cref{sec:prelims-milp}.
In our implementation, we impose monotonicity only for pairs $X \subseteq Y$ with $|Y|=|X|+1$, which is equivalent to~(iii) by transitivity.

Similarly, for computing $\sharpw$, we use the following mixed integer linear program implicit in \cref{def:sharpw}:
\begin{align*}
        \noalign{$\max \fwidth(H)$}\\
        &\text{s.t. }&\text{(i) }&\subf(X) \in [0, |V(H)|]&\forall X \subseteq V(H)\\
        &&\text{(ii) }&\subf(\emptyset) = 0&\\
        &&\text{(iii) }&\subf(X) \le \subf(Y)&\forall X \subseteq Y \subseteq V(H)\\
        &&\text{(iv) }&\subf(e) \le 1&\forall e \in E(H)\\
        &&\text{(v) }&\subf(X \cup Y) + \subf(X \cap Y) \le \subf(X) + \subf(Y)&\forall X, Y \subseteq V(H), e \in E(H), X \cap Y \subseteq e.
\end{align*}

For computing $\mwidth$, we use the following mixed integer linear program implicit in \cref{def:mwidth5,lm:forest-increments}.
For a pattern $H$, its variables are $d_a$ for every $a \in V(H)$ and $\psi(X)$ for every $X \subseteq V(H)$:
\begin{align*}
    \noalign{$\max \funwidth{\psi}(H)$}\\
    &\text{s.t. }&\text{(i) }&d_a \in [0, 1]&\forall a \in V(H)\\
    &&\text{(ii) }&\psi(X) \in [0, |V(H)|]&\forall X \subseteq V(H)\\
    &&\text{(iii) }&\psi(\emptyset) = 0&\\
    &&\text{(iv) }&\psi(X) \le \psi(Y)&\forall X \subseteq Y \subseteq V(H)\\
    &&\text{(v) }&\psi(X \cup \set{a}) \le \psi(X) + (1 - d_a)&\forall X \subseteq V(H), a \in V(H) \setminus X\\
    &&\text{(vi) }&\psi(X \cup \set{a}) \le \psi(X) + d_b&\forall X \subseteq V(H), a \in V(H) \setminus X, b \in X, \set{a, b} \in E(H).
\end{align*}
For fixed $d$, \cref{lm:forest-increments} says that the pointwise largest feasible function $\psi$ is $\hmfun_d$.
Since $\funwidth{\psi}(H)$ is monotone in the values of $\psi$, the optimum is $\max_{d \in [0,1]^{V(H)}}\funwidth{\hmfun_d}(H)=\dslogwidth(H)$, which equals $\mwidth(H)$ by \cref{lm:m-equiv-3}.

The graph database \cite{BDM} lists all connected graphs with a given number of nodes or edges.
Our tool includes from it all connected graphs on up to eight nodes, and all connected graphs on up to twelve edges.
For iterating over all clique-separator-free patterns with a fixed number of nodes or edges, we filter out graphs that have clique separators.
(Note that a disconnected graph has a clique separator of size zero.)

\subsection{Optimizations}\label{sec:implementation:optimizations}

The two main bottlenecks for the running time of the software tool are the enumeration of minimal tree decompositions and the MILP solvers.
The tool is written in Python; to speed up the former, the enumeration of tree decompositions is implemented in C++.
For the latter, in this section, we describe three ideas to reduce the cost of the MILPs: discarding bags whose value cannot affect the answer, reusing bounds between the width notions, and using the degree exponents found for \mwidth as a hint for solving other MILPs.

\subparagraph{Discarding bags.}
In all three width MILPs, the objective is the maximum of $\funwidth{f}(H)$ over feasible functions $f$.
Write $L \coloneqq \clemb(H)$, which is a lower bound on each of the three widths by \cref{lm:width-hierarchy}.
Suppose that for some bag $B$, we have $f(B) \le L$ for every feasible $f$.
Deleting $B$ cannot affect whether the maximum bag value of a tree decomposition exceeds $L$.
Consequently, if $W'$ is the optimum after deleting all such bags, the original width is $\max\{L,W'\}$, where we take the maximum of an empty collection of bags to be zero.

As a simple example, the constraints give $f(B) \le |B|$, so bags of size at most $L$ can be discarded.
We use a stronger bound based on the fractional edge cover number $\rho^*(B)$ of $B$ in $H$ from \cref{sec:intro-fec}, that is, the minimum total nonnegative weight assigned to the edges of $H[B]$ and the nodes of $B$ such that each node of $B$ is covered by weight at least one~\cite{AtseriasGM13}.

\begin{lemma}\label{lm:filter-out-fec-easy-bags}
    Let $H$ be a pattern and $B$ be a subset of $V(H)$.
    For every feasible solution $d, \psi$ of the mixed integer linear program for $\mwidth$, we have $\psi(B) \le \rho^*(B)$.
    Furthermore, for each $\subf \in \sharpmodset(H)$, we have $\subf(B) \le \rho^*(B)$.
\end{lemma}

\begin{proof}
    Let $d,\psi$ be feasible for the $\mwidth$ MILP.
    Order $B=\{a_1,\ldots,a_k\}$ arbitrarily, put $B_i=\{a_1,\ldots,a_i\}$, and define
    \[
        y_{a_i}\coloneqq \psi(B_i)-\psi(B_{i-1}).
    \]
    Monotonicity and the singleton increment constraint give $0\le y_{a_i}\le 1-d_{a_i}\le1$.
    For every edge $\{a_i,a_j\}$ with $i<j$, the leaf increment constraint also gives $y_{a_j}\le d_{a_i}$.
    Thus $y_{a_i}+y_{a_j}\le1$.
    These are exactly the feasibility constraints in the dual of the fractional edge-cover program: a node has weight at most one, and the endpoints of an edge have total weight at most one.
    By weak duality,
    \[
        \psi(B)=\sum_{i=1}^k y_{a_i}\le\rho^*(B).
    \]

    Furthermore, $\sharpmodset(H)\subseteq\msubmodset(H)$.
    By \cref{lm:msub-increments}, every $\subf\in\sharpmodset(H)\subseteq\msubmodset(H)$ is feasible for the $\mwidth$ MILP with $d_a=1-\subf(\set{a})$.
    Hence the bound already proved applies.
\end{proof}

Using \cref{lm:filter-out-fec-easy-bags}, we apply the following optimization:
We first compute $\clemb(H)$ and $\rho^*(B)$ for all $B \subseteq V(H)$.
While generating tree decompositions, we delete every bag $B$ satisfying $\rho^*(B) \le \clemb(H)$ and retain only minimal collections of the remaining bags under $\succeq$.
We call these collections \emph{interesting tree decompositions}.%
\footnote{This optimization need not be valid for $\dslwidth$, so we disable it when \nolinkurl{--check_worst_m_width_degrees_for_ds2_width} is set.}
(Technically, interesting tree decompositions are not tree decompositions.
They are collections of bags to which some ``easy'' bags can be added to form a valid tree decomposition.)%

\subparagraph{Reusing bounds.}
Another optimization we use is adding hints to the solver about the range of the objective function.
We use \cref{lm:width-hierarchy,lm:width-hierarchy-ds} to bound the width notion we want to compute from above and below with some other already computed graph parameters.
Out of the width notions, we start by computing \mwidth as its mixed integer linear program is the simplest one.
Let $d, \psi$ be an optimal solution to the full mixed integer linear program for $\mwidth$.
We have $\funwidth{\hmfun_d}(H) = \funwidth{\psi}(H) = \mwidth(H)$.
We save the vector $d$ from the optimal solution and call it the \emph{worst degree exponents}.

As can be seen from \cref{sec:examples} and from experiments, situations in which $\subw \neq \mwidth$ or $\sharpw \neq \mwidth$ are rather rare.
At the same time, in our experiments, computing $\subw$ and $\sharpw$ took significantly more time than computing $\mwidth$.
If we have $\clemb = \mwidth$, \cref{lm:width-hierarchy} immediately gives $\clemb = \subw = \sharpw = \mwidth$, and we do not need to run the mixed integer linear programs for $\subw$ and $\sharpw$.

\subparagraph{Using the worst degree exponents.}
Even if $\clemb < \mwidth$, sometimes we can still use some heuristics to avoid computing $\subw$ and $\sharpw$ directly.
In particular, due to \cref{lm:m-equiv-1}, we have $\hmfun_d \in \msubmodset(H)$ for the worst degree exponents $d$.
There is no guarantee that $\hmfun_d \in \sharpmodset(H)$ holds, but if it does, then $\sharpw(H) \ge \funwidth{\hmfun_d}(H) = \mwidth(H)$.
In this case, we immediately have $\mwidth(H) = \sharpw(H)$, and we do not need to run the mixed integer linear program.
The same optimization works for $\subw$ if $\hmfun_d \in \submodset(H)$.

If this membership test fails, we run our final heuristic.
We keep only the singleton values of $\hmfun_d$ and let the solver choose the remaining values.
More precisely, we run the MILP for $\sharpw$ with the additional constraints $\subf(\set{a})=1-d_a$ for all $a\in V(H)$.
Restricting a maximization problem gives a lower bound, so a feasible solution of width $\mwidth(H)$ proves $\sharpw(H)=\mwidth(H)$.
Finally, if these heuristics fail, we run the unrestricted MILP.
The same procedure works for $\subw$.

The user can apply the same idea to large patterns by supplying candidate degree exponents via \verb+--fix_degrees+.
For \mwidth, this restricts the vector $d$; for $\subw$ and $\sharpw$, the corresponding singleton values can be used as above.
The resulting values are lower bounds, and certify equality only when they match an independently established upper bound.

\section{Examples} \label{sec:examples}

From \cref{lm:width-hierarchy,lm:width-hierarchy-ds}, we have
\[\clemb(H) \le \subw(H) \le \sharpw(H) \le \mwidth(H) \le \dslwidth(H)\]
for every pattern $H$ and integer $\ell \ge 1$.
In this section, we show examples for which each of these inequalities is strict and investigate their minimality in terms of the number of nodes, edges, and parameter value.
All the above graph parameters except for $\dslwidth(H)$ split over clique separators due to \cref{lm:everything-splits-over-clique-separators}.
Thus, when searching for \emph{minimal} examples separating two of these parameters, we may restrict the search to patterns without clique separators.
For the last comparison, \cref{lm:ds-clique-refinement} allows the same restriction when we require a strict gap for every constant number of degree ranges.
To verify that an example is minimal, we run a computer search over all smaller patterns, using the database of all connected unlabeled graphs from~\cite{BDM} on up to $8$ nodes and with up to $12$ edges, from which we discard graphs with clique separators.
Thus, all our minimality claims with respect to the number of nodes or edges are computer-aided.

\subsection{Examples of \texorpdfstring{$\clemb(H) < \subw(H)$}{clemb(H) < subw(H)}} \label{sec:examples:clemb-vs-subw}

Let $K_{2k} \setminus \PM_k$ for $k \ge 2$ be the complete graph on $2k$ nodes from which a perfect matching $\PM_k$, that is, $k$ disjoint edges, is removed.
We determine all graph parameters for such a pattern.

\begin{lemma}\label{lm:params-for-overline-pm}
    For every $k \ge 2$, we have $\subw(K_{2k} \setminus \PM_k) = \dswidth{2}(K_{2k} \setminus \PM_k) = k - 1 / 2$ and $\clemb(K_{2k} \setminus \PM_k) = 3k / 4$.
\end{lemma}

\begin{proof}
    Let $H \coloneqq K_{2k} \setminus \PM_k$.
    Write $a_1, a_2, \ldots, a_k, b_1, b_2, \ldots b_k$ for the nodes of $H$ such that $\set{a_j, b_j} \notin E(H)$ for all $j \in \fragmentcc{1}{k}$.
    We first prove $\dswidth{2}(H) \le k - 1 / 2$.
    Let the degree-splitting exponent for each node be $1 / 2$.
    For $k = 2$, we have $H = C_4$.
    If some node is high-degree, use the two three-node bags containing that node and its opposite node; cover each bag by the high-degree node alone and an edge, at cost $1/2+1$.
    Otherwise, every node is low-degree, and each three-node bag is a path of cost $1+1/2$.
    Now consider the case $k \ge 3$.
    Consider some $i \in \set{\lo, \hi}^{V(H)}$.
    Pick a tree decomposition of $H$ with two bags: $V(H) \setminus \set{b_1}$ and $V(H) \setminus \set{a_1}$.
    We show that for the function $f_d^i$ from \cref{def:dswidth}, we have $f_d^i(V(H) \setminus \set{b_1}) \le k - 1 / 2$.
    We consider two cases.
    If $i_{a_1} = \hi$, we choose a forest in which $a_1$ is an isolated node, $a_j$ is adjacent to $b_{j+1}$ for $j \in \fragmentco{2}{k}$, and $a_k$ is adjacent to $b_2$.
    It consists of $k-1$ edges and a high-degree isolated node, at cost $(k-1)+1/2=k-1/2$.
    Otherwise, if $i_{a_1} = \lo$, we remove the edge $\set{a_2, b_3}$ from the forest in the previous case and add edges $\set{a_1, a_2}$ and $\set{a_1, b_3}$.
    It consists of $k-2$ edges and a three-node path centered at a low-degree node, at cost $(k-2)+(1+1/2)=k-1/2$.
    Exchanging $a_1$ and $b_1$ gives the same bound for $V(H) \setminus \set{a_1}$.
    It follows that $\funwidth{\hf_d^i}(H) \le k - 1 / 2$, and thus $\dswidth{2}(H) \le k - 1 / 2$.

    We now show $\subw(H) \ge k - 1 / 2$.
    Every graph of treewidth $t$ has a node of degree at most $t$.
    As every node of $H$ has degree $2k-2$, it follows that $\tw(H) \ge 2k - 2$.
    Thus by \cref{lm:tw-vs-widths} we have $\subw(H) \ge k - 1 / 2$.
    Together with $\subw(H) \le \dswidth{2}(H)$ from \cref{lm:width-hierarchy,lm:width-hierarchy-ds}, this proves both equalities.

    We now show $\clemb(H) \ge 3k / 4$.
    We define an embedding $\psi$ of $K_{3k}$ into $H$ by taking two copies of each singleton $\set{a_j}$ and one copy of each pair $\set{b_j,b_{j+1}}$ (including $\set{b_k, b_1}$).
    It is easy to verify that $\psi$ is a valid clique embedding.
    Each $a_j$ has two singleton preimages and each $b_j$ belongs to two pair images. Hence, every edge has weak depth at most four.
    Consequently, $\clemb(H) \ge 3k / \wed(\psi) \ge 3k / 4$.

    It remains to show $\clemb(H) \le 3k / 4$.
    Consider a clique embedding $\psi \colon K_{\ell} \to 2^{V(H)}$ and write $w \coloneqq \wed(\psi)$.
    The two singletons $\set{a_j}$ and $\set{b_j}$ cannot both be images, because they do not touch.
    By exchanging the names $a_j,b_j$ within each pair, we may therefore assume that no image is a singleton $\set{b_j}$ for every $j$.
    Partition the $\ell$ images into two subsets $A$ and $B$: set $A$ consists of images that intersect $\set{\row{a}{k}}$, and elements of $B$ are contained in $\set{\row{b}{k}}$. (Images are counted with multiplicity.)
    Every image in $B$ contains at least two nodes.

    We bound $|A|$ and $|B|$ by summing weak depths over two sets of edges.
    Each image in $A$ meets at least $k-1$ edges of the clique on $\set{\row{a}{k}}$, so
    \[
        (k-1)|A| \le \binom{k}{2}w, \qquad\text{and hence}\qquad 2|A| \le kw.
    \]
    Among the $k(k-1)$ edges $\set{a_{j},b_{j'}}$ with $j\ne j'$, each image in $A$ meets at least $k-1$ edges and each image in $B$ meets at least $2(k-1)$ edges. Thus,
    \[
        (k-1)(|A|+2|B|) \le k(k-1)w, \qquad\text{and hence}\qquad |A|+2|B| \le kw.
    \]
    Adding the first bound to twice the second gives $4\ell=4(|A|+|B|) \le 3kw$.
    Therefore, $\ell/\wed(\psi) \le 3k/4$, as required.
\end{proof}

For $k = 3$, the pattern $K_6 \setminus \PM_3$ is the octahedron, and it is the smallest pattern separating $\clemb$ and $\subw$ in terms of the number of nodes.

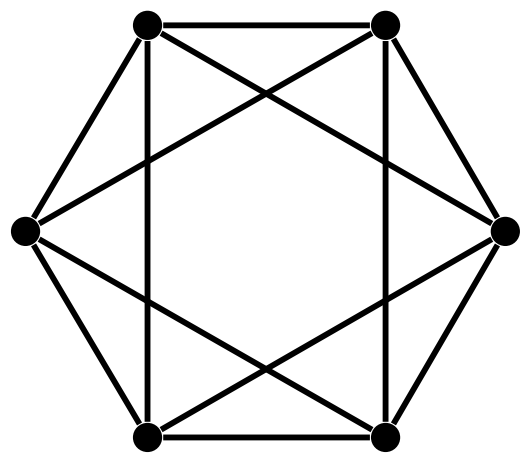
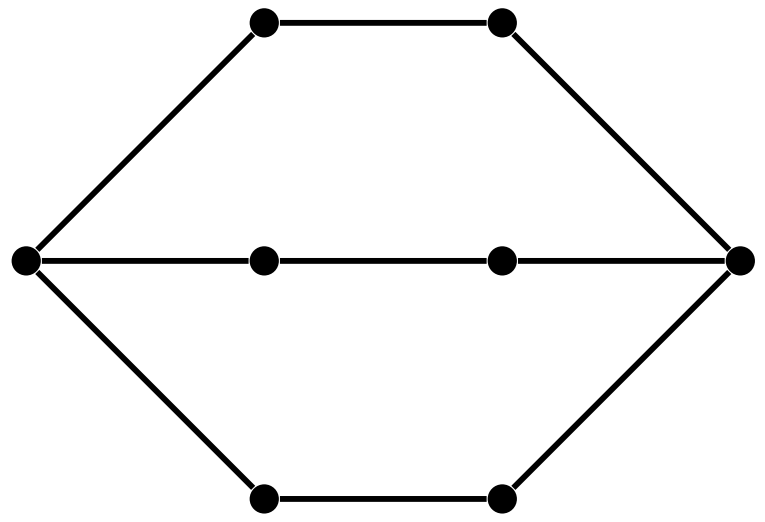
\begin{figure}
    \begin{subfigure}{0.48\textwidth}
        \begin{center}
            \begin{tikzpicture}[x=1.4pt, y=1.4pt, every node/.style={circle, fill, inner sep=0pt, minimum size=5.5pt}]
    \node (n0) at (16.4,55.4) {};
    \node (n1) at (48.4,55.4) {};
    \node (n2) at (16.4,0) {};
    \node (n3) at (48.4,0) {};
    \node (n4) at (64.5,27.7) {};
    \node (n5) at (0,27.7) {};
    \foreach \a/\b in {0/1, 0/2, 0/4, 0/5, 1/3, 1/4, 1/5, 2/3, 2/4, 2/5, 3/4, 3/5} {
        \draw[line width=1.1pt] (n\a) -- (n\b);
    }
\end{tikzpicture}
        \end{center}
        \caption{A pattern with the smallest number of nodes satisfying $\clemb < \subw$. This is the octahedron graph (or alternatively $K_6 \setminus \PM_3$). We have $\clemb = 2 + 1 / 4$ and $\subw = 2 + 1 / 2$.}
        \label{fig:clemb-vs-subw:pm3-bar}
    \end{subfigure}
    \hfill
    \begin{subfigure}{0.48\textwidth}
        \begin{center}
            \begin{tikzpicture}[x=1.4pt, y=1.4pt, every node/.style={circle, fill, inner sep=0pt, minimum size=5.5pt}]
    \node (n0) at (0,32) {};
    \node (n1) at (32,64) {};
    \node (n2) at (64,64) {};
    \node (n3) at (96,32) {};
    \node (n4) at (32,32) {};
    \node (n5) at (64,32) {};
    \node (n6) at (32,0) {};
    \node (n7) at (64,0) {};
    \foreach \a/\b in {0/1, 0/4, 0/6, 1/2, 2/3, 3/5, 3/7, 4/5, 6/7} {
        \draw[line width=1.1pt] (n\a) -- (n\b);
    }
\end{tikzpicture}
        \end{center}
        \caption{A pattern with the smallest number of edges and the smallest values of $\clemb$ and $\subw$ satisfying $\clemb < \subw$. This is $P(3, 3, 3)$. We have $\clemb = 2 - 1 / 9$ and $\subw = 2$.}
        \label{fig:clemb-vs-subw:p333}
    \end{subfigure}
    \caption{The smallest examples of patterns with $\clemb < \subw$.} \label{fig:clemb-vs-subw}
\end{figure}

\begin{calemma}\label{lm:ca-clemb-subw-octahedron}
    The octahedron $K_6 \setminus \PM_3$ satisfies $\clemb(K_6 \setminus \PM_3) = 2 + 1/4 < 2 + 1/2 = \subw(K_6 \setminus \PM_3)$, and it is the only pattern on at most $6$ nodes satisfying $\clemb < \subw$.
    See \cref{fig:clemb-vs-subw:pm3-bar}.
\end{calemma}

\begin{proof}
    The values of $\clemb$ and $\subw$ follow from \cref{lm:params-for-overline-pm} for $k = 3$.
    As explained at the beginning of \cref{sec:examples}, for the uniqueness it suffices to consider patterns without clique separators.
    An exhaustive computer search computes $\clemb$ and $\subw$ for all such patterns on at most $6$ nodes and finds that $K_6 \setminus \PM_3$ is the only one satisfying $\clemb < \subw$.
    This can be verified by running:
\begin{verbatim}
python width-calculator.py --clemb --subw -n 2-6 -vv
\end{verbatim}
\end{proof}

\begin{calemma}\label{lm:ca-clemb-subw-boat}
    The pattern $P(3, 3, 3)$ satisfies $\clemb(P(3, 3, 3)) = 2 - 1/9 < 2 = \subw(P(3, 3, 3))$, and it is the only pattern with at most $9$ edges satisfying $\clemb < \subw$.
    Moreover, every pattern $H$ with $\clemb(H) < \subw(H)$ satisfies $\subw(H) \ge 2$ and $\clemb(H) \ge 2 - 1/9$.
    See \cref{fig:clemb-vs-subw:p333}.
\end{calemma}

\begin{proof}
    The values $\clemb(P(3, 3, 3)) = 2 - 1/9$ and $\subw(P(3, 3, 3)) = 2$ were shown in~\cite{FanKZ23}.
    As explained at the beginning of \cref{sec:examples}, for the uniqueness it suffices to consider patterns without clique separators.
    An exhaustive computer search computes $\clemb$ and $\subw$ for all such patterns with at most $9$ edges and finds that $P(3, 3, 3)$ is the only one satisfying $\clemb < \subw$.
    This can be verified by running:
\begin{verbatim}
python width-calculator.py --clemb --subw -m 1-9 -vv
\end{verbatim}
    Finally, let $H$ be a pattern with $\clemb(H) < \subw(H)$.
    By \cite[Theorem D.1]{BG24}, every pattern with $\subw < 2$ satisfies $\clemb = \subw$, so $\subw(H) \ge 2$.
    By \cite[Theorem D.1 and Corollary 1.9]{BG24}, it follows that $\clemb(H) \ge 2$ or $H$ contains $P(3, 3, 3)$ as an induced minor.
    In the latter case, $\clemb(H) \ge \clemb(P(3, 3, 3)) = 2 - 1/9$ by \cref{lm:clemb-of-induced-minors}.
\end{proof}

\subsection{Examples of \texorpdfstring{$\subw(H) < \sharpw(H)$}{subw(H) < \#subw(H)}} \label{sec:examples:subw-vs-sharpw}

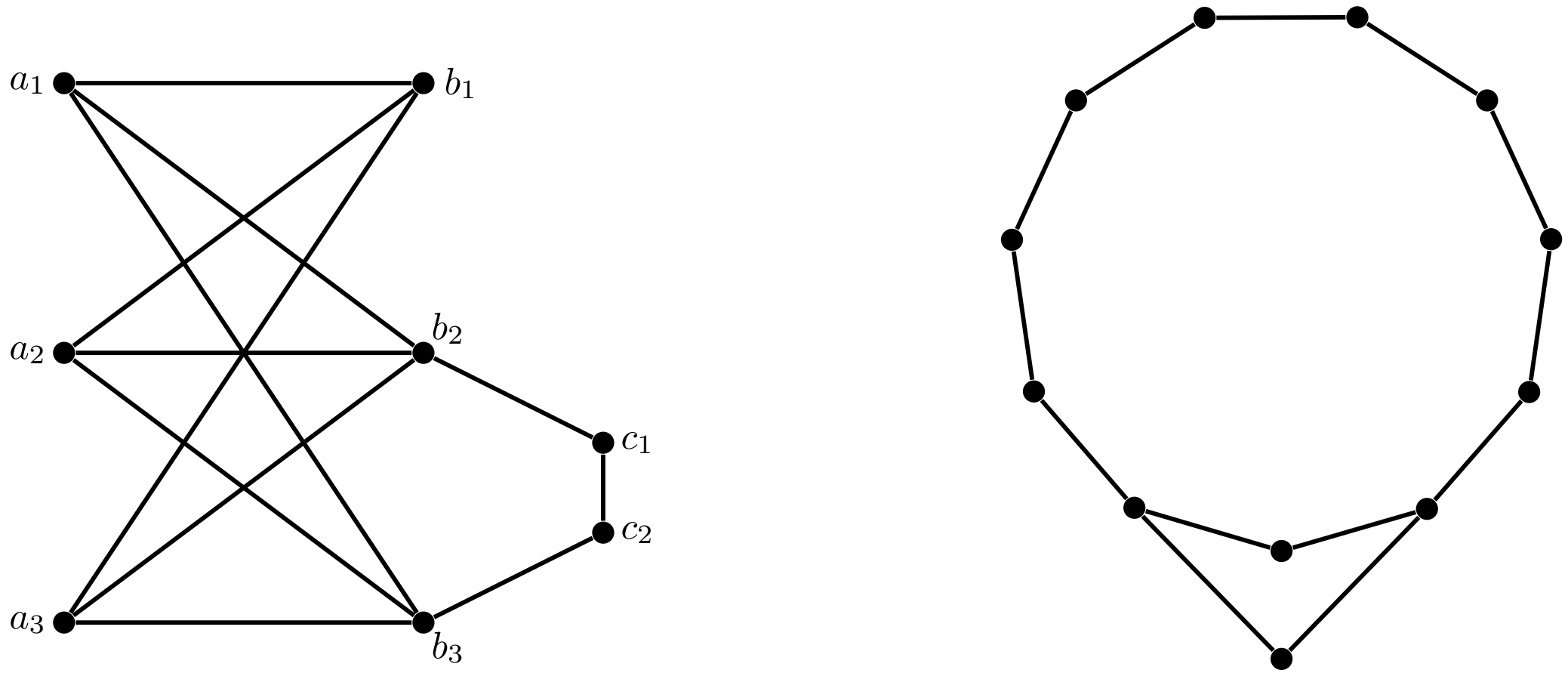
\begin{figure}
    \begin{subfigure}{0.48\textwidth}
        \begin{center}
            {\patternlabelstrue\begin{tikzpicture}[x=1.4pt, y=1.4pt, every node/.style={circle, fill, inner sep=0pt, minimum size=5.5pt}]
    \node (n0) at (0,96) {};
    \node (n1) at (0,48) {};
    \node (n2) at (0,0) {};
    \node (n3) at (64,96) {};
    \node (n4) at (64,48) {};
    \node (n5) at (64,0) {};
    \node (n6) at (96,32) {};
    \node (n7) at (96,16) {};
    \foreach \a/\b in {0/3, 0/4, 0/5, 1/3, 1/4, 1/5, 2/3, 2/4, 2/5, 4/6, 5/7, 6/7} {
        \draw[line width=1.1pt] (n\a) -- (n\b);
    }
    \ifpatternlabels
        \node[left=3.5pt, fill=none, minimum size=0pt, font=\small] at (n0) {$a_1$};
        \node[left=3.5pt, fill=none, minimum size=0pt, font=\small] at (n1) {$a_2$};
        \node[left=3.5pt, fill=none, minimum size=0pt, font=\small] at (n2) {$a_3$};
        \node[right=3.5pt, fill=none, minimum size=0pt, font=\small] at (n3) {$b_1$};
        \node[above right=2pt, fill=none, minimum size=0pt, font=\small] at (n4) {$b_2$};
        \node[below right=2pt, fill=none, minimum size=0pt, font=\small] at (n5) {$b_3$};
        \node[right=3.5pt, fill=none, minimum size=0pt, font=\small] at (n6) {$c_1$};
        \node[right=3.5pt, fill=none, minimum size=0pt, font=\small] at (n7) {$c_2$};
    \fi
\end{tikzpicture}}
        \end{center}
        \caption{A pattern with the smallest number of nodes and edges satisfying $\subw < \sharpw$. We have $\subw = 2 + 1 / 4$ and $\sharpw = 2 + 3 / 11$.}
        \label{fig:subw-vs-sharpw:min-vert}
    \end{subfigure}
    \hfill
    \begin{subfigure}{0.48\textwidth}
        \begin{center}
            \begin{tikzpicture}[x=1.4pt, y=1.4pt, every node/.style={circle, fill, inner sep=0pt, minimum size=5.5pt}]
    \node (n0) at (34.3,114.1) {};
    \node (n1) at (61.5,114.2) {};
    \node (n2) at (84.6,99.4) {};
    \node (n3) at (96,74.7) {};
    \node (n4) at (92.1,47.5) {};
    \node (n5) at (73.9,26.7) {};
    \node (n6) at (48,19.2) {};
    \node (n7) at (48,0) {};
    \node (n8) at (21.8,26.9) {};
    \node (n9) at (3.9,47.6) {};
    \node (n10) at (0,74.6) {};
    \node (n11) at (11.4,99.4) {};
    \foreach \a/\b in {0/1, 0/11, 1/2, 2/3, 3/4, 4/5, 5/6, 5/7, 6/8, 7/8, 8/9, 9/10, 10/11} {
        \draw[line width=1.1pt] (n\a) -- (n\b);
    }
\end{tikzpicture}
        \end{center}
        \caption{A pattern with the smallest value of $\subw$ satisfying $\subw < \sharpw$. This is $P(9, 2, 2)$. We have $\subw = 2 - 1 / 8$ and $\sharpw \ge 2 - 1 / 9$.}
        \label{fig:subw-vs-sharpw:p922}
    \end{subfigure}
    \caption{The smallest examples of patterns with $\subw < \sharpw$.} \label{fig:subw-vs-sharpw}
\end{figure}

\begin{calemma}\label{lm:ca-subw-sharpw-nodes-edges}
    The pattern $H$ depicted in \cref{fig:subw-vs-sharpw:min-vert}, which has $8$ nodes and $12$ edges, satisfies $\subw(H) = 2 + 1/4 < 2 + 3/11 = \sharpw(H)$.
    In contrast, all patterns on at most $7$ nodes and all patterns with at most $11$ edges satisfy $\subw = \sharpw$.
\end{calemma}

\begin{proof}
    As explained at the beginning of \cref{sec:examples}, for the nonexistence claims it suffices to consider patterns without clique separators.
    An exhaustive computer search computes $\subw$ and $\sharpw$ for all such patterns on at most $8$ nodes.
    It finds no pattern with at most $7$ nodes satisfying $\subw < \sharpw$, and it finds patterns with $8$ nodes satisfying $\subw < \sharpw$, among them the pattern $H$ with the values stated above.
    This can be verified by running:
\begin{verbatim}
python width-calculator.py --subw --sharp_subw -vv -n 2-8
\end{verbatim}
    A second exhaustive computer search computes $\subw$ and $\sharpw$ for all patterns without clique separators with at most $11$ edges and finds none satisfying $\subw < \sharpw$.
    This can be verified by running:
\begin{verbatim}
python width-calculator.py --subw --sharp_subw -vv -m 1-11
\end{verbatim}
(Note that running these commands takes a long time.)
\end{proof}

\begin{table}[htbp]
    \centering
    \renewcommand{\arraystretch}{1.25}
    \begin{tabular}{c|p{0.8\textwidth}}
        $\subw$ & Clique-separator-free patterns \\
        \hline
        $1$ & $K_2$ \\
        $2 - 1/2$ & $K_3$, $C_4$ \\
        $2 - 1/3$ & $C_5$, $C_6$, $K_{2,3}$ \\
        $2 - 1/4$ & $C_7$, $C_8$, $K_{2,4}$ \\
        $2 - 1/5$ & $C_9$, $C_{10}$, $K_{2,5}$, $P(3,2,2)$, $P(4,2,2)$ \\
        $2 - 1/6$ & $C_{11}$, $C_{12}$, $K_{2,6}$, $P(5,2,2)$, $P(6,2,2)$, $P(3,3,2)$ \\
        $2 - 1/7$ & $C_{13}$, $C_{14}$, $K_{2,7}$, $P(7,2,2)$, $P(8,2,2)$, $P(3,2,2,2)$, $P(4,2,2,2)$ \\
        $2 - 1/8$ & $C_{15}$, $C_{16}$, $K_{2,8}$, $P(9,2,2)$, $P(10,2,2)$, $P(5,2,2,2)$, $P(6,2,2,2)$, $P(4,3,2)$, $P(5,3,2)$ \\
    \end{tabular}
    \caption{All clique-separator-free patterns with $\subw \le 2 - 1/8$, according to \cite[Section~1.4 and Theorem~D.1]{BG24}.}
    \label{tab:bg24-list}
\end{table}

To determine the smallest value of $\subw$ for which the two parameters differ, we combine the classification of all clique-separator-free patterns with $\subw \le 2 - 1/8$ from~\cite{BG24}, shown in \cref{tab:bg24-list}, with computations of our tool.

\begin{catheorem}\label{thm:ca-subw-sharpw-min-value}
    Every pattern $H$ with $\subw(H) < 2 - 1/8$ satisfies $\subw(H) = \sharpw(H)$.
    In contrast, $\subw(P(9, 2, 2)) = 2 - 1/8 < 2 - 1/9 \le \sharpw(P(9, 2, 2))$.
    Hence, $2 - 1/8$ is the smallest value of $\subw$ for a pattern with $\subw < \sharpw$.
    See \cref{fig:subw-vs-sharpw:p922}.
\end{catheorem}

\begin{proof}
    Let $H$ be a pattern with $\subw(H) < 2 - 1/8$.
    By \cref{lm:everything-splits-over-clique-separators}, it suffices to show $\subw(H') = \sharpw(H')$ for every $H' \in D(H)$ (see \cref{def:clique-separator-decomp}).
    Each such $H'$ is clique-separator-free and satisfies $\subw(H') \le \subw(H) < 2 - 1/8$, so it appears in one of the first seven rows of \cref{tab:bg24-list}.
    If $H'$ is a cycle, then $\subw(H') = \sharpw(H')$ by \cite{AlonYZ97} and \cite[Lemma 17]{FanKZ23}.
    If $H' = P(7, 2, 2)$ or $H' = P(8, 2, 2)$, then $\subw(H') = \sharpw(H') = \mwidth(H')$ by \cref{m-width-of-p822,lm:width-hierarchy}.
    For each of the remaining patterns, our tool computes $\subw$ and $\sharpw$ and finds that the two values are equal.
    This can be verified by running the following command for each pattern individually:
\begin{verbatim}
python width-calculator.py --subw --sharp_subw -a "[edge list]"
\end{verbatim}

    The value $\subw(P(9, 2, 2)) = 2 - 1/8$ is given in \cref{tab:bg24-list}.
    For the lower bound on $\sharpw(P(9, 2, 2))$, we run our tool with the fixed degree exponents listed in the command below.
    See \cref{sec:implementation:optimizations} on why this gives a lower bound on $\sharpw(P(9, 2, 2))$.
\begin{verbatim}
python width-calculator.py --sharp_subw -a "P(9, 2, 2)" \
--fix_degrees 1/9 1/9 1/9 1/9 2/9 2/9 2/9 2/9 1/9 1/9 5/9 5/9
\end{verbatim}
\end{proof}

Furthermore, the following lemma shows that all three parameters $\clemb$, $\subw$, and $\sharpw$ can differ on a single pattern.

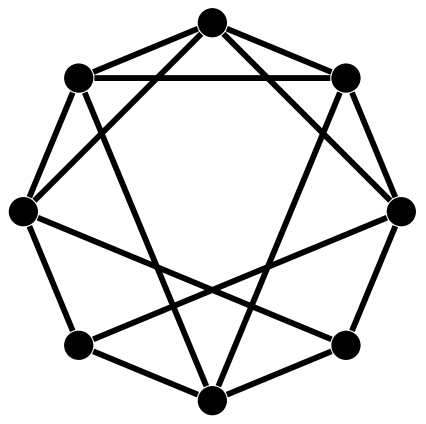
\begin{figure}[htbp]
    \begin{center}
        \begin{tikzpicture}[scale=1, every node/.style={circle, fill, inner sep=0pt, minimum size=5.5pt}]
    \foreach \v/\k in {1/0, 4/1, 7/2, 2/3, 6/4, 3/5, 5/6, 8/7} {
        \node (n\v) at ({90 + 45 * \k}:1.25) {};
    }
    \foreach \a/\b in {1/4, 1/5, 1/7, 1/8, 2/5, 2/6, 2/7, 3/5, 3/6, 3/7, 4/6, 4/7, 4/8, 5/8, 6/8} {
        \draw[line width=1.1pt] (n\a) -- (n\b);
    }
\end{tikzpicture}
    \end{center}
    \caption{A pattern with $\clemb < \subw < \sharpw$. We have $\clemb = 2 + 3/8$, $\subw = 2 + 5/9$, and $\sharpw = 2 + 4/7$.}
    \label{fig:clemb-subw-sharpw-example}
\end{figure}

\begin{calemma}\label{lm:ca-clemb-subw-sharpw-example}
    The pattern depicted in \cref{fig:clemb-subw-sharpw-example} satisfies $\clemb(H) = 2 + 3 / 8 < \subw(H) = 2 + 5 / 9 < \sharpw(H) = 2 + 4 / 7$.
\end{calemma}

\begin{proof}
    Our tool computes the three parameters of $H$.
    This can be verified by running the following command with the pattern's edge list:
\begin{verbatim}
python width-calculator.py --clemb --subw --sharp_subw -a "[edge list]"
\end{verbatim}
\end{proof}

\subsection{Examples of \texorpdfstring{$\sharpw(H) < \mwidth(H)$}{\#subw(H) < m-width(H)}} \label{sec:examples:sharpw-vs-mwidth}

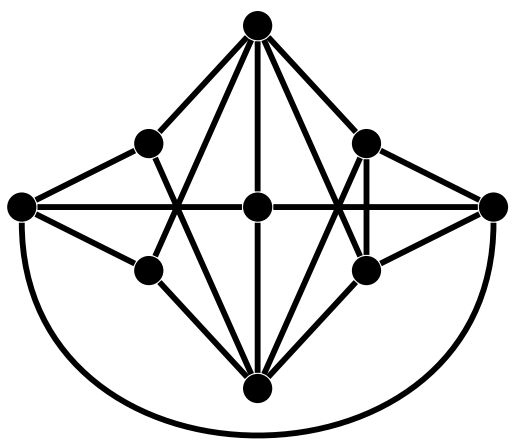
\begin{figure}[htbp]
    \begin{center}
        \begin{tikzpicture}[scale=0.6, every node/.style={circle, fill, inner sep=0pt, minimum size=5.5pt}]
    \node (n8) at (0,2) {};
    \node (n9) at (0,-2) {};
    \node (n2) at (0,0) {};
    \node (n3) at (-1.2,0.7) {};
    \node (n4) at (-1.2,-0.7) {};
    \node (n1) at (1.2,0.7) {};
    \node (n5) at (1.2,-0.7) {};
    \node (n6) at (-2.6,0) {};
    \node (n7) at (2.6,0) {};
    \foreach \u in {1,2,3,4,5} {\draw[line width=1.1pt] (n8) -- (n\u); \draw[line width=1.1pt] (n9) -- (n\u);}
    \foreach \a/\b in {6/2, 6/3, 6/4, 7/1, 7/2, 7/5, 1/5} {\draw[line width=1.1pt] (n\a) -- (n\b);}
    \draw[line width=1.1pt] (n6) .. controls (-2.6,-3.3) and (2.6,-3.3) .. (n7);
\end{tikzpicture}
    \end{center}
    \caption{A pattern with the smallest number of nodes satisfying $\sharpw < \mwidth$. We have $\sharpw = 8/3$ and $\mwidth = 27/10$.}
    \label{fig:sharpw-mwidth-example}
\end{figure}

\begin{calemma}\label{lm:ca-sharpw-mwidth-example}
    The pattern depicted in \cref{fig:sharpw-mwidth-example} has $9$ nodes and satisfies $\sharpw(H) = 8 / 3 < 27 / 10 = \mwidth(H)$.
    In contrast, all patterns on at most $8$ nodes satisfy $\sharpw = \mwidth$.
\end{calemma}

\begin{proof}
    Our tool computes $\sharpw(H)$ and $\mwidth(H)$.
    This can be verified by running the following command with the pattern's edge list:
\begin{verbatim}
python width-calculator.py --sharp_subw --m_width -a "[edge list]"
\end{verbatim}
    As explained at the beginning of \cref{sec:examples}, for the second claim it suffices to consider patterns without clique separators.
    An exhaustive computer search computes $\sharpw$ and $\mwidth$ for all such patterns on at most $8$ nodes and finds none satisfying $\sharpw < \mwidth$.
    This can be verified by running:
\begin{verbatim}
python width-calculator.py --sharp_subw --m_width -vv -n 2-8
\end{verbatim}
\end{proof}

We leave open the questions of finding such patterns with the smallest number of edges and the smallest value of $\sharpw$.

\subsection{Examples of \texorpdfstring{$\mwidth(H) < \dslwidth(H)$}{m-width(H) < degree-splitting width(H)}} \label{sec:examples:mwidth-vs-dslwidth}

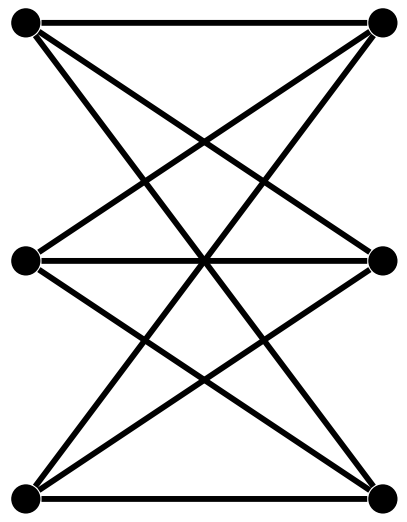
\begin{figure}
    \begin{center}
        \begin{tikzpicture}[x=1.4pt, y=1.4pt, every node/.style={circle, fill, inner sep=0pt, minimum size=5.5pt}]
    \node (n0) at (0,64) {};
    \node (n1) at (0,32) {};
    \node (n2) at (0,0) {};
    \node (n3) at (48,64) {};
    \node (n4) at (48,32) {};
    \node (n5) at (48,0) {};
    \foreach \a/\b in {0/3, 0/4, 0/5, 1/3, 1/4, 1/5, 2/3, 2/4, 2/5} {
        \draw[line width=1.1pt] (n\a) -- (n\b);
    }
\end{tikzpicture}
    \end{center}

    \caption{A pattern with the smallest number of nodes and edges satisfying $\mwidth < \dslwidth$ for all $\ell \ge 1$. This is $K_{3, 3}$. We have $\mwidth = 2$ and $\dslwidth \ge 2 + 1/(144 \ell)$.}
    \label{fig:k33}
\end{figure}

\begin{catheorem}\label{thm:ca-k33-min-nodes-edges}
    The pattern $K_{3, 3}$ satisfies $\mwidth(K_{3, 3}) = 2 < \dslwidth(K_{3, 3})$ for every $\ell \ge 1$.
    In contrast, every pattern $H$ on at most $5$ nodes satisfies $\mwidth(H) = \dswidth{2}(H)$, and every pattern $H$ with at most $8$ edges satisfies $\mwidth(H) = \dslwidth(H)$ for some $\ell \ge 1$.
    Hence, $K_{3, 3}$ has the smallest number of nodes and the smallest number of edges among patterns satisfying $\mwidth < \dslwidth$ for every $\ell \ge 1$.
    See \cref{fig:k33}.
\end{catheorem}

\begin{proof}
    The first claim holds by \cref{lm:k33-widths,lm:k33-ds-width}.

    Now let $H$ be a pattern on at most five nodes.
    The inequality $\mwidth(H) \le \dswidth{2}(H)$ holds by \cref{lm:width-hierarchy-ds}, so it suffices to find splitting exponents showing that $\dswidth{2}(H) \le \mwidth(H)$.
    Our tool finds degrees attaining \mwidth and verifies whether splitting by them is sufficient to upper bound $\dswidth{2}(H)$ by $\mwidth(H)$.
    This heuristic works for all patterns on at most five nodes except for $K_{2, 3}$; this can be verified by running:
\begin{verbatim}
python width-calculator.py --m_width \
--check_worst_m_width_degrees_for_ds2_width \
-n 2-5 --allow_clique_separators \
--remove_graph "K_{2,3}" -vv
\end{verbatim}
    For $K_{2, 3}$, there still exist splitting degrees attaining $\mwidth$, which give the common value $\mwidth(K_{2,3}) = \dswidth{2}(K_{2,3}) = 5/3$.
    This can be verified by running:
\begin{verbatim}
python width-calculator.py --m_width -a "K_{2,3}" \
--fix_degrees 1/3 1/3 2/3 2/3 2/3 \
--check_worst_m_width_degrees_for_ds2_width
\end{verbatim}

    Now let $H$ be a pattern with at most eight edges.
    We first show that every clique-separator-free pattern $H'$ with at most eight edges satisfies $\mwidth(H') = \dswidth{2}(H')$.
    For all these patterns except $C_8$, $K_{2,3}$, and $P(3,3,2)$, a computer search produces splitting degree certificates proving $\mwidth=\dswidth{2}$.
    These certificates can be verified by running:
\begin{verbatim}
python width-calculator.py --m_width \
--check_worst_m_width_degrees_for_ds2_width \
-m 1-8 --remove_graph "C_8" "K_{2,3}" "P(3,3,2)" -vv
\end{verbatim}
    For $C_8$, the cycle bounds of \cite{AlonYZ97} and \cite[Lemma 17]{FanKZ23}, recalled in the proof of \cref{thm:ca-subw-sharpw-min-value}, already give $\mwidth(C_8)=\dswidth{2}(C_8)=7/4$.
    The case of $K_{2,3}$ is handled by the explicit splitting degrees above.
    For $P(3,3,2)$, we have $\mwidth(P(3,3,2))=2-1/6$.
    This can be verified by running:
\begin{verbatim}
python width-calculator.py --m_width -a "P(3,3,2)"
\end{verbatim}
    Moreover, $\dswidth{2}(P(3,3,2))\le2-1/6$.
    This upper bound can be verified by running:
\begin{verbatim}
python width-calculator.py --m_width -a "P(3,3,2)" \
--fix_degrees 1/6 1/6 1/6 1/3 1/2 1/2 2/3 \
--check_worst_m_width_degrees_for_ds2_width
\end{verbatim}
    Since $\mwidth \le \dswidth{2}$ by \cref{lm:width-hierarchy-ds}, this gives $\mwidth(P(3,3,2)) = \dswidth{2}(P(3,3,2))$.

    Every pattern in $D(H)$ (see \cref{def:clique-separator-decomp}) is clique-separator-free and has at most eight edges, so it satisfies $\mwidth = \dswidth{2}$.
    By \cref{lm:ds-clique-refinement}, $H$ therefore satisfies $\mwidth(H)=\dslwidth(H)$ for some big enough constant $\ell$.
\end{proof}

We leave open the question of finding such a pattern with the smallest value of $\mwidth$.

\subsection{Open Problems} \label{sec:example:open-problems}

In this section we discuss minimal patterns $H$ for which the fastest currently known algorithms do not match the corresponding conditional lower bounds.

\subparagraph{Conditional lower bounds.}
We use two fine-grained complexity hypotheses, both about finding cliques in dense graphs.
In the \emph{$k$-Clique Listing} problem, the input is an $n$-node graph $G$, and the task is to find all $k$-cliques in $G$.
The \cliquelistconj states that there are no constants $k \ge 3$ and $\varepsilon > 0$ such that $k$-Clique Listing can be solved in $\tOh(n^{k-\varepsilon} + t)$ time where $t$ is the number of $k$-cliques in $G$.
The \combcliqueconj states that, for every constant $k \ge 3$ and $\varepsilon > 0$, no \emph{combinatorial} algorithm detects a $k$-clique in an unweighted $n$-node graph in time $\Oh(n^{k-\varepsilon})$.
Here, as usual, ``combinatorial'' informally excludes algorithms based on fast matrix multiplication.
We allow randomized algorithms in both hypotheses.

Historically, lower bounds for variants of subgraph isomorphism were also shown under the 3XOR, 3SUM, and Zero-Clique Hypotheses~\cite{JoglekarR18,BG24}.
These hypotheses were recently refuted~\cite{AlmanW26}, and for 3SUM there is even a refutation that does not require fast matrix multiplication~\cite{ChanIMY26}.
Therefore, we do not use these hypotheses.
The result of~\cite{AlmanW26} does not seem to immediately break the \cliquelistconj.

The following result of Bringmann and Gorbachev~\cite{BG24} based on the result of Fan, Koutris, and Zhao~\cite{FanKZ23} states conditional lower bounds based on clique embeddings.

\begin{fact}[Lower Bounds via $\clemb$, Extension of {\cite[Lemma 2.3]{BG24}}] \label{fct:all-H-iso-clique-hardness}
    Let $H$ be a pattern and $\varepsilon > 0$.
    \Hlistiso cannot be solved in time $\tOh(m^{\clemb(H) - \varepsilon} + t)$ assuming the \cliquelistconj.
    \Henumiso cannot be solved in preprocessing time $\Oh(m^{\clemb(H) - \varepsilon})$ and delay $\polylog(m)$, assuming the \cliquelistconj.
    \Hcountiso cannot be solved in time $\Oh(m^{\clemb(H) - \varepsilon})$ using a \emph{combinatorial algorithm} assuming the \combcliqueconj.\footnote{The lower bound for combinatorial counting algorithms is not stated in \cite[Lemma 2.3]{BG24}; it trivially follows by applying the reduction from \cite[Lemma 5.3]{BG24} to counting instead of \Hminiso.}
    \lipicsEnd
\end{fact}

\begin{figure}
    \begin{subfigure}[t]{0.48\textwidth}
        \centering
        \begin{tikzpicture}[x=1.4pt, y=1.4pt, every node/.style={circle, fill, inner sep=0pt, minimum size=5.5pt}]
    \node (n0) at (16.4,55.4) {};
    \node (n1) at (48.4,55.4) {};
    \node (n2) at (16.4,0) {};
    \node (n3) at (48.4,0) {};
    \node (n4) at (64.5,27.7) {};
    \node (n5) at (0,27.7) {};
    \foreach \a/\b in {0/1, 0/2, 0/4, 0/5, 1/3, 1/4, 1/5, 2/3, 2/4, 2/5, 3/4, 3/5} {
        \draw[line width=1.1pt] (n\a) -- (n\b);
    }
\end{tikzpicture}
        \caption{A pattern with the smallest number of nodes such that optimal time complexities of \listH and \enumH assuming the \cliquelistconj are unknown. This is $K_6 \setminus \PM_3$. The conditionally optimal time complexity exponent lies in $[2 + 1 / 4, 2 + 1 / 2]$.}
        \label{fig:open-problems:pm3-bar}
    \end{subfigure}
    \hfill
    \begin{subfigure}[t]{0.48\textwidth}
        \centering
        \begin{tikzpicture}[x=1.4pt, y=1.4pt, every node/.style={circle, fill, inner sep=0pt, minimum size=5.5pt}]
    \node (n0) at (0,32) {};
    \node (n1) at (32,64) {};
    \node (n2) at (64,64) {};
    \node (n3) at (96,32) {};
    \node (n4) at (32,32) {};
    \node (n5) at (64,32) {};
    \node (n6) at (32,0) {};
    \node (n7) at (64,0) {};
    \foreach \a/\b in {0/1, 0/4, 0/6, 1/2, 2/3, 3/5, 3/7, 4/5, 6/7} {
        \draw[line width=1.1pt] (n\a) -- (n\b);
    }
\end{tikzpicture}
        \caption{A pattern with the smallest number of edges and the smallest values of $\clemb$ and $\subw$ such that optimal time complexities of \listH and \enumH assuming the \cliquelistconj are unknown. This is $P(3, 3, 3)$. The conditionally optimal time complexity exponent lies in $[2 - 1 / 9, 2]$.}
        \label{fig:open-problems:p333}
    \end{subfigure}

    \bigskip
    \begin{subfigure}[t]{0.48\textwidth}
        \centering
        \begin{tikzpicture}[x=1.4pt, y=1.4pt, every node/.style={circle, fill, inner sep=0pt, minimum size=5.5pt}]
    \node (n0) at (0,72) {};
    \node (n1) at (0,56) {};
    \node (n2) at (0,32) {};
    \node (n3) at (0,0) {};
    \node (n4) at (32,64) {};
    \node (n5) at (32,32) {};
    \node (n6) at (32,0) {};
    \foreach \a/\b in {0/1, 0/4, 0/5, 1/5, 1/6, 2/4, 2/5, 2/6, 3/4, 3/5, 3/6} {
        \draw[line width=1.1pt] (n\a) -- (n\b);
    }
\end{tikzpicture}
        \caption{A further pattern for which the optimal time complexities of \listH and \enumH assuming the \cliquelistconj are unknown. The conditionally optimal time complexity exponent lies in $[2 + 1 / 4, 2 + 2 / 7]$.}
        \label{fig:open-problems:min-edges}
    \end{subfigure}
    \hfill
    \begin{subfigure}[t]{0.48\textwidth}
        \centering
        \begin{tikzpicture}[x=1.4pt, y=1.4pt, every node/.style={circle, fill, inner sep=0pt, minimum size=5.5pt}]
    \node (n0) at (20.2,0) {};
    \node (n1) at (52.2,0) {};
    \node (n2) at (10.1,30.4) {};
    \node (n3) at (36.2,48.9) {};
    \node (n4) at (0,56) {};
    \node (n5) at (20.2,56) {};
    \node (n6) at (10.1,81.6) {};
    \node (n7) at (62.3,30.4) {};
    \node (n8) at (52.2,56) {};
    \node (n9) at (72.4,56) {};
    \node (n10) at (62.3,81.6) {};
    \foreach \a/\b in {0/1, 0/2, 1/7, 2/3, 2/4, 2/5, 3/7, 4/6, 5/6, 6/10, 7/8, 7/9, 8/10, 9/10} {
        \draw[line width=1.1pt] (n\a) -- (n\b);
    }
\end{tikzpicture}
        \caption{A further pattern for which the optimal time complexities of \listH and \enumH assuming the \cliquelistconj are unknown. The conditionally optimal time complexity exponent lies in $[2 + 1 / 6, 2 + 3 / 16]$.}
        \label{fig:open-problems:min-exponent}
    \end{subfigure}
    \caption{Patterns underlying open problems for \listH and \enumH.} \label{fig:open-problems}
\end{figure}
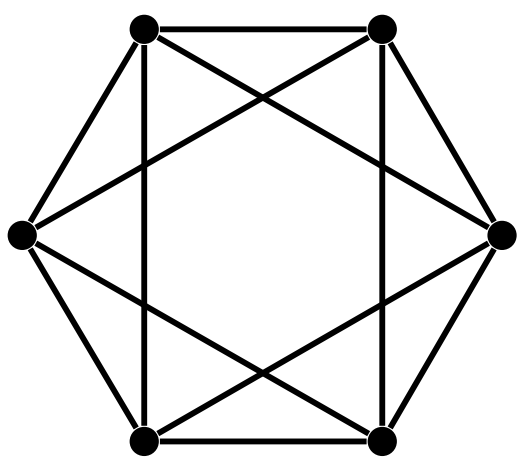
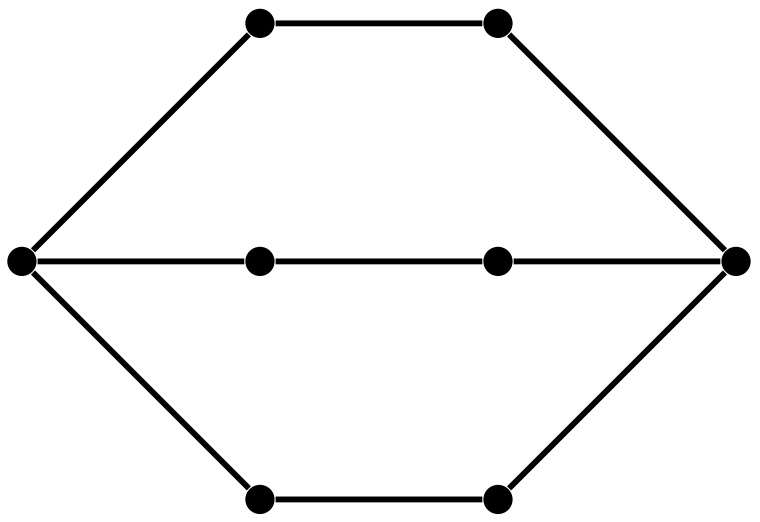
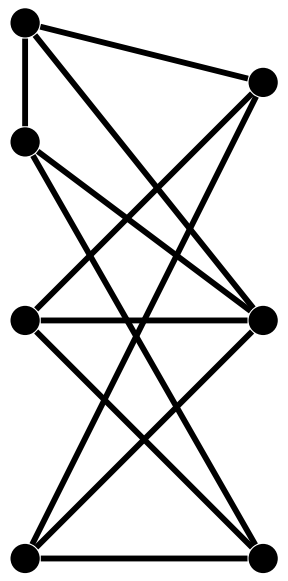
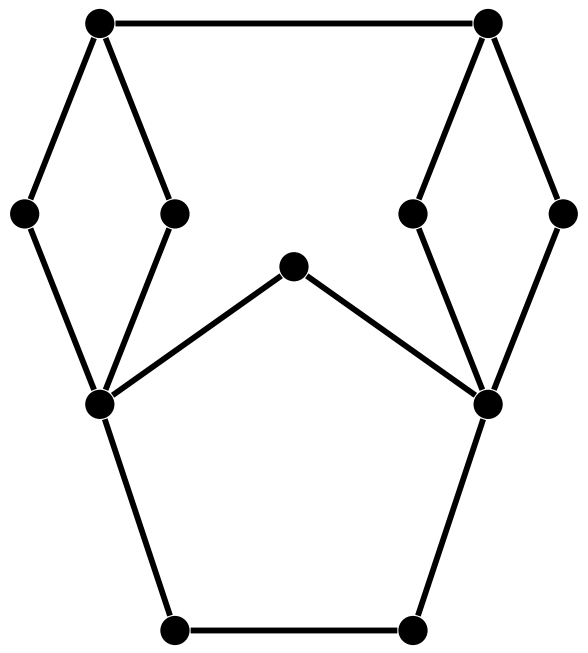

For patterns $H$ with $\subw(H) = \clemb(H)$, the PANDA framework~\cite{KhamisNS24} gives conditionally optimal algorithms for \listH and \enumH (via \cref{lm:all-reductions}) assuming the \cliquelistconj due to \cref{fct:all-H-iso-clique-hardness}.
Hence, by \cref{lm:ca-clemb-subw-octahedron}, $K_6 \setminus \PM_3$ is the smallest open problem in terms of the number of nodes for these versions of subgraph isomorphism.
See \cref{fig:open-problems:pm3-bar}.
Similarly, by \cref{lm:ca-clemb-subw-boat}, $P(3, 3, 3)$ is the smallest open problem in terms of the number of edges, and it also has the smallest values of $\clemb$ and $\subw$ among all open problems.
See \cref{fig:open-problems:p333}.
Historically, \cite[Lemma 2.7]{BG24} showed that \listiso{P(3, 3, 3)} and \enumiso{P(3, 3, 3)} are quadratically hard assuming the 3SUM Hypothesis.
Since this hypothesis was recently refuted~\cite{AlmanW26}, this is no longer evidence of hardness.

We also point out two further open problems.
They were open already before the 3SUM Hypothesis was refuted, so faster algorithms for them do not necessarily require the new techniques of~\cite{AlmanW26,ChanIMY26}.

\begin{calemma}\label{lm:ca-open-problems}
    The pattern $H_1$ depicted in \cref{fig:open-problems:min-edges} satisfies $\clemb(H_1) = 9 / 4 < 16 / 7 = \subw(H_1)$.
    The pattern $H_2$ depicted in \cref{fig:open-problems:min-exponent} satisfies $\clemb(H_2) = 2 + 1 / 6 < 2 + 3 / 16 = \subw(H_2)$.
\end{calemma}

\begin{proof}
    Our tool computes $\clemb$ and $\subw$ of both patterns.
    This can be verified by running the following command for the edge list of each pattern:
\begin{verbatim}
python width-calculator.py --clemb --subw -a "[edge list]"
\end{verbatim}
    (Note that running this command for $H_2$ may take several minutes.)
\end{proof}

Optimal time complexities for \countH are wide open due to the existence of fast-matrix-multiplication-based algorithms for this problem variant.
For combinatorial counting, \cref{fct:all-H-iso-clique-hardness} rules out a combinatorial algorithm in time $\Oh(m^{\clemb(H) - \Omega(1)})$ for any pattern $H$ under the \combcliqueconj.
Consequently, according to \cite[Theorem~B.1]{BG24}, the smallest open problem for combinatorial counting in terms of the time complexity exponent is \countiso{P(3, 3, 3)}.
See \cref{fig:open-problems:p333}.
Note that even before the 3SUM Hypothesis was refuted, no quadratic hardness for \countiso{P(3, 3, 3)} was known.

Furthermore, \cref{sec:examples:subw-vs-sharpw} proves
\[
    \sharpw(P(9,2,2))\ge2-1/9>2-1/8=\subw(P(9,2,2)).
\]
Nevertheless, \cite[Lemma 3.1]{BG24} combined with \cref{fct:counting-reduction} shows that \countiso{P(9, 2, 2)} can be solved combinatorially in time $\Oh(m^{2 - 1 / 8})$.
The same phenomenon occurs for the pattern $H$ depicted in \cref{fig:subw-vs-sharpw:min-vert}: it has $\sharpw = 2 + 3 / 11 > 2 + 1 / 4 = \subw$, but \countH can still be solved in time $\tOh(m^{2 + 1 / 4})$: in \cref{lm:better-counting-algo-for-specific-graph} in \cref{sec:appendix-m-width}, we give an algorithm that solves \Henciso in this time, which suffices by \cref{fct:counting-reduction}.
We are not aware of any patterns $H$ for which known techniques do not suffice to solve \countH combinatorially in time $\tOh(m^{\subw(H)})$.
We leave it as an exciting open problem to find a general algorithm for \encH (and thus for \countH) working in time $\tOh(m^{\subw(H)})$ or to show a counterexample.

\paragraph*{AI Disclosure}
The majority of the work on this project was done in 2024 without the use of AI.
This includes almost all of the code and the first version of the paper.
In later stages of the project, Large Language Models (GPT-5.5 and Claude Opus 5.5) were used to fix some bugs in the code and to make further minor improvements.
For the paper, AI was used to help turn the first draft into the final version.
In \cref{sec:preliminaries} and the subsequent sections, it was used for rewriting and for minor simplifications of proofs.
For \cref{lm:tw-vs-clemb,lm:expander}, it proposed new citations that allowed us to improve the results of these lemmas.
For the introduction, the authors wrote a high-level draft, AI filled in the details, and the authors subsequently edited the result extensively.
All figures were drawn with the help of AI.
The authors verified the correctness of all AI-generated content and take full responsibility for the contents of the paper.

\bibliography{refs}

\appendix

\section{Deferred Proofs from \texorpdfstring{\cref{sec:m-width,sec:high-low,sec:examples}}{Sections 3, 4, and 9}} \label{sec:appendix-m-width}

\mequivthree*

\begin{proof}
    We compare the LP from \cref{def:mwidth_general}, which is used for \cref{def:mwidth1}, with the two forest increments from \cref{lm:forest-increments}.
    Write $h=|V(H)|$.
    We first prove an upper bound $\mwidth(H)\le\dslogwidth(H)$ and then a lower bound $\mwidth(H)\ge\dslogwidth(H)$.

    \subparagraph{Upper bound.}
    Fix $m\ge2$, a host graph $G\in\cG_m(H)$, and a degree configuration $c$.
    We may assume that every edge relation of $G(c)$ is nonempty, since otherwise $q_{G(c)}=0$.
    For every $a\in V(H)$, let
    \[
        D_a\coloneqq\max_{e\ni a}2^{c(e,a)},
        \qquad d_a\coloneqq\log_m D_a\in[0,1],
        \qquad \varepsilon_m\coloneqq\log_m2.
    \]
    Choose an edge $e$ attaining the maximum in the expression defining $D_a$.
    Every node in $V_a(c)$ has at least $D_a$ neighbors in the corresponding edge relation of the original graph $G$, so $|V_a(c)|D_a\le m$.
    In every other incident relation the degree in $G$ is less than $2D_a$.
    Passing to $G(c)$ can only decrease these degrees.
    Consequently,
    \[
        d(\emptyset,\set{a},E_e(G(c)))\le1-d_a,
        \qquad
        d(\set{b},\set{a,b},E_{\set{a,b}}(G(c)))\le d_b+\varepsilon_m.
    \]

    Let $s$ be any feasible solution of $\prog_{G(c)}(B)$.
    Condition~(iv) allows us to add an isolated node $a$ at cost at most $1-d_a$, or a leaf with neighbor $b$ at cost at most $d_b+\varepsilon_m$.
    Building a forest $F$ one node at a time therefore gives
    \[
        s_{V(F)}\le \val_d(F)+|E(F)|\varepsilon_m
        \le \val_d(F)+h\varepsilon_m.
    \]
    By monotonicity, the same bound holds for $s_B$ whenever $F$ covers $B$.
    Minimizing over covering forests yields
    \[
        q_{G(c)}(B)\le\hmfun_d(B)+h\varepsilon_m.
    \]
    Taking the corresponding widths and maximizing over $c$ and $G$ gives
    \[
        \max_{G\in\cG_m(H)}\mwidthH{H}(G)\le\dslogwidth(H)+h\log_m2.
    \]
    Letting $m\to\infty$ proves $\mwidth(H)\le\dslogwidth(H)$.

    \subparagraph{Lower bound.}
    Fix $d\in[0,1]^{V(H)}$ and let $m$ tend to infinity through powers of two.
    Set
    \[
        D_a\coloneqq2^{\floor{\log_2(m^{d_a})}},
        \qquad d'_a\coloneqq\log_mD_a,
        \qquad n_a\coloneqq m/D_a.
    \]
    All $D_a$ and $n_a$ are integers, and $0\le d_a-d'_a<\log_m2$.
    We construct a host graph $G_m$ with $n_a$ nodes in each part $V_a$.
    For an edge $e=\set{a,b}$, use the complete bipartite graph between $V_a$ and $V_b$ if $D_aD_b>m$.
    Otherwise, partition $V_a$ into $m/(D_aD_b)$ groups of size $D_b$ and $V_b$ into the same number of groups of size $D_a$, and put a complete bipartite graph between each pair of corresponding groups.

    Every node in $V_a$ then has exactly $\min\set{D_a,n_b}$ neighbors in $V_b$, and
    $|E_e(G_m)|=\min\set{m,n_an_b}\le m$.
    In particular, all nodes of each part have the same degree in each incident relation, so some degree configuration retains the whole graph.
    We now compute its nonzero extension bounds for an edge $e=\set{a,b}$.
    Every node of $V_a$ has a neighbor in $V_b$, so the projection of $E_e(G_m)$ onto $a$ consists of all $n_a = m/D_a$ nodes of $V_a$.
    By symmetry, every node of $V_b$ has exactly $\min\set{D_b,n_a}$ neighbors in $V_a$.
    Finally, $|E_e(G_m)| = \min\set{m, n_an_b}$.
    Therefore, we obtain
    \begin{align*}
        d(\emptyset,\set{a},E_e(G_m))&=\log_m n_a=1-d'_a,\\
        d(\set{b},e,E_e(G_m))&=\log_m\min\set{D_b,n_a}=\min\set{d'_b,1-d'_a},\\
        d(\emptyset,e,E_e(G_m))&=\log_m\min\set{m,n_an_b}=\min\set{1,2-d'_a-d'_b}.
    \end{align*}
    The first two lines are precisely the isolated-node and leaf increments of \cref{lm:forest-increments}.
    The last line adds no constraint: adding $a$ as an isolated node and then $b$ either as a leaf or as an isolated node costs
    $(1-d'_a)+\min\set{d'_a,1-d'_b}=\min\set{1,2-d'_a-d'_b}$.
    If one or both endpoints are already in $Z$, the corresponding increment is only smaller.
    These cases exhaust Condition~(iv).
    Therefore, by \cref{lm:forest-increments},
    \[
        q_{G_m}(B)=\hmfun_{d'}(B)
        \qquad\text{for every }B\subseteq V(H).
    \]

    Each forest cost is affine in $d$, and there are finitely many forests and tree decompositions.
    Hence $\funwidth{\hmfun_d}(H)$ is continuous in $d$.
    Since $d'\to d$, we obtain
    \[
        \mwidth(H)\ge
        \limsup_{\substack{m\to\infty\\m\text{ a power of }2}}\mwidthH{H}(G_m)
        \ge\funwidth{\hmfun_d}(H).
    \]
    Taking the supremum over $d$ proves $\mwidth(H)\ge\dslogwidth(H)$.
\end{proof}

We now prove \cref{lm:mwidth-submodular}.
By \cref{lm:m-equiv-3}, it suffices to show that \dslogwidth and the submodular formulation of \mwidth coincide.
We first show that the monotonic closures of the forest costs belong to $\msubmodset(H)$.

\begin{lemma}\label{lm:m-equiv-1}
    For every pattern $H$ and every $d\in[0,1]^{V(H)}$, we have $\hmfun_d \in \msubmodset(H)$.
\end{lemma}

\begin{proof}
    By \cref{lm:forest-increments}, $\hmfun_d$ is monotone, normalized, and edge-dominated.
    Let $Y$ be contained in an edge.
    Order its at most two nodes with those in $X$ first, and add the remaining nodes to $X$ in that order.
    If the first node $a$ needs to be added, its increment is at most $1-d_a=\hmfun_d(\set{a})$.
    When the second node $b$ needs to be added after $a$, the two increment bounds give
    \[
        \hmfun_d(Z\cup\set{b})-\hmfun_d(Z)\le\min\set{1-d_b,d_a}
        =\hmfun_d(\set{a,b})-\hmfun_d(\set{a})\qquad\text{for } a\in Z,\ b\notin Z.
    \]
    Summing these increments gives $\hmfun_d(X\cup Y)-\hmfun_d(X)\le \hmfun_d(Y)-\hmfun_d(X\cap Y)$.
    Hence $\hmfun_d\in\msubmodset(H)$.
\end{proof}

With \cref{lm:m-equiv-1} at hand, we can prove \cref{lm:mwidth-submodular}.

\mwidthsubmodular*

\begin{proof}
    By \cref{lm:m-equiv-3}, it suffices to show $\dslogwidth(H) = \mpwidth(H)$.
    Let $\psi\in\msubmodset(H)$ and set $d_a\coloneqq1-\psi(\set{a})$ for every $a \in V(H)$.
    Monotonicity and edge domination give $d_a\in[0,1]$.
    By \cref{lm:msub-increments}, we have $\psi(X\cup\set{a})\le\psi(X)+1-d_a$ for every $a \notin X$, and $\psi(X\cup\set{a})\le\psi(X)+d_b$ for every $a\notin X$ and $b\in X$ with $\set{a,b}\in E(H)$.
    Thus \cref{lm:forest-increments} yields $\psi\le\hmfun_d$ entrywise.
    Taking widths and then suprema proves $\mpwidth(H)\le\dslogwidth(H)$.
    Conversely, $\hmfun_d\in\msubmodset(H)$ for every $d\in[0,1]^{V(H)}$ by \cref{lm:m-equiv-1}, and taking the supremum over $d$ proves $\dslogwidth(H)\le\mpwidth(H)$.
\end{proof}

The following lemma determines the widths of the patterns $P(7, 2, 2)$ and $P(8, 2, 2)$, which we use in the proof of \cref{thm:ca-subw-sharpw-min-value}.

\begin{lemma}\label{m-width-of-p822}
    For every $L\in\set{7,8}$, we have $\mwidth(P(L,2,2))=\subw(P(L,2,2))=2-1/7$.
\end{lemma}

\begin{proof}
    The lower bounds follow from \cite[Lemma 2.8]{BG24} and \cref{lm:width-hierarchy}.
    It remains to prove the upper bounds on $\mwidth$.
    Fix $L \in \set{7, 8}$ and write $H = P(L, 2, 2)$.
    Let $a_0,a_1,\ldots,a_L$ be the nodes of the path of length $L$, and let $c_1,c_2$ be the internal nodes of the two paths of length two between $a_0$ and $a_L$.
    Fix arbitrary $d_v \in [0,1]$ for every $v \in V(H)$, and assume without loss of generality that $d_{c_1} \le d_{c_2}$.
    We consider multiple cases and in each case construct a tree decomposition whose every bag $B$ satisfies $\mfun_d(B) \le 13/7$.
    This suffices because $\hmfun_d \le \mfun_d$.

    If $d_{a_0} > 1/7$, delete $a_0$ from $H$, take a width-one tree decomposition of the resulting tree, and add $a_0$ to every bag.
    Every resulting bag can be covered by an edge and the isolated node $a_0$, and thus has value at most $1+(1-d_{a_0}) < 13/7$.
    The case $d_{a_L} > 1/7$ is symmetric.
    Hence, from now on assume
    \begin{equation}\label{eq:p822-endpoints}
        d_{a_0},d_{a_L} \le 1/7.
    \end{equation}

    If $d_{c_1} > 4/7$, then also $d_{c_2} > 4/7$.
    Delete $c_1$ and $c_2$, take a width-one tree decomposition of the resulting path, and add $c_1$ and $c_2$ to every bag.
    Covering each bag by its path edge and the two isolated nodes gives value at most
    $1+(1-d_{c_1})+(1-d_{c_2}) < 13/7$.
    We may therefore assume
    \begin{equation}\label{eq:p822-short-path}
        d_{c_1} \le 4/7.
    \end{equation}

    Suppose that $d_{a_i} \ge 2/7$ for some $i \in \fragmentcc{1}{L-1}$.
    Delete $a_i$ and use the tree decomposition with bags $\set{a_0,c_1,c_2}$ and $\set{a_L,c_1,c_2}$, together with the adjacent pairs on the two remaining parts of the long path.
    Add $a_i$ to every bag.
    A bag arising from a long-path edge has value at most $1+(1-d_{a_i}) \le 12/7$.
    Each of the other two bags can be covered by the path with center $a_0$ or $a_L$, respectively, and the isolated node $a_i$; by \eqref{eq:p822-endpoints}, its value is at most
    $1+d_{a_0}+1-d_{a_i} \le 13/7$, or symmetrically $1+d_{a_L}+1-d_{a_i} \le 13/7$.
    We may thus assume
    \begin{equation}\label{eq:p822-interior}
        d_{a_i} < 2/7 \qquad\text{for every } i \in \fragmentcc{1}{L-1}.
    \end{equation}

    Next suppose that $d_{a_i} > 1/7$ for some $i \in \set{1,2}$, and choose the smallest such $i$.
    Consider the path of bags
    \[
        \set{a_0,a_L,c_1,c_2},
        \quad \set{a_0,\ldots,a_i,a_L,c_1},
        \quad \set{a_i,a_{L-1},a_L},\ldots,
        \set{a_i,a_{i+1},a_{i+2}}.
    \]
    The first bag is covered by the path $a_Lc_1a_0c_2$ and has value at most
    $1+d_{c_1}+d_{a_0} \le 12/7$.
    The second is covered by the path $a_Lc_1a_0a_1\cdots a_i$ and has value at most
    $1+d_{c_1}+\sum_{j=0}^{i-1}d_{a_j} \le 1+4/7+i/7 \le 13/7$.
    Every remaining bag is covered by one edge and the isolated node $a_i$, and hence has value at most $1+(1-d_{a_i}) < 13/7$.
    The cases $d_{a_i}>1/7$ for some $i \in \set{L-2,L-1}$ are symmetric.
    We may therefore finally assume
    \begin{equation}\label{eq:p822-near-endpoints}
        d_{a_1},d_{a_2},d_{a_{L-2}},d_{a_{L-1}} \le 1/7.
    \end{equation}

    Consider the two-bag tree decomposition
    \[
        \set{a_0,a_1,\ldots,a_4,c_1,c_2},
        \qquad
        \set{a_4,a_5,\ldots,a_L,c_1,c_2}.
    \]
    Each bag induces a tree, which we use as its covering forest.
    By \eqref{eq:p822-endpoints}, \eqref{eq:p822-interior}, and \eqref{eq:p822-near-endpoints}, the first bag has value
    \[
        1+2d_{a_0}+d_{a_1}+d_{a_2}+d_{a_3} \le 13/7.
    \]
    The second bag has value
    \[
        1+2d_{a_L}+\sum_{j=5}^{L-1}d_{a_j} \le 13/7;
    \]
    for $L=8$, use $d_{a_5}<2/7$ and $d_{a_6},d_{a_7}\le1/7$, while for $L=7$ both terms in the sum are at most $1/7$.
    This covers all cases and proves $\mwidth(H) \le 13/7$.
\end{proof}

The following lemma shows that the pattern $H$ from \cref{fig:subw-vs-sharpw:min-vert}, which satisfies $\subw(H) = 2 + 1/4 < 2 + 3/11 = \sharpw(H)$, can nevertheless be counted in time $\tOh(m^{\subw(H)})$ (see \cref{sec:example:open-problems}).

\begin{lemma}\label{lm:better-counting-algo-for-specific-graph}
    For the pattern $H$ depicted in \cref{fig:subw-vs-sharpw:min-vert}, there is an algorithm that solves \Henciso in time $\tOh(m^{2 + 1 / 4})$.
\end{lemma}

\begin{proof}
    The graph $H$ consists of $K_{3,3}$ with parts $\{a_1,a_2,a_3\}$ and $\{b_1,b_2,b_3\}$, together with the path $b_2,c_1,c_2,b_3$.
    Assume $m\ge2$ and apply degree uniformization as described before \cref{def:mwidth5}.
    Fix one resulting graph $G_i$ and write $d_f=\log_m 2^{i_f}\in[0,1]$ for every $f\in V(H)$.
    Its part $\subp{f}$ has $\Oh(m^{1-d_f})$ nodes, each of degree $\Oh(m^{d_f})$ in $G_i$.
    Permuting $a_1,a_2,a_3$ if necessary, assume $d_{a_1}\le d_{a_2}\le d_{a_3}$.

    Write $S=d_{a_1}+d_{a_2}+d_{a_3}$.
    We first consider two tree decompositions and solve the problem for all graphs $G_i$ for which these tree decompositions are sufficient.
    The first tree decomposition has bags $\set{a_j,b_1,b_2,b_3}$ for $j\in[3]$ and $\set{b_2,c_1,c_2,b_3}$.
    Cover each of the first three bags either by its star, as in \cref{fig:count-td-1}, or by the path $b_2,a_j,b_3$ and the isolated node $b_1$, as in \cref{fig:count-td-2}.
    The last bag is covered by two edges.
    Choosing the cheaper forest for each bag gives total time
    \[
        \Oh\bigl(m^2+m^{\min\set{1+2d_{a_3},\,2+d_{a_3}-d_{b_1}}}\bigr).
    \]
    In the second decomposition, every bag contains $a_1,a_2,a_3$, together with either $b_1$ or one edge of the path $b_2,c_1,c_2,b_3$; see \cref{fig:count-td-3}.
    List that singleton or edge and choose the three $a$-nodes independently, in total time $\Oh(m^{4-S})$.
    If $\min\set{1+2d_{a_3},2+d_{a_3}-d_{b_1}, 4 - S}\le9/4$, we are done.
    Otherwise,
    \begin{equation}\label{eq:counting-remaining-degrees}
        d_{a_3}>5/8,\qquad d_{b_1}<d_{a_3}-1/4,\qquad S<7/4,
    \end{equation}
    and hence $d_{a_1}+d_{a_2}-d_{a_3}=S-2d_{a_3}<1/2$.

    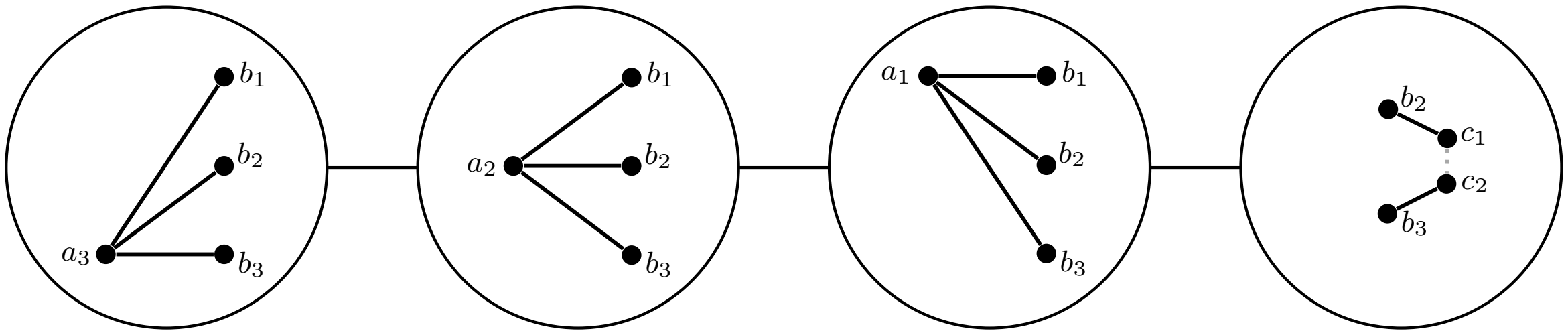
\begin{figure}
        \begin{center}
            \begin{tikzpicture}[scale=0.8,
        vtx/.style={circle, fill, inner sep=0pt, minimum size=5.5pt},
        lab/.style={inner sep=0pt, font=\footnotesize},
        forest/.style={line width=1.1pt},
        covered/.style={line width=1.1pt, black!34, dotted},
        bag/.style={line width=0.8pt}]
    \foreach \x in {0, 5.08, 10.16, 15.24} {\draw[bag] (\x,0) circle (1.98);}
    \foreach \x in {0, 5.08, 10.16} {\draw[bag] (\x+1.98,0) -- (\x+3.1,0);}
    \begin{scope}[shift={(0,0)}]
        \node[vtx] (a3) at (-0.75,-1.07) {}; \node[lab, anchor=base east] at (-0.91,-1.15) {$a_3$};
        \node[vtx] (b1) at (0.71,1.12) {}; \node[lab, anchor=base west] at (0.89,1.05) {$b_1$};
        \node[vtx] (b2) at (0.71,0.02) {}; \node[lab, anchor=base west] at (0.86,0.05) {$b_2$};
        \node[vtx] (b3) at (0.71,-1.07) {}; \node[lab, anchor=base west] at (0.87,-1.3) {$b_3$};
        \foreach \p/\q in {a3/b1, a3/b3, a3/b2} {\draw[forest] (\p) -- (\q);}
    \end{scope}
    \begin{scope}[shift={(5.08,0)}]
        \node[vtx] (a2) at (-0.8,0.02) {}; \node[lab, anchor=base east] at (-0.98,-0.05) {$a_2$};
        \node[vtx] (b1) at (0.66,1.11) {}; \node[lab, anchor=base west] at (0.84,1.05) {$b_1$};
        \node[vtx] (b2) at (0.66,0.02) {}; \node[lab, anchor=base west] at (0.81,0.04) {$b_2$};
        \node[vtx] (b3) at (0.66,-1.08) {}; \node[lab, anchor=base west] at (0.82,-1.3) {$b_3$};
        \foreach \p/\q in {a2/b2, a2/b1, a2/b3} {\draw[forest] (\p) -- (\q);}
    \end{scope}
    \begin{scope}[shift={(10.16,0)}]
        \node[vtx] (a1) at (-0.76,1.13) {}; \node[lab, anchor=base east] at (-0.95,1.08) {$a_1$};
        \node[vtx] (b1) at (0.7,1.13) {}; \node[lab, anchor=base west] at (0.88,1.06) {$b_1$};
        \node[vtx] (b2) at (0.7,0.03) {}; \node[lab, anchor=base west] at (0.84,0.06) {$b_2$};
        \node[vtx] (b3) at (0.7,-1.06) {}; \node[lab, anchor=base west] at (0.86,-1.28) {$b_3$};
        \foreach \p/\q in {a1/b1, a1/b2, a1/b3} {\draw[forest] (\p) -- (\q);}
    \end{scope}
    \begin{scope}[shift={(15.24,0)}]
        \node[vtx] (b2) at (-0.16,0.72) {}; \node[lab, anchor=base west] at (-0.02,0.75) {$b_2$};
        \node[vtx] (b3) at (-0.17,-0.57) {}; \node[lab, anchor=base west] at (-0.01,-0.79) {$b_3$};
        \node[vtx] (c1) at (0.57,0.36) {}; \node[lab, anchor=base west] at (0.72,0.33) {$c_1$};
        \node[vtx] (c2) at (0.56,-0.2) {}; \node[lab, anchor=base west] at (0.73,-0.26) {$c_2$};
        \foreach \p/\q in {b2/c1, c2/b3} {\draw[forest] (\p) -- (\q);}
        \foreach \p/\q in {c1/c2} {\draw[covered] (\p) -- (\q);}
    \end{scope}
\end{tikzpicture}
        \end{center}
        \caption{The first tree decomposition from \cref{lm:better-counting-algo-for-specific-graph}, with stars covering its first three bags. Here and later, solid black lines inside a bag represent the edges from the forest used to create a submaterialization of the bag, and dotted gray edges represent all the other edges that the bag covers.}
        \label{fig:count-td-1}
    \end{figure}
    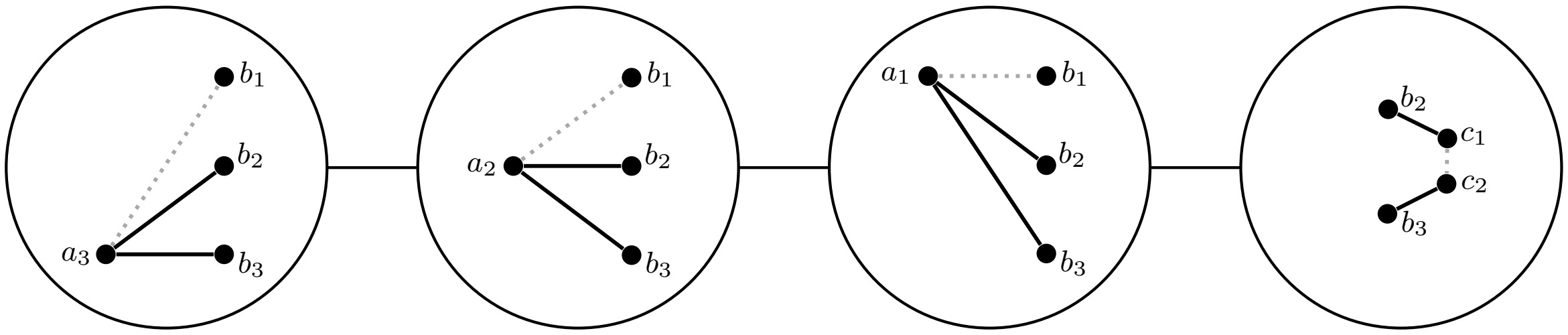
\begin{figure}
        \begin{center}
            \begin{tikzpicture}[scale=0.8,
        vtx/.style={circle, fill, inner sep=0pt, minimum size=5.5pt},
        lab/.style={inner sep=0pt, font=\footnotesize},
        forest/.style={line width=1.1pt},
        covered/.style={line width=1.1pt, black!34, dotted},
        bag/.style={line width=0.8pt}]
    \foreach \x in {0, 5.08, 10.16, 15.24} {\draw[bag] (\x,0) circle (1.98);}
    \foreach \x in {0, 5.08, 10.16} {\draw[bag] (\x+1.98,0) -- (\x+3.1,0);}
    \begin{scope}[shift={(0,0)}]
        \node[vtx] (a3) at (-0.75,-1.07) {}; \node[lab, anchor=base east] at (-0.91,-1.15) {$a_3$};
        \node[vtx] (b1) at (0.71,1.12) {}; \node[lab, anchor=base west] at (0.89,1.05) {$b_1$};
        \node[vtx] (b2) at (0.71,0.02) {}; \node[lab, anchor=base west] at (0.86,0.05) {$b_2$};
        \node[vtx] (b3) at (0.71,-1.07) {}; \node[lab, anchor=base west] at (0.87,-1.3) {$b_3$};
        \foreach \p/\q in {a3/b3, a3/b2} {\draw[forest] (\p) -- (\q);}
        \foreach \p/\q in {a3/b1} {\draw[covered] (\p) -- (\q);}
    \end{scope}
    \begin{scope}[shift={(5.08,0)}]
        \node[vtx] (a2) at (-0.8,0.02) {}; \node[lab, anchor=base east] at (-0.98,-0.05) {$a_2$};
        \node[vtx] (b1) at (0.66,1.11) {}; \node[lab, anchor=base west] at (0.84,1.05) {$b_1$};
        \node[vtx] (b2) at (0.66,0.02) {}; \node[lab, anchor=base west] at (0.81,0.04) {$b_2$};
        \node[vtx] (b3) at (0.66,-1.08) {}; \node[lab, anchor=base west] at (0.82,-1.3) {$b_3$};
        \foreach \p/\q in {a2/b2, a2/b3} {\draw[forest] (\p) -- (\q);}
        \foreach \p/\q in {a2/b1} {\draw[covered] (\p) -- (\q);}
    \end{scope}
    \begin{scope}[shift={(10.16,0)}]
        \node[vtx] (a1) at (-0.76,1.13) {}; \node[lab, anchor=base east] at (-0.95,1.08) {$a_1$};
        \node[vtx] (b1) at (0.7,1.13) {}; \node[lab, anchor=base west] at (0.88,1.06) {$b_1$};
        \node[vtx] (b2) at (0.7,0.03) {}; \node[lab, anchor=base west] at (0.84,0.06) {$b_2$};
        \node[vtx] (b3) at (0.7,-1.06) {}; \node[lab, anchor=base west] at (0.86,-1.28) {$b_3$};
        \foreach \p/\q in {a1/b2, a1/b3} {\draw[forest] (\p) -- (\q);}
        \foreach \p/\q in {a1/b1} {\draw[covered] (\p) -- (\q);}
    \end{scope}
    \begin{scope}[shift={(15.24,0)}]
        \node[vtx] (b2) at (-0.16,0.72) {}; \node[lab, anchor=base west] at (-0.02,0.75) {$b_2$};
        \node[vtx] (b3) at (-0.17,-0.57) {}; \node[lab, anchor=base west] at (-0.01,-0.79) {$b_3$};
        \node[vtx] (c1) at (0.57,0.36) {}; \node[lab, anchor=base west] at (0.72,0.33) {$c_1$};
        \node[vtx] (c2) at (0.56,-0.2) {}; \node[lab, anchor=base west] at (0.73,-0.26) {$c_2$};
        \foreach \p/\q in {b2/c1, c2/b3} {\draw[forest] (\p) -- (\q);}
        \foreach \p/\q in {c1/c2} {\draw[covered] (\p) -- (\q);}
    \end{scope}
\end{tikzpicture}
        \end{center}
        \caption{Alternative forests for the first tree decomposition from \cref{lm:better-counting-algo-for-specific-graph}.}
        \label{fig:count-td-2}
    \end{figure}
    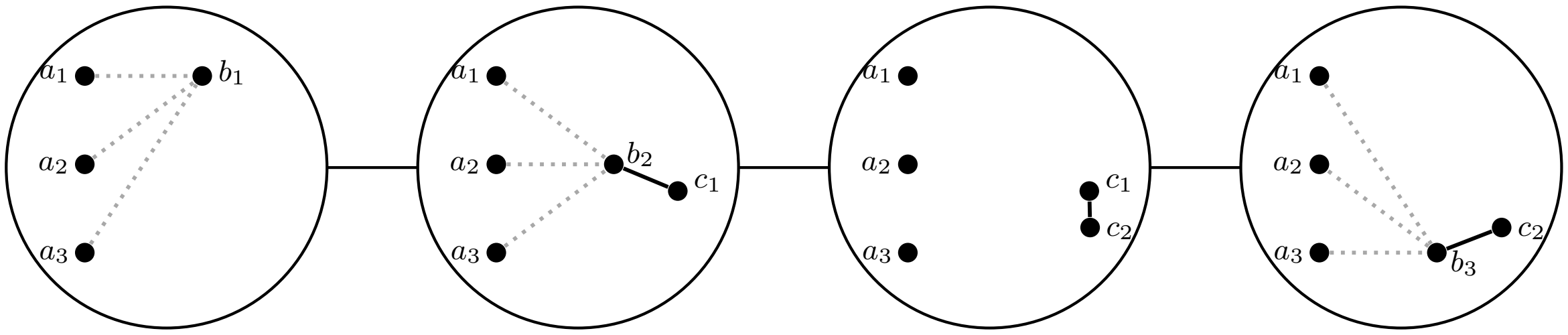
\begin{figure}
        \begin{center}
            \begin{tikzpicture}[scale=0.8,
        vtx/.style={circle, fill, inner sep=0pt, minimum size=5.5pt},
        lab/.style={inner sep=0pt, font=\footnotesize},
        forest/.style={line width=1.1pt},
        covered/.style={line width=1.1pt, black!34, dotted},
        bag/.style={line width=0.8pt}]
    \foreach \x in {0, 5.08, 10.16, 15.24} {\draw[bag] (\x,0) circle (1.98);}
    \foreach \x in {0, 5.08, 10.16} {\draw[bag] (\x+1.98,0) -- (\x+3.1,0);}
    \begin{scope}[shift={(0,0)}]
        \node[vtx] (a1) at (-1.01,1.13) {}; \node[lab, anchor=base east] at (-1.18,1.1) {$a_1$};
        \node[vtx] (a2) at (-1.01,0.04) {}; \node[lab, anchor=base east] at (-1.19,-0.03) {$a_2$};
        \node[vtx] (a3) at (-1.01,-1.05) {}; \node[lab, anchor=base east] at (-1.18,-1.13) {$a_3$};
        \node[vtx] (b1) at (0.44,1.13) {}; \node[lab, anchor=base west] at (0.63,1.07) {$b_1$};
        \foreach \p/\q in {a3/b1, a1/b1, a2/b1} {\draw[covered] (\p) -- (\q);}
    \end{scope}
    \begin{scope}[shift={(5.08,0)}]
        \node[vtx] (a1) at (-1.01,1.13) {}; \node[lab, anchor=base east] at (-1.18,1.1) {$a_1$};
        \node[vtx] (a2) at (-1.01,0.04) {}; \node[lab, anchor=base east] at (-1.19,-0.03) {$a_2$};
        \node[vtx] (a3) at (-1.01,-1.05) {}; \node[lab, anchor=base east] at (-1.18,-1.13) {$a_3$};
        \node[vtx] (b2) at (0.44,0.04) {}; \node[lab, anchor=base west] at (0.59,0.06) {$b_2$};
        \node[vtx] (c1) at (1.23,-0.29) {}; \node[lab, anchor=base west] at (1.42,-0.25) {$c_1$};
        \foreach \p/\q in {b2/c1} {\draw[forest] (\p) -- (\q);}
        \foreach \p/\q in {a3/b2, a1/b2, a2/b2} {\draw[covered] (\p) -- (\q);}
    \end{scope}
    \begin{scope}[shift={(10.16,0)}]
        \node[vtx] (a1) at (-1.01,1.13) {}; \node[lab, anchor=base east] at (-1.18,1.1) {$a_1$};
        \node[vtx] (a2) at (-1.01,0.04) {}; \node[lab, anchor=base east] at (-1.19,-0.03) {$a_2$};
        \node[vtx] (a3) at (-1.01,-1.05) {}; \node[lab, anchor=base east] at (-1.18,-1.13) {$a_3$};
        \node[vtx] (c1) at (1.23,-0.29) {}; \node[lab, anchor=base west] at (1.42,-0.25) {$c_1$};
        \node[vtx] (c2) at (1.24,-0.74) {}; \node[lab, anchor=base west] at (1.43,-0.84) {$c_2$};
        \foreach \p/\q in {c1/c2} {\draw[forest] (\p) -- (\q);}
    \end{scope}
    \begin{scope}[shift={(15.24,0)}]
        \node[vtx] (a1) at (-1.01,1.13) {}; \node[lab, anchor=base east] at (-1.18,1.1) {$a_1$};
        \node[vtx] (a2) at (-1.01,0.04) {}; \node[lab, anchor=base east] at (-1.19,-0.03) {$a_2$};
        \node[vtx] (a3) at (-1.01,-1.05) {}; \node[lab, anchor=base east] at (-1.18,-1.13) {$a_3$};
        \node[vtx] (b3) at (0.44,-1.05) {}; \node[lab, anchor=base west] at (0.6,-1.28) {$b_3$};
        \node[vtx] (c2) at (1.24,-0.74) {}; \node[lab, anchor=base west] at (1.43,-0.84) {$c_2$};
        \foreach \p/\q in {c2/b3} {\draw[forest] (\p) -- (\q);}
        \foreach \p/\q in {a3/b3, a1/b3, b3/a2} {\draw[covered] (\p) -- (\q);}
    \end{scope}
\end{tikzpicture}
        \end{center}
        \caption{The second tree decomposition from \cref{lm:better-counting-algo-for-specific-graph}, with $a_1,a_2,a_3$ isolated in each covering forest.}
        \label{fig:count-td-3}
    \end{figure}

    We now split by the number of common neighbors of a \emph{pair} of nodes, as in \cite[Lemma 3.1]{BG24}.
    Materialize the path $b_2,a_1,b_3$ and group its tuples by the pair in $\subp{b_2}\times\subp{b_3}$.
    This takes time $\tOh(m^{1+d_{a_1}})\subseteq\tOh(m^2)$ and gives a dictionary of all common neighbors in $\subp{a_1}$.
    A pair is \emph{low-degree} if it has fewer than $\tau\coloneqq m^{d_{a_1}-1/4}$ such neighbors, and \emph{high-degree} otherwise.
    Each low-degree pair has fewer than $\tau$ choices for its $a_1$-node, and there are $\Oh(m^{1+d_{a_1}}/\tau)\subseteq\Oh(m^{1+1/4})$ high-degree pairs.
    We construct one partial encoding for each of the two cases.

    \begin{enumerate}
        \item \textbf{Low-degree pairs.}
            Use bags $\set{a_1,a_2,a_3,b_1}$, $\set{a_1,a_2,a_3,b_2,b_3}$, and $\set{b_2,c_1,c_2,b_3}$, connected in a path as in \cref{fig:count-td-4}.
            Cover the first bag by the path $a_1,b_1,a_2$ and the isolated node $a_3$, in time $\Oh(m^{2+d_{b_1}-d_{a_3}})\subseteq\Oh(m^{7/4})$ by \eqref{eq:counting-remaining-degrees}.
            For the middle bag, generate the paths $b_2,a_2,b_3$ in time $\Oh(m^{1+d_{a_2}})$.
            For every low-degree pair $(v_{b_2}, v_{b_3})$ generated this way, append its fewer than $\tau$ common neighbors in $\subp{a_1}$ and all $\Oh(m^{1-d_{a_3}})$ choices in $\subp{a_3}$.
            Including the time to generate the initial path tuples, the cost is
            \[
                \tOh\bigl(m^{1+d_{a_2}}+m^{(1+d_{a_2})+(d_{a_1}-1/4)+(1-d_{a_3})}\bigr)
                =\tOh\bigl(m^{1+d_{a_2}}+m^{7/4+S-2d_{a_3}}\bigr)
                \subseteq\tOh(m^{2+1/4}),
            \]
            because $S-2d_{a_3}<1/2$.
            The last bag $\{b_2,c_1,c_2,b_3\}$ again takes time $\Oh(m^2)$.
            Restricting the middle bag to low-degree pairs makes this partial encoding contain exactly the low-degree $H$-subgraphs.
            \begin{figure}
                \begin{center}
                    \begin{tikzpicture}[scale=0.8,
        vtx/.style={circle, fill, inner sep=0pt, minimum size=5.5pt},
        lab/.style={inner sep=0pt, font=\footnotesize},
        forest/.style={line width=1.1pt},
        covered/.style={line width=1.1pt, black!34, dotted},
        bag/.style={line width=0.8pt}]
    \foreach \x in {0, 5.08, 10.16} {\draw[bag] (\x,0) circle (1.98);}
    \foreach \x in {0, 5.08} {\draw[bag] (\x+1.98,0) -- (\x+3.1,0);}
    \begin{scope}[shift={(0,0)}]
        \node[vtx] (a1) at (-1.01,1.13) {}; \node[lab, anchor=base east] at (-1.18,1.1) {$a_1$};
        \node[vtx] (a2) at (-1.01,0.04) {}; \node[lab, anchor=base east] at (-1.19,-0.03) {$a_2$};
        \node[vtx] (a3) at (-1.01,-1.05) {}; \node[lab, anchor=base east] at (-1.18,-1.13) {$a_3$};
        \node[vtx] (b1) at (0.44,1.13) {}; \node[lab, anchor=base west] at (0.63,1.07) {$b_1$};
        \foreach \p/\q in {a1/b1, a2/b1} {\draw[forest] (\p) -- (\q);}
        \foreach \p/\q in {a3/b1} {\draw[covered] (\p) -- (\q);}
    \end{scope}
    \begin{scope}[shift={(5.08,0)}]
        \node[vtx] (a1) at (-0.81,1.13) {}; \node[lab, anchor=base east] at (-0.97,1.09) {$a_1$};
        \node[vtx] (a2) at (-0.81,0.04) {}; \node[lab, anchor=base east] at (-0.98,-0.03) {$a_2$};
        \node[vtx] (a3) at (-0.81,-1.06) {}; \node[lab, anchor=base east] at (-0.97,-1.13) {$a_3$};
        \node[vtx] (b2) at (0.65,0.04) {}; \node[lab, anchor=base west] at (0.8,0.06) {$b_2$};
        \node[vtx] (b3) at (0.65,-1.06) {}; \node[lab, anchor=base west] at (0.81,-1.28) {$b_3$};
        \foreach \p/\q in {a2/b2, a2/b3} {\draw[forest] (\p) -- (\q);}
        \foreach \p/\q in {a1/b2, a1/b3, a3/b2, a3/b3} {\draw[covered] (\p) -- (\q);}
        \draw[line width=0.4pt, rotate around={1:(0.76,-0.51)}] (0.76,-0.51) ellipse [x radius=0.63, y radius=1.03];
        \draw[line width=0.8pt, red!54.5!black, -{Triangle[length=6pt, width=4.4pt]}] (0.66,0.51) .. controls (0.38,1.08) and (-0.08,1.3) .. (-0.7,1.16);
    \end{scope}
    \begin{scope}[shift={(10.16,0)}]
        \node[vtx] (b2) at (-0.15,0.74) {}; \node[lab, anchor=base west] at (0,0.76) {$b_2$};
        \node[vtx] (b3) at (-0.15,-0.59) {}; \node[lab, anchor=base west] at (0.01,-0.81) {$b_3$};
        \node[vtx] (c1) at (0.64,0.4) {}; \node[lab, anchor=base west] at (0.82,0.45) {$c_1$};
        \node[vtx] (c2) at (0.64,-0.27) {}; \node[lab, anchor=base west] at (0.84,-0.37) {$c_2$};
        \foreach \p/\q in {b2/c1, c2/b3} {\draw[forest] (\p) -- (\q);}
        \foreach \p/\q in {c1/c2} {\draw[covered] (\p) -- (\q);}
    \end{scope}
\end{tikzpicture}
                \end{center}
                \caption{Tree decomposition for low-degree pairs $(v_{b_2}, v_{b_3})$ from \cref{lm:better-counting-algo-for-specific-graph}.}
                \label{fig:count-td-4}
            \end{figure}
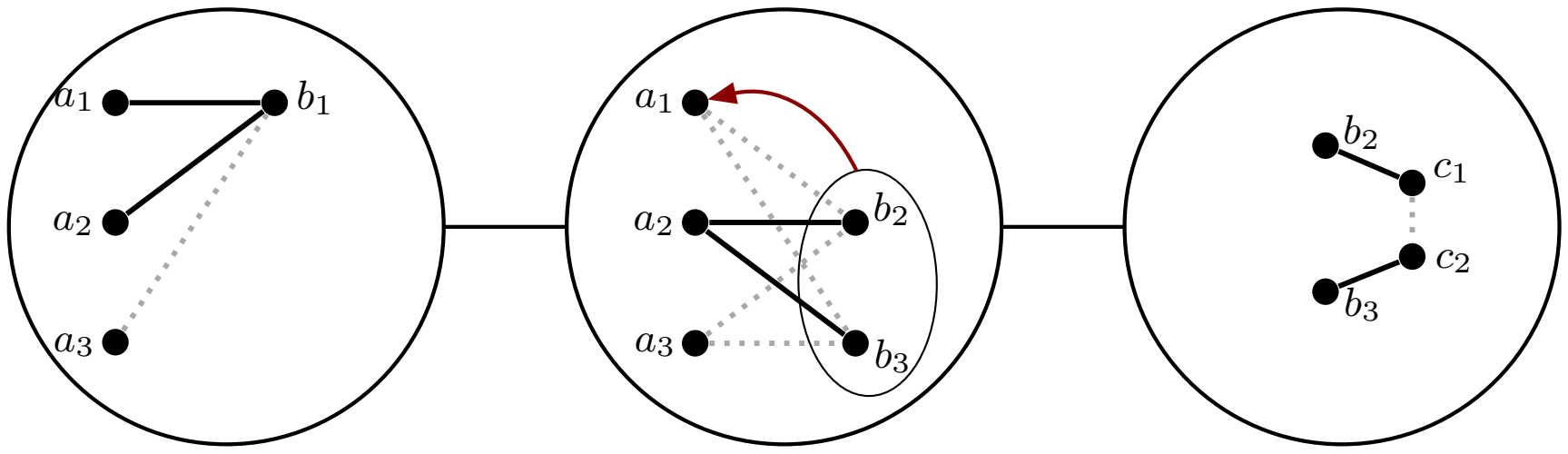
        \item \textbf{High-degree pairs.}
            The graph $H-\{b_2,b_3\}$ is a forest.
            Take a tree decomposition with one bag for each of the forest edges and add $b_2$ and $b_3$ to every bag.
            Materialize each bag by taking the product of the $\Oh(m^{1+1/4})$ high-degree pairs and the $\Oh(m)$ host edges for the other two nodes.
            This takes time $\Oh(m^{2+1/4})$ per bag and encodes exactly the high-degree $H$-subgraphs.
    \end{enumerate}

    Thus each $G_i$ has an encoding computable in time $\tOh(m^{2+1/4})$.
    Every $H$-subgraph lies in exactly one degree-uniformized graph $G_i$ and, in the last case, has exactly one of the two kinds of pairs.
    Taking the union of the encodings therefore gives a full $H$-encoding of $G$.
    There are only $\Oh(\log^{|V(H)|} m)$ graphs $G_i$, so the total time is $\tOh(m^{2+1/4})$.
\end{proof}

\end{document}